\documentclass[prd,showpacs,altaffilletter,twocolumn,
groupaddress,floatfix,nofootinbib]{revtex4} 

\usepackage{amsfonts,amsmath,graphicx,bbold}

\usepackage{ifpdf}

\def\sfrac#1#2{{\textstyle\frac{#1}{#2}}}
\newcommand{\vnr}{\ensuremath{v_{\operatorname{rel}}}}
\newcommand{\LQCD}{\Lambda_{\mathrm{QCD}}}
\newcommand{\dgamma}{\hat{\gamma}}
\newcommand{\bs}[1]{\ensuremath{{\boldsymbol{#1}}}}

\newcommand{\vegas}{\textsc{vegas}}
\newcommand{\hippy}{\textsc{HiPPy}}
\newcommand{\hpsrc}{\textsc{HPsrc}}
\newcommand{\taylur}{\textsc{TaylUR}}

\newcommand{\cambridge}{Department of Applied Mathematics and
Theoretical Physics, University of Cambridge, Centre for Mathematical
Sciences, Cambridge CB3 0WA, United Kingdom}

\newcommand{\edinburgh}{SUPA, School of Physics and Astronomy,
University of Edinburgh, King's Buildings, Mayfield Road, Edinburgh
EH9 3JZ, United Kingdom}

\newcommand{\cornell}{Laboratory for Elementary-Particle Physics,
Cornell University, Ithaca, NY 14853, U.S.A.}

\newcommand{\zeuthen}{Deutsches Elektronen-Synchroton DESY,
Platanenallee 6, 15738 Zeuthen, Germany}

\newcommand{\glasgow}{SUPA, Department of Physics and Astronomy,
University of Glasgow, Glasgow G12 8QQ, United Kingdom}

\begin{document}

\title{Moving NRQCD for heavy-to-light form factors on the lattice}

\author{R.~R.~\surname{Horgan}} 
\author{L.~\surname{Khomskii}}
\author{S.~\surname{Meinel}}
\author{M.~\surname{Wingate}}
\affiliation{\cambridge}

\author{K.~M.~\surname{Foley}}
\author{G.~P.~\surname{Lepage}}
\affiliation{\cornell}

\author{G.~M.~\surname{von Hippel}}
\affiliation{\zeuthen}

\author{A.~\surname{Hart}}
\author{E.~H.~\surname{M\"{u}ller}}
\affiliation{\edinburgh}

\author{C.~T.~H.~\surname{Davies}}
\author{A.~\surname{Dougall}}
\author{K.~Y.~\surname{Wong}}
\affiliation{\glasgow}

\collaboration{HPQCD Collaboration}
\noaffiliation

\pacs{12.38.Bx, 12.38.Gc, 12.39.Hg, 13.20.He, 14.40.Nd, 14.65.Fy}
\preprint{DAMTP-2008-113}
\preprint{Edinburgh 2008/27}

\begin{abstract} 
We formulate Non-Relativistic Quantum Chromodynamics (NRQCD) on a lattice
which is boosted relative to the usual discretization frame.  Moving NRQCD
(mNRQCD) allows us to treat the momentum for the heavy quark arising from the
frame choice exactly.  We derive mNRQCD through $\mathcal{O}(1/m^2,\vnr^4)$,
as accurate as the NRQCD action in present use, both in the continuum and on
the lattice with $\mathcal{O}(a^4)$ improvements.  We have carried out
extensive tests of the formalism through calculations of two-point correlators
for both heavy-heavy (bottomonium) and heavy-light ($B_s$) mesons in 2+1
flavor lattice QCD and obtained nonperturbative determinations of energy shift
and external momentum renormalization.  Comparison to perturbation theory at
$\mathcal{O}(\alpha_s)$ is also made.  The results demonstrate the
effectiveness of mNRQCD.  In particular we show that the decay constants of
heavy-light and heavy-heavy mesons can be calculated with small systematic
errors up to much larger momenta than with standard NRQCD.
\end{abstract}

\maketitle

\section{Introduction}
\label{sec:intro}

The Cabibbo-Kobayashi-Maskawa (CKM) matrix is the focus of
intense study; an inconsistency between independent determinations of
CKM matrix elements from different physical processes would be
evidence for new physics beyond the Standard Model. While experimental
measurements of exclusive semileptonic decays have reached good
precision and will be improved further by LHCb, determinations of CKM
matrix elements from the decay rates are complicated by the need for
precise theoretical calculations in nonperturbative quantum
chromodynamics (QCD). Lattice QCD provides a first-principles approach
to these calculations and it is important to reduce systematic and
statistical errors as far as possible.

For example, the decay $B \rightarrow \pi \ell\nu$
\cite{Athar:2003yg,Hokuue:2006nr,Aubert:2006px} can be used to determine the
CKM matrix element $V_{ub}$ while the rare decays $B \rightarrow
K^*\gamma,~K^{(*)}\ell^+\ell^-$
\cite{Nakao:2004th,Yarritu:2008cy,Eisenhardt:2008zz,Bachmann:2008zz,Hicheur:2008hm}
provide excellent opportunities to study contributions from new physics, as
the flavor-changing neutral current $b\rightarrow s$ is
loop-suppressed in the Standard Model. In both cases, a nonperturbative
calculation of the hadronic form factors is required.

These form factors are a function of the momentum transfer squared, $q^2$,
where $q=p_B-p_F$ is the difference between the four-momenta of the $B$ meson
and the meson in the final state. If this meson is light compared to the $B$
meson, the recoil momentum at small values of $q^2$ can be very large.
Unfortunately, current lattice QCD calculations of these form
factors work well only for low recoil momenta, \textit{i.e.}\ large $q^2$
\cite{Dalgic:2006dt,Shigemitsu:2004ft,Okamoto:2004xg,Bailey:2008wp},
while for $B \rightarrow K^*\gamma$ one has $q^2=0$ and experimental data
for $B\rightarrow \pi \ell \nu$ covers the full $q^2$ range
\cite{Athar:2003yg,Hokuue:2006nr,Aubert:2006px}.

By computing at just one or a few points with large $q^2$, one might be able
to reduce the error on $|V_{ub}|$ from $B\to\pi\ell\nu$, where the shape of
the form factor is now being measured precisely by experiment \cite{Athar:2003yg,Hokuue:2006nr,Aubert:2006px}.
However, the form factors governing the rare $b\to s$ decays
are not well-determined and must be computed using lattice QCD.  Given the
propensity for models of new physics to introduce new sources of
flavor-changing neutral currents, it is desirable to have new tools
to reduce the errors on the Standard Model calculations of
differential cross sections for rare decays.

In this paper we present a technique for extending lattice QCD calculations of the decays of mesons
containing one heavy quark to lower $q^2$ values than has hitherto been possible
by reducing the discretization errors owing to the large recoil of the final state meson.

The formalism that we describe and put to the test in subsequent sections is a
generalization of Non-Relativistic QCD (NRQCD)
\cite{Caswell:1985ui,Lepage:1992tx}.
The NRQCD formalism, which has already had considerable success in the study
of heavy-quark systems, relies on the fact that fluctuations in the heavy
quark momentum within a heavy meson are small compared with the mass of the
meson itself. The Lagrangian of NRQCD is expressed as a sum over operators
whose importance is governed by power-counting rules; in dimensionless units
the operators are ordered in powers of $g, \vnr$ for heavy-heavy mesons and in
powers of $\alpha_s, \LQCD/m$ for heavy-light mesons, where $\alpha_s =
g^2/(4\pi)$ is the strong coupling constant, $\vnr \sim |\bs{p}|/m$ is the
relative internal velocity of the heavy quarks and $m$ is the heavy quark
mass.\footnote{These rules are frequently referred to as NRQCD and HQET power
  counting schemes, respectively.  Note that the choice of NRQCD as a lattice
  action is compatible with both schemes.  See Sec.~\ref{sec:power-counting}
  below.}  For NRQCD, the heavy meson is usually taken to be at rest in the
lattice frame. This is appropriate for calculations of the mass spectrum of
heavy-light and heavy-heavy mesons and for zero-recoil or low-recoil
decays. However, for the heavy-to-light decays of the $B$-meson cited above,
outside the low recoil region the momentum of the light meson in the final
state becomes comparable to the inverse lattice spacing.  Consequently the
calculation is sensitive to lattice artifacts which lead to large systematic
errors.

It is therefore better to give the $B$ meson a non-zero momentum in the
opposite direction, thereby reducing the final meson's momentum at a given
$q^2$. To substantially reduce the momentum of the final meson, the momentum
of the $B$ meson has to be very large, so that NRQCD would no longer be able
to describe the $b$ quark inside it due to relativistic and lattice spacing
errors.  However, we note that fluctuations of momentum of the heavy quark
inside the $B$ meson are much smaller than the momentum of the meson
itself. Therefore, to reduce errors, instead of discretizing the momentum of
the $b$ quark itself, we choose to discretize its fluctuations inside the
moving $B$ meson. The formalism which achieves this goes by the name of moving
NRQCD (mNRQCD) in which the expansion is about the state where the heavy quark
is moving with a velocity $v$, the frame velocity; this formalism was
introduced briefly in
\cite{Sloan:1997fc};
%
%
%
Earlier, related approaches were proposed in 
\cite{Mandula:1991ds,Hashimoto:1995in}.

The remainder of the paper is structured as follows. In
Section~\ref{sec:frame-choice} we discuss the choice of the optimal reference
frame for the lattice calculations. We give an explicit derivation of the
continuum mNRQCD action in Section~\ref{sec:mnrqcd_derivation}. We explain how
the theory is discretized in Section~\ref{sec:lattice-mnrqcd}. In
Section~\ref{sec:renorm} we develop perturbative methods for mNRQCD and
explain how to derive the renormalization of parameters due to radiative
corrections. We give 1-loop results for the heavy quark renormalization
constants. The construction of decay currents is discussed in section
\ref{sec:current_construction}.
Then, in section~\ref{sec:simulation_results} we present the results of
nonperturbative calculations based on two-point correlators for heavy-heavy
and heavy-light mesons in mNRQCD. These include the spectrum, renormalization
constants and decay constants for various values
of the frame velocity $v$.

The perturbative and nonperturbative renormalization constants are compared in
Section~\ref{sec:comparison_NP_PT}.
We summarize and discuss our results in
Section~\ref{sec:conclusions}.

In the Appendices we specify some notation (Appendix~\ref{app:notation}),
describe the removal of time derivatives in
the $\mathcal{O}(\LQCD^2/m^2)$ mNRQCD Hamiltonian (Appendix~\ref{app:2nd_time_der_removal}),
give explicit expressions for the lattice derivative operators
(Appendix~\ref{app:latt_deriv}) and tadpole improvement corrections
(Appendix~\ref{app:tadpoles}) and present further perturbative results for a
set of simpler actions (Appendix~\ref{app:results-simple-actions}). We
comment on the poles of the Symanzik-improved gluon action in
Appendix~\ref{app:improved_gluon_poles}.

Preliminary versions of this work have been presented in
Refs.~\cite{Foley:2002qv,Dougall:2004hw,Dougall:2005zh,Foley:2005fx,
Meinel:2007eh,Meinel:2008th}.
%
%
%
%
\section{Minimizing Errors}
\label{sec:frame-choice}

We start by parametrizing the 4-momentum of the $b$ quark as
\begin{equation*}
p = m\: u+k
\end{equation*}
where $m$ is the mass of the $b$-quark, and $u$ a 4-velocity.  In traditional
(non-moving) NRQCD one has $u = (1,0,0,0)$, and a non-relativistic expansion
in the residual 3-momentum $\bs{k}$ is performed.  In other words, the
heavy-quark mass term is removed from the Lagrangian. Thus, the 3-momentum
$\bs{p}$, which is equal to $\bs{k}$ in this case, has to be small to prevent
large relativistic errors as well as discretization errors on the lattice.

In moving NRQCD, we generalize this to other frames of reference, removing the
momentum $m\, u$ with an arbitrary 4-velocity $u$ from the Lagrangian, and
again performing a non-relativistic expansion in the residual 3-momentum
$\bs{k}$. The relativistic energy of the heavy quark is
$E=\sqrt{\mathstrut{\bs{p}}^2+m^2}=\sqrt{(m\bs{u}+\bs{k})^2+m^2}$.  Taylor
expansion for small $|\bs{k}|$ gives
\begin{equation*}
E= \gamma m+\bs{v}\cdot\bs{k}
+\frac{\bs{k}^2-(\bs{v}\cdot\bs{k})^2}{2\gamma m}+\cdots
\end{equation*}
where we write $u=(u_0,\bs{u})=(\gamma,\gamma\bs{v})$ with the 3-velocity
$\bs{v}$ and $\gamma=(1-\bs{v}^2)^{-1/2}$. Discarding the constant term
$\gamma m$, we expect that the $\mathcal{O}(1/m)$ ``kinetic'' part of the
continuum mNRQCD Hamiltonian in momentum space will be given by
\begin{equation*}
H_0=\bs{v}\cdot\bs{k}+\frac{\bs{k}^2-(\bs{v}\cdot\bs{k})^2}{2\gamma m}.
\end{equation*}
Of course, the size of $\bs{k}$ and the associated relativistic and
discretization errors depend on the choice of $u$. The standard choice is
$u=p_B/M_B$, the 4-velocity of the $B$ meson.  Then, the residual momentum
$\bs{k}$ is small compared to $\bs{p}_B$ and the non-relativistic expansion in
$\bs{k}$ is a good approximation even for $B$ mesons at moderately high
velocities.

\subsubsection{Discretization errors}

One of the main applications of the mNRQCD approach is to the heavy-to-light
weak decay of a $B$-meson to a final state including a light meson. As
discussed in the introduction the size of the discretization errors in a
lattice calculation depends on the momentum of the final state meson; states
with spatial momenta comparable to the inverse lattice spacing can be affected
by lattice artifacts.  Nevertheless, one wishes to compute matrix elements
over the whole physical kinematic range, including the large recoil regime
where the final state has large momentum relative to the $B$ meson.  With
mNRQCD we attempt to reduce discretization errors by choosing a non-zero frame
velocity $v$, so that the final state meson can have moderate spatial momentum
in the lattice frame, even as we explore large recoil kinematics.
  
If the $B$ meson is at rest, the residual momentum $\bs{k}$ has a distribution
with width of the order $\LQCD$ and the residual energy has a distribution
with width of the order $\LQCD^2/(2m)\ll\LQCD$. Note that the momentum
$\bs{p}_\mathrm{spec}$ of the light quark in the $B$~meson (the ``spectator
quark'') is of the same order by momentum conservation.

For a $B$ meson moving with velocity $\bs{v}$, the momentum distribution is
boosted to approximately $\gamma \LQCD$.  Let us now consider a decay
$B\rightarrow F$ where $F$ denotes the light meson in the final state and the
4-momenta are
\begin{eqnarray*}
p_B&=&(\gamma M_B, \:\gamma M_B \bs{v}),\\
p_F&=&(\sqrt{M_F^2+|\bs{p}_F|^2}, \:\bs{p}_F)
\end{eqnarray*}
where $\bs{v}$ is antiparallel to $\bs{p}_F$. For a given value of
$q^2=(p_B-p_F)^2$, we shall determine the optimal velocity of the $B$ meson
which minimizes discretization errors. The discretization errors are
determined by the momenta carried by the quarks (and gluons) and are typically
proportional to $(a\times\mathrm{momentum})^2$ where $a$ is the lattice
spacing.  The full mNRQCD action described in this paper has no tree-level
$\mathcal{O}(a^2\bs{k}^2)$-errors, but has $\mathcal{O}(\alpha_s a^2\bs{k}^2)$
errors due to radiative corrections. The same is true for highly improved
light quark actions such as ASQTAD
\cite{Orginos:1998ue,Lepage:1998vj,Orginos:1999cr} 
or HISQ 
\cite{Follana:2006rc}.
Assuming that the constants of proportionality for the discretization errors
are the same, discretization errors are minimal if all quarks involved in the
decay have momenta of the same size.

\begin{figure}
\begin{center}
\includegraphics[width=\linewidth]{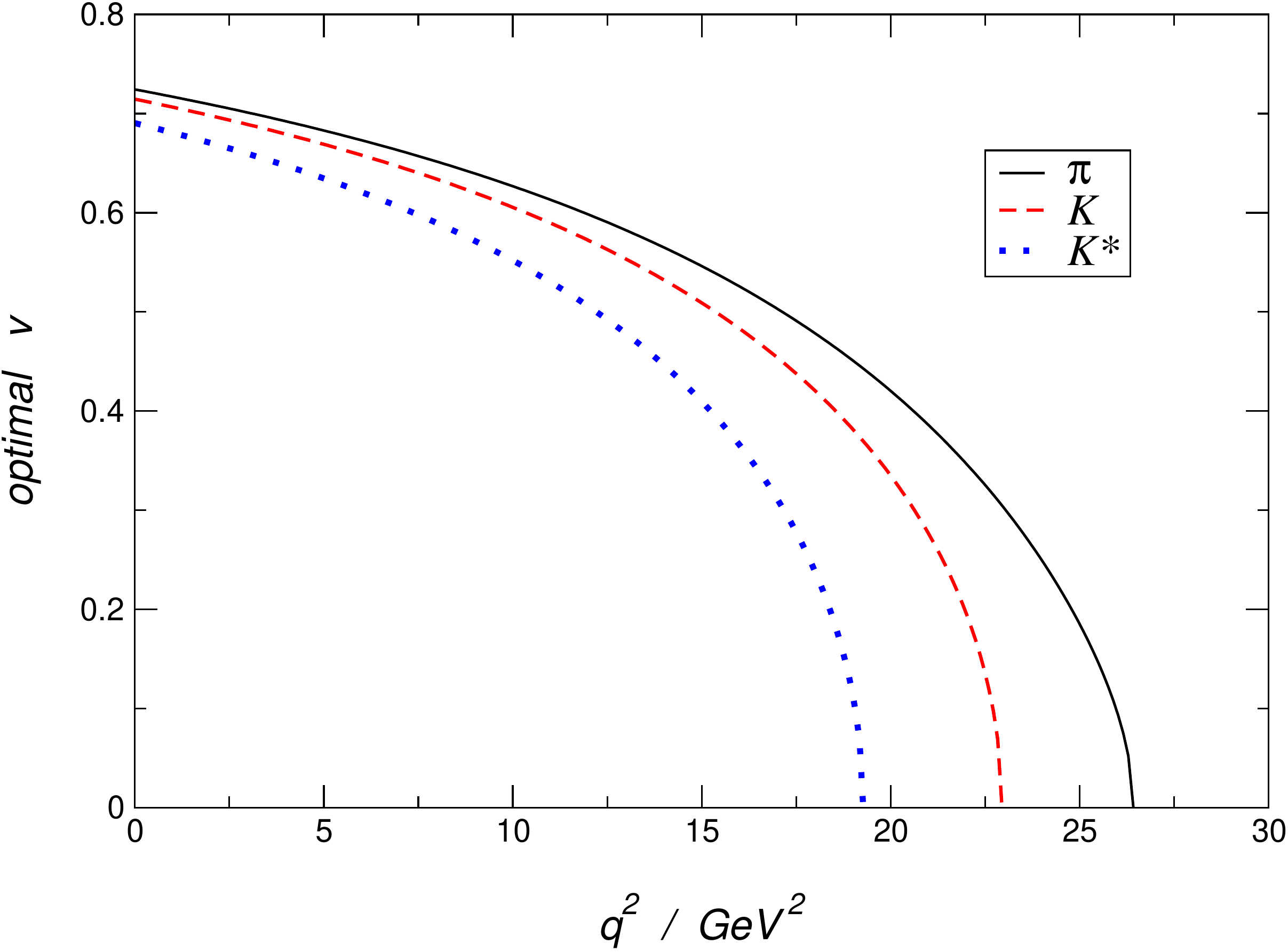}
\end{center}
\caption{Estimate of optimal velocity, minimizing discretization errors, for
 $B\to F$ form factors (see text) as a function of $q^2$ for the $\pi$, $K$
 and $K^*$ light mesons.}
\label{fig:plot_v_opt}
\end{figure}

The increase in discretization errors for the quarks in the $B$ meson
due to the boost of the momentum distribution when going from zero velocity
($\gamma=1$) to a non-zero velocity $\bs{v}$ is proportional to
\begin{equation}
\gamma^2\LQCD^2-\LQCD^2. \label{eq:B_discr_errors}
\end{equation}
Assuming that the quarks in the light meson share the momentum equally, each
carrying momentum of order $\bs{p}_F/2$, we expect that the increase in
relative discretization errors for the light meson when going from zero
momentum to $\bs{p}_F$ is proportional to
\begin{equation}
\left(\sfrac12|\bs{p}_F|\right)^2.  \label{eq:pi_discr_errors}
\end{equation}
The total error is the sum of these terms with some coefficients that we
presume are of order unity.  Noting that $(p_B-p_F)^2=q^2$, we choose $v$ to
minimize the total error to give the optimal $v$ as a function of $q^2$.
Investigation with different reasonable choices of coefficients shows that the
minimum error is compatible with setting the two above terms equal. The result
is plotted in Figure~\ref{fig:plot_v_opt} for the $\pi$, $K$ and $K^*$ light
mesons.  We find that at maximum recoil a velocity of $|\bs{v}|\approx0.7$
would minimize discretization errors. Of course this is only a very crude
estimate, and the optimal velocity depends on the details of the lattice
computation.

\subsubsection{Statistical errors}

Lattice calculations are not only limited by discretization errors, but also
by statistical errors. Unfortunately these increase exponentially when going
to lower $q^2$. Consider for instance the $B$-meson two-point function with
momentum $\bs{p}_B$, denoted as $\langle B^\dagger(\bs{p}_B,0)
B(\bs{p}_B,\tau) \rangle$, which is required in the form factor computations
alongside the pion two-point function and the $B\rightarrow F$ three-point
function. The variance in the correlator is  \cite{Lepage:1989hd}
\begin{eqnarray}
\nonumber \sigma^2(\tau)&=&\left\langle [B^\dagger(\bs{p}_B,0) B(\bs{p}_B,\tau)]
[B^\dagger(\bs{p}_B,0) B(\bs{p}_B,\tau)]^\dagger \right\rangle\\
&&-\left|\langle B^\dagger(\bs{p}_B,0) B(\bs{p}_B,\tau) \rangle\right|^2. \label{eq:noise}
\end{eqnarray}
The correlator in the first line of (\ref{eq:noise}) couples to the
combination of a heavy-heavy (HH) meson at rest and a pion at rest, so for
large Euclidean time $\tau$, it will decay like $\exp(-(M_{HH}+M_\pi)\tau)$.
However, the second line is simply the square of the two-point function, which
will decay like $\exp(-2E_B(\bs{p}_B)\tau)$ where $E_B(\bs{p}_B)$ is the
energy of a $B$ meson with momentum $\bs{p}_B$. Since
$M_{HH}+M_\pi<2E_B(\bs{p}_B)$, the variance will be dominated by the first
line at large $\tau$. This means that the signal-to-noise ratio approaches
zero exponentially fast in Euclidean time $\tau$,
\begin{equation}
\frac{\langle B^\dagger(\bs{p}_B,0) B(\bs{p}_B,\tau) \rangle}{\sigma(\tau)}
\propto e^{-\left(E_B(\bs{p}_B)-\sfrac12M_{HH}-\sfrac12M_\pi\right)\tau},
\label{eq:noise2}
\end{equation}
and at fixed $\tau$ it decreases as the momentum $\bs{p}_B$ increases.  A
similar analysis can be performed for the $B\rightarrow F$ three-point
function and for the light meson two-point function. At lower $q^2$, the
momenta $\bs{p}_B$, $\bs{p}_F$ and the corresponding energies are larger and
hence the signal decays faster, while the variance is independent of $q^2$.
(For an example with heavy-light correlators, see \cite{Gregory:2008sk}.)

The above argument illustrates that using mNRQCD to extend the kinematic range
of calculations requires the efficient use of techniques for reducing
statistical noise. Already progress has been made reducing
statistical errors using stochastic sources \cite{Davies:2007vb}.
Nevertheless, calculations at lower $q^2$ will undoubtedly require increased
computational effort.  In view of the opportunity for rare $b\to s$ 
decays to discover or further constrain non-Standard Model physics, via
$B\to K^*\gamma, K^{(*)}\ell^+\ell^-$ for example, such effort is 
worthwhile.

\subsubsection{Heavy-quark expansion of the current}

Even in continuum mNRQCD systematic errors for
heavy-to-light decays increase when going to lower $q^2$, since the
convergence of the heavy-quark expansion for the current mediating the decay gets worse. 
The heavy-quark expansion requires that all momentum scales for the light degrees of 
freedom are small compared to the mass of the heavy quark, which is approximately 
equal to the mass of the $B$ meson, $M_B$. In the low-recoil regime, the only relevant
scale is $\LQCD\ll M_B$, but for large recoil the momentum of the light meson
in the $B$ rest frame is large.

\begin{figure}
\begin{center}
\includegraphics[width=\linewidth]{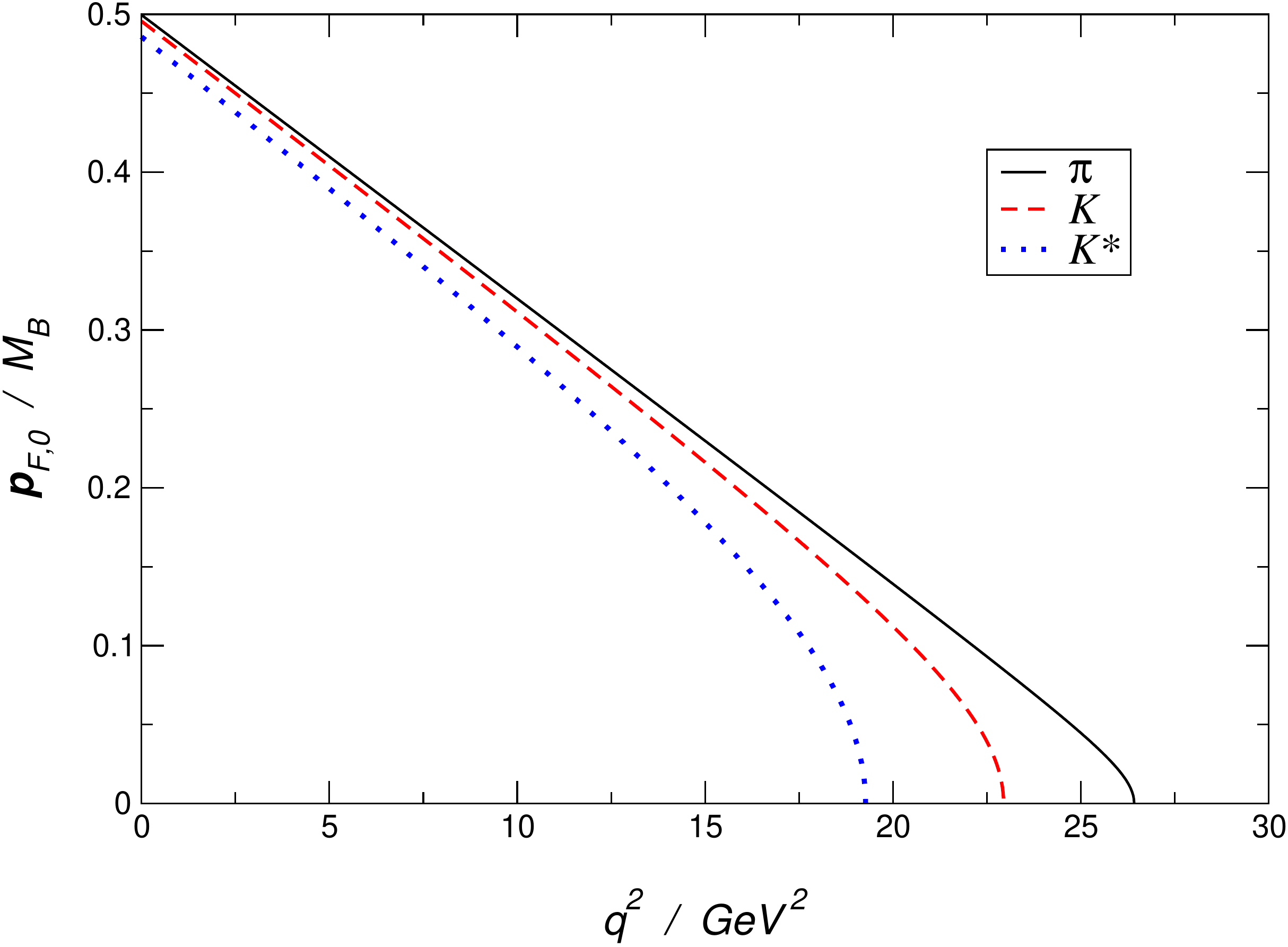}
\end{center}
\caption{The ratio $|\bs{p}_{F,0}|/M_B$ as a function of $q^2$ for the $\pi$,
$K$ and $K^*$ light mesons.}
\label{fig:HQET_convergence}
\end{figure}

The light meson energy in the $B$ rest frame can be written in a
Lorentz-invariant way as
\begin{equation}
E_{F,0}=\frac{p_B\cdot p_F}{M_B}=\frac{M_B^2+M_F^2-q^2}{2M_B} \,.
\end{equation}
The light meson momentum in this frame is then
$|\bs{p}_{F,0}|=\sqrt{E_{F,0}^2-M_F^2}$. In Fig. \ref{fig:HQET_convergence},
we plot the ratio $|\bs{p}_{F,0}|/M_B$ as a function of $q^2$ for for the
$\pi$, $K$ and $K^*$ light mesons. This ratio becomes almost $0.5$ at $q^2=0$,
which has to be compared to $\LQCD/M_B \approx 0.1$ in the low-recoil limit.

%
%
%
%
%
\section{Derivation of \lowercase{m}NRQCD}

\label{sec:mnrqcd_derivation}

\subsection{Continuum \lowercase{m}NRQCD}

To derive the mNRQCD action in Minkowski space, we work in two frames, the
optimal frame with coordinates $x$ and the rest frame of the $B$ meson with
coordinates $x'$. The two frames are related by a Lorentz boost with velocity
$\bs{v}$,
\begin{equation*}
x=\Lambda x'.
\end{equation*}
For the explicit form of $\Lambda=\Lambda(\bs{v})$, see Appendix
\ref{app:notation}. We denote the physical (full QCD Dirac spinor) heavy quark
field in the two frames by $\Psi(x)$ and $\Psi'(x')$. They are related by the
spinorial representation of the boost,
\begin{align*}
\Psi(x) & = S(\Lambda)\Psi'(x'),\\
\overline{\Psi}(x) & = \overline{\Psi'}(x')\overline{S}(\Lambda) \,.
\end{align*}
The spinorial boost matrix $S(\Lambda)$ is defined in Appendix
\ref{app:notation}. The Dirac Lagrangian for $\Psi'$ is
\begin{equation}
\mathcal{L}'(x') = \overline{\Psi'}(x') (i\dgamma\cdot D'-m)\Psi'(x').
\label{eq:Dirac_Lagrangian}
\end{equation}
(The hat simply distinguishes a Dirac spin matrix from the $\gamma$ of the
Lorentz transformation. Our convention for these matrices is given
in Appendix~\ref{app:notation}.)  Since the heavy quark is approximately at
rest in this frame, we can approximate this Lagrangian very well by the
standard NRQCD Lagrangian. One approach to constructing this Lagrangian is by
writing down all possible operators that are allowed by the symmetries of the
theory. This approach is described for example in \cite{Lepage:1992tx} and
\cite{Braaten:1996ix} and has the advantage that it includes operators which
only arise at higher loop order as, for example, four-quark operators. By
matching to full QCD one finds, however, that these are suppressed by
$\alpha_s^2$ and will play no role in our analysis.

\subsubsection{FWT transformation}
We use a Foldy-Wouthuysen-Tani (FWT) transformation to derive
the Lagrangian order by order in $1/m$ via field redefinitions, since this
automatically generates the correct tree level coefficients of all operators.
For a detailed description of the method, see \cite{Korner:1991kf}.
The transformation can be written as
\begin{equation}
\Psi'(x')=T'_{\scriptscriptstyle\mathrm{FWT}}\:e^{-im{x'}^0\dgamma^0}
\:\tilde{\Psi}'(x')  \label{eq:TFWT0}
\end{equation}
which defines the transformed field $\tilde{\Psi}'$.
(A corresponding transformation defines $\overline{\Psi'}(x')$). The factor
$e^{-im{x'}^0\dgamma^0}$ removes the additive mass term from the Lagrangian and
$T'_{\scriptscriptstyle\mathrm{FWT}}$ is given by
\begin{align}
\nonumber T'_{\scriptscriptstyle\mathrm{FWT}} & =\exp\left[\frac{1}{2m}
\left(i\bs{\dgamma}\cdot\bs{D'}\right)\right]\\
\nonumber & \times \exp\left[\frac{1}{2m^2}
\left(-\frac{ig}{2}\dgamma^0\bs{\dgamma}\cdot \bs{E'}\right)\right] \\
\nonumber & \times \exp\left[\frac{1}{2m^3}\left(\frac{g}{4}\bs{\dgamma}\cdot
({D'_0}^\mathrm{ad} \bs{E'})+\frac{1}{3}\left(i\bs{\dgamma}\cdot\bs{D'}\right)^3
\right)\right] \\
& \times \left[1+\mathcal{O}(1/m^4)\right]. \label{eq:TFWT}
\end{align}
(The chromoelectric and chromomagnetic components of the gluon field strength
tensor are defined by ${E}_k={F}_{0k}$, ${B}_j=-\frac12\epsilon_{jkl}{F}_{kl}\:\:$
in Minkowski space). The resulting Lagrangian is
\begin{eqnarray}
\nonumber\mathcal{L}'&=&\overline{\tilde{\Psi}'}\bigg[i\dgamma^0{D}'_0+
\frac{\bs{D'}^2}{2m}+\frac{g}{2m}\bs{\Sigma}\!\cdot\!\bs{B'}\\
\nonumber&+&\frac{g}{8m^2}\dgamma^0\left(  {\bs{D'}}^{\mathrm{ad}}
\!\!\cdot\!{\bs{E'}}+i\bs{\Sigma}\cdot
\left(\bs{D'}\!\times\!{\bs{E'}}\!-\!\bs{E'}\!\times\!\bs{D'}\right) \right)
\bigg]\tilde{\Psi}'\\
&+&\mathcal{O}(1/m^3),
\label{eqn:NRQCD_Lagrangian_1msqr}
\end{eqnarray}
with
\begin{equation*}
\Sigma^j\equiv\left(\begin{array}{cc} \sigma^j & 0 \\ 0 & \sigma^j
\end{array}\right).
\end{equation*}
Note that in (\ref{eqn:NRQCD_Lagrangian_1msqr}) the adjoint derivative
${\bs{D'}}^{\mathrm{ad}}$ acts on $\bs{E'}$ only, whereas the standard
derivatives $\bs{D'}$ act on all quantities to their right.

As a result of the FWT transformation, all operators in the new Lagrangian
commute with $\dgamma^0$, that is, the quark and antiquark components are
decoupled to this order.

The next step is to re-express the Lagrangian,
(\ref{eqn:NRQCD_Lagrangian_1msqr}), in terms of quantities in the frame $x$
(which we will put onto the lattice). To this end, we \textit{define} a new
field $\tilde{\Psi}(x)$ via the trivial transformation law
\begin{align}
\nonumber\tilde{\Psi}(x)&\equiv\tilde{\Psi}'(x')\\
\overline{\tilde{\Psi}}(x)&\equiv\overline{\tilde{\Psi}'}(x')
\label{eq:frame_change}
\end{align}
Note that in order to preserve the commutativity with $\dgamma^0$ we do not
include the spinorial boost matrix $S(\Lambda)$ in
(\ref{eq:frame_change}). This is in contrast to the standard continuum
``moving HQET'' Lagrangian.

Under the change of coordinates $x=\Lambda x'$, derivative operators in
the Lagrangian and FWT transformation transform like
\begin{equation}
D'_\mu=\Lambda^\nu_{\:\:\:\mu}D_\nu. \label{eq:derivative_transformation_law}
\end{equation}
The transformation law for the gluon field strength tensor,
\begin{equation*}
F'_{\mu\nu}(x')=\Lambda^\rho_{\:\:\:\mu}\Lambda^\kappa_{\:\:\:\nu}F_{\rho\kappa}(x)
\end{equation*}
leads to the following transformation for the chromoelectric and
chromomagnetic components:
\begin{align}
\bs{E'}(x') & = \gamma\biggl(\bs{E}(x)+\bs{v}\times\bs{B}(x) - 
\frac{\gamma}{\gamma+1}\bs{v}\bigl(\bs{v}\cdot\bs{E}(x)\bigr)\biggr)\,,
\nonumber \\ 
\bs{B'}(x') & = \gamma\biggl(\bs{B}(x)-\bs{v}\times\bs{E}(x)-
\frac{\gamma}{\gamma+1}\bs{v}\bigl(\bs{v}\cdot\bs{B}(x)\bigr)\biggr) \,.
\label{eq:Btransf-mink}
\end{align}
Using (\ref{eq:frame_change}), (\ref{eq:derivative_transformation_law})
and (\ref{eq:Btransf-mink}), the Lagrangian (\ref{eqn:NRQCD_Lagrangian_1msqr})
can be expressed entirely in the new frame with coordinates $x$. Note
that Lorentz invariance can be used to simplify the transformation in
the following way: $x'^0 = u' \cdot x' = u \cdot x$, where $u'
= (1,\bs{0})$ and $u=(u^0,\bs{u})=(\gamma,\gamma \bs{v})$. Similarly,
$D'_0 = u' \cdot D' = u\cdot D$ and $\bs{D'}^2 = ( u \cdot D)^2 - D^2$.
The term with the adjoint derivative of the chromoelectric field can be
written as ${\bs{D'}}^{\mathrm{ad}}\!\cdot\!{\bs{E'}}=D^\mathrm{ad}_\mu
u_\nu F^{\mu\nu}.$ The other occurrences of the field strengths are simply
replaced by (\ref{eq:Btransf-mink}), but we will not insert this expression
explicitly for the sake of legibility. The Lagrangian becomes
\begin{eqnarray}
\nonumber\mathcal{L}&=&\overline{\tilde{\Psi}}\bigg[i\dgamma^0 u\cdot D
+ \frac{(u \cdot D)^2-D^2}{2m} +\frac{g}{2m}\bs{\Sigma}\!\cdot\!\bs{B'}\\
\nonumber&&+\;\frac{g}{8m^2}\dgamma^0\left(  D^\mathrm{ad}_\mu u_\nu
F^{\mu\nu}+i\epsilon_{jkl}\Sigma^j\Lambda^\mu_{\:\:\:k}
\left\{D_\mu,\: E'_l\right\}\right) \bigg]\tilde{\Psi}\\
&&+\;\mathcal{O}(1/m^3).
\label{eqn:NRQCD_Lagrangian_1msqr_moving}
\end{eqnarray}
%
%
\subsubsection{\label{h_order}Removing time derivatives in the Hamiltonian}
Note that the operators of order $1/m$ and $1/m^2$ in
(\ref{eqn:NRQCD_Lagrangian_1msqr_moving}) now contain time
derivatives. In the following, we will show how these can be removed
via further field redefinitions to ensure that in the lattice
computations the propagator can be obtained by solving an initial
value problem using a time evolution equation.

It is convenient to write the Lagrangian
(\ref{eqn:NRQCD_Lagrangian_1msqr_moving}) in the following form,
\begin{equation}
\mathcal{L}=\gamma\:\overline{\tilde{\Psi}}\left[O_0+\frac{1}{\gamma m}O_1
+\frac{1}{(\gamma m)^2}O_2\right]\tilde{\Psi}+\mathcal{O}(1/m^3),
\label{eq:Lagrangian-O}
\end{equation}
with
\begin{eqnarray*}
O_0&=&i\dgamma^0(D_0 + \bs{v}\cdot\bs{D}),\\
O_1&=&\frac12\left((u\cdot D)^2-D^2\right)
+\frac{g}{2}\bs{\Sigma}\!\cdot\!\bs{B'},\\
O_2&=&\frac{g}{8}\gamma\:\dgamma^0\left(D^\mathrm{ad}_\mu u_\nu F^{\mu\nu}
+i\epsilon_{jkl}\Sigma^j\Lambda^\mu_{\:\:\:k}\left\{D_\mu,\: E'_l\right\}\right).
\end{eqnarray*}
We start by removing the time derivatives in $O_1$. To see how this can
be done, we note that any field redefinition of the form
\begin{eqnarray*}
\tilde{\Psi}&=&\exp\left(\frac{1}{\gamma m}U\right)\tilde{\Psi}_{(1)},\\
\overline{\tilde{\Psi}}&=&\overline{\tilde{\Psi}}_{(1)}\exp
\left(\frac{1}{\gamma m}U\right) \label{eqn:td_trans_1}
\end{eqnarray*}
will result in
\begin{align*}
\mathcal{L}&=\gamma\:\overline{\tilde{\Psi}}_{(1)}\left[O_0
+\frac{1}{\gamma m}O_{(1)1}+\frac{1}{(\gamma m)^2}O_{(1)2}\right]
\tilde{\Psi}_{(1)}\\
&+\mathcal{O}(1/m^3)
\end{align*}
with the new operators
\begin{eqnarray}
\nonumber O_{(1)1}&=&O_1+\left\{U,\:O_0\right\},\\
\nonumber O_{(1)2}&=&O_2+\left\{U,\:O_1\right\}+U O_0 U
+\frac{1}{2}\left\{U^2,\:O_0\right\}.\\
\label{eq:O12}
\end{eqnarray}
Thus, we need to write $O_1=O_{(1)1}-\left\{U,\:O_0\right\}$ with some
operator $U$ such that $O_{(1)1}$ does not contain time derivatives.
This is indeed possible:
\begin{eqnarray*}
O_1&\!=\!&\frac12\left[\gamma^2D_0^2+\gamma^2\!\left\{D_0,\:\bs{v}
\!\cdot\!\bs{D}\right\}+\gamma^2(\bs{v}\!\cdot\!\bs{D})^2\!
-\!D_0^2+\bs{D}^2\right]\\
&\!+\!&\frac{g}{2}\bs{\Sigma}\!\cdot\!\bs{B'}\\
&\!=\!&\underbrace{\frac12\left[\bs{D}^2-(\bs{v}
\!\cdot\!\bs{D})^2\right]+\frac{g}{2}\bs{\Sigma}\!
\cdot\!\bs{B'}}_{\equiv\:\:O_{(1)1}}\\
&\!+\!&\underbrace{\frac12\left[(\gamma^2\!-\!1) D_0^2
+\gamma^2\left\{D_0,\:\bs{v}\!\cdot\!\bs{D}\right\}
+(\gamma^2\!+\!1)(\bs{v}\!\cdot\!\bs{D})^2\right]}_{=
\:\:-\:\left\{U,\:O_0\right\}},
\end{eqnarray*}
and we can now read off the operator $U$:
\begin{equation}
U=\frac i4\dgamma^0\left[(\gamma^2-1) D_0+(\gamma^2+1)
\bs{v}\cdot\bs{D}\right]. \label{eq:U}
\end{equation}
The next step is to remove the time derivatives (other than the adjoint time
derivative, which acts on the gluon field strength only) in the new
operator $O_{(1)2}$, given in (\ref{eq:O12}). Similarly to before,
we use a field redefinition
\begin{eqnarray}
\nonumber \tilde{\Psi}_{(1)}&=&\exp\left(\frac{1}{(\gamma m)^2}V\right)
\tilde{\Psi}_{(2)},\\
\overline{\tilde{\Psi}}_{(1)}&=&\overline{\tilde{\Psi}}_{(2)}
\exp\left(\frac{1}{(\gamma m)^2}V\right), \label{eqn:td_trans_2}
\end{eqnarray}
now with an extra power of $1/(\gamma m)$, so that the lower order terms
are unaffected. The derivation of the operator $V$ is given in Appendix
\ref{app:2nd_time_der_removal}.
%
%
\subsubsection{mNRQCD Lagrangian}
%
%
Finally, we rescale the fields
\begin{eqnarray}
\nonumber \tilde{\Psi}_{(2)}&=&\frac{1}{\sqrt{\gamma}}\Psi_v,\\
\overline{\tilde{\Psi}}_{(2)}&=&\frac{1}{\sqrt{\gamma}}\overline{\Psi}_v,
\label{eq:rescale}
\end{eqnarray}
to remove the factor of $\gamma$ in front of $\mathcal{L}$. We arrive at the following
result for the tree-level moving NRQCD Lagrangian in Minkowski space:
\begin{eqnarray}
\nonumber\mathcal{L}&=&\overline{\Psi}_v\bigg[i\dgamma^0D_0
+ i\dgamma^0\bs{v}\!\cdot\!\bs{D}+\frac{\bs{D}^2
-(\bs{v}\!\cdot\!\bs{D})^2}{2\gamma m}\\
\nonumber&&\hspace{6ex}+\frac{g}{2\gamma m}\bs{\Sigma}\!\cdot\!{\bs{B'}}\\
\nonumber&&\hspace{6ex}+\frac{i}{4\gamma^2 m^2}\dgamma^0
\left(\left\{\bs{v}\!\cdot\!{\bs{D}},\:\bs{D}^2\right\}
-2(\bs{v}\!\cdot\!\bs{D})^3\right)\\
\nonumber&&\hspace{6ex}+\frac{g}{8 m^2}\dgamma^0
\left(\bs{D}^{\mathrm{ad}}\cdot\bs{E}-\bs{v}
\cdot(\bs{D}^{\mathrm{ad}}\times\bs{B})\right)\\
\nonumber&&\hspace{6ex}+\frac{ig}{8\gamma m^2}\dgamma^0
\:\bs{\Sigma}\cdot\left(\bs{D}\times{\bs{E'}}
-{\bs{E'}}\times\bs{D} \right)\\
\nonumber&&\hspace{6ex}-\frac{ig}{8(\gamma+1)m^2}
\dgamma^0\left\{\bs{v}\cdot\bs{D},\:\:\bs{\Sigma}
\cdot(\bs{v}\times{\bs{E'}}) \right\}\\
\nonumber&&\hspace{6ex}+\frac{(2-\bs{v}^2)g}{16 m^2}
\dgamma^0\left(D^{\mathrm{ad}}_0-\bs{v}\cdot\bs{D}^{\mathrm{ad}}\right)
\left(\bs{v}\cdot\bs{E}\right)\\
\nonumber&&\hspace{6ex}+\frac{ig}{4\gamma^2 m^2}\dgamma^0
\left\{\bs{v}\cdot\bs{D},\:\:\bs{\Sigma}\cdot{\bs{B'}} \right\}
\bigg]\Psi_v\\
&&+\mathcal{O}(1/m^3).
\label{eq:mNRQCD_1msqr}
\end{eqnarray}
As before, all terms commute with $\dgamma^0$. We can therefore introduce 2-component
fields $\psi_v(x)$ and $\xi_v(x)$,
\begin{equation*}
\Psi_v=\left(\begin{array}{c}{\psi_v} \\ {\xi_v} \end{array}\right),
\hspace{1cm} \overline{\Psi}_v=\left( \psi_v^\dag,\:\:\:\: -\xi_v^\dag \right),
\end{equation*}
to explicitly separate the Lagrangian into the quark and antiquark pieces:
\begin{align}
\nonumber\mathcal{L}=\psi_v^\dag&\left[iD_0 + i\bs{v}\!\cdot\!\bs{D}
+\frac{\bs{D}^2-(\bs{v}\!\cdot\!\bs{D})^2}{2\gamma m}
+\frac{g}{2\gamma m}\bs{\sigma}\!\cdot\!{\bs{B'}}\right]\psi_v \\
\nonumber+\xi_v^\dag&\left[iD_0 + i\bs{v}\!\cdot\!\bs{D}
-\frac{\bs{D}^2-(\bs{v}\!\cdot\!\bs{D})^2}{2\gamma m}
-\frac{g}{2\gamma m}\bs{\sigma}\!\cdot\!{\bs{B'}}\right]\xi_v\\
&+\mathcal{O}(1/m^2).
\label{eq:mNRQCD_1m_2comp}
\end{align}
Terms with odd powers of $1/m$ (i.e.\ those without a factor of $\dgamma^0$ in
(\ref{eq:mNRQCD_1msqr})) appear with the opposite sign in the antiquark
Lagrangian.

Note that we have chosen a particular notation convention for the 2-component
antiquark field: $\xi_v$ creates an antiquark whereas $\psi_v$ annihilates a
quark. While the quark and antiquark terms in (\ref{eq:mNRQCD_1m_2comp}) take
a similar form, dictated by charge conjugation invariance, it should be borne
in mind that $\psi_v$ and $\xi_v$ have these different interpretations when
constructing the heavy quark and antiquark Green functions.
As an aside, note that our new result (\ref{eq:mNRQCD_1msqr}) differs slightly
at order $1/m^2$ from the one given in Refs.~\cite{Foley:2004rp, Foley:2005fx,
Dougall:2005zh}.

Let us now summarize the tree-level relation between the full QCD field
$\Psi(x)$ and the moving NRQCD two-component fields $\psi_v(x)$, $\xi_v(x)$:
\begin{equation}
\Psi(x)\:=\:S(\Lambda)\:\:T_{\scriptscriptstyle \mathrm{FWT}}
\:\:e^{-im\:u\cdot x\:\dgamma^0}\:\:A_{\scriptscriptstyle D_t}
\:\:\frac{1}{\sqrt{\gamma}}\left(\begin{array}{c}{\psi_v}(x) \\
{\xi_v}(x) \end{array}\right)
\label{eq:MNRQCD_field_redef}
\end{equation}
where $T_{\scriptscriptstyle\mathrm{FWT}}$ is the FWT transformation
(\ref{eq:TFWT}) expressed in the frame $x$, \textit{i.e.}\
\begin{equation*}
T_{\scriptscriptstyle\mathrm{FWT}}=\exp\left(\frac{i\dgamma^j
\Lambda^\mu_{\:\:\:j}D_\mu}{2m}\!\right) \exp\left(\frac{ig\bs{\dgamma}
\!\cdot\! \bs{E'}\dgamma^0}{(2m)^2}\right)\times...
\end{equation*}
and
\begin{equation*}
A_{\scriptscriptstyle D_t}=\exp\left(\frac{U}{\gamma m}\right)
\exp\left(\frac{V}{(\gamma m)^2}\right)\times...
\end{equation*}
removes the unwanted time derivatives in the Lagrangian ($U$ and $V$ were
defined in equations (\ref{eq:U}) and (\ref{eq:V}), respectively).

The field redefinition (\ref{eq:MNRQCD_field_redef}) can be used to obtain
tree-level expressions for currents containing the heavy quark in
calculations of decay constants and form factors, as discussed briefly
in section \ref{sec:current_construction}.
%
%
\subsection{Power counting}
\label{sec:power-counting}
When deriving the mNRQCD Lagrangian in the previous section, we were formally
expanding in powers of $1/m$. As is well known from heavy-quark effective theory,
for heavy--light systems such as $B$ mesons, the expansion really is in $\LQCD/m$
with the QCD scale $\LQCD \sim 500$~MeV. The Lorentz transformation does not
affect the power counting and thus the Lagrangian (\ref{eq:mNRQCD_1msqr}) is
complete through order $(\LQCD/m)^2$.
\begin{table}
\begin{tabular}{ccccccc}
\hline\hline
 \\[-2ex]
\multicolumn{2}{l}{$\overline{Q}Q$ rest frame} &\hspace{2ex} & \multicolumn{2}{l}{lattice frame} & &
$v\rightarrow 1$\\
 \\[-2ex]
\hline
 \\[-2.4ex]
$D'_t$&   $m\vnr^2$ & &$D_t$& $\gamma m(\vnr^2+v\vnr)$ &
&$\gamma m\vnr$\\
$\bs{D'}$&   $m\vnr$      & &$\bs{D}$&$\gamma m(\vnr+v\vnr^2)$ & & $\gamma m\vnr$\\
$g\bs{E'}$&  $m^2\vnr^3$  & &$g\bs{E}$ & $\gamma m^2(\vnr^3\!-\!v\vnr^4\! -
\frac{\gamma v^2\vnr^3}{\gamma+1})$& &$\gamma m^2\vnr^3$\\
$g\bs{B'}$&  $m^2\vnr^4$  & &$g\bs{B}$&$\gamma m^2(\vnr^4\!+\!v\vnr^3 \!-
\frac{\gamma v^2\vnr^4}{\gamma+1})$& &$\gamma m^2\vnr^3$\\
\hline
 \\[-2.4ex]
$D'\!\cdot\! D'$& $m^2\vnr^2$ & &$D\!\cdot\! D$&$m^2\vnr^2$
$(\neq \gamma^2m^2\vnr^2)$& &$m^2\vnr^2$\\
$u'\!\cdot\! D'$& $m^2\vnr^2$ & &$u\!\cdot\! D$&$m^2\vnr^2$
$(\neq \gamma^2m^2\vnr^2)$& &$m^2\vnr^2$\\
\hline
\hline
\end{tabular}
\caption{Power counting rules appropriate for mNRQCD with heavy-heavy mesons.
In the large velocity limit (last column), the Lorentz boost removes the
differences in order found for NRQCD, giving $D_t\sim\bs{D}$ and
$\bs{E}\sim\bs{B}$. In the last two rows note that the na\"{\i}ve power
counting rules can give the wrong counting (see text).}
\label{table:powercounting}
\end{table}
For heavy-heavy mesons such as the $\Upsilon$, the situation is more
complicated. In the frame where the meson is at rest, the power counting
is governed by powers of $\vnr$, the small non-relativistic internal
velocity of the heavy quarks inside the meson \cite{Lepage:1992tx}.
For $\Upsilon$ systems, one has $\vnr^2 \sim 0.1$. It turns out that
all terms of the Lagrangian (\ref{eqn:NRQCD_Lagrangian_1msqr}) are of
order $\vnr^4$ or lower, but one term of order $\vnr^4$ is missing.
By expanding the expression for the relativistic kinetic energy in
powers of the residual momentum $\bs{k}$,
\begin{eqnarray*}
E_{\mathrm{kin}}-m&=&\sqrt{\bs{k}^2+m^2}- m\\
 &=&\frac{\bs{k}^2}{2m}-\frac{\bs{k}^4}{8m^3}+\frac{\bs{k}^6}{16m^5}-...
\end{eqnarray*}
and replacing $\bs{k}$ by the operator $-i\bs{D}$, we see that we must
include the operator $\bs{D}^4/(8m^3)$ into (\ref{eqn:NRQCD_Lagrangian_1msqr})
in order to obtain accuracy to order $\vnr^4$. The corresponding term in
the moving NRQCD Lagrangian can be obtained in the same way,
\begin{eqnarray*}
 E_{\mathrm{kin}}-\gamma m&=&\sqrt{(\gamma m \bs{v}+\bs{k})^2+m^2}-\gamma m\\
 &=&\bs{v}\cdot\bs{k}+\frac{1}{2\gamma m}\left(\bs{k}^2-(\bs{v}\cdot\bs{k})^2\right)\\
&&+\frac{1}{4\gamma^2m^2}\big(-\{\bs{v}\cdot\bs{k},\:\bs{k}^2\}+2(\bs{v}\cdot\bs{k})^3\big)\\
&&+\frac{1}{8\gamma^3m^3}\big(-\bs{k}^4+3\left\{\bs{k}^2,\:(\bs{v}\cdot\bs{k})^2 \right\}\\
&&\hspace{12ex}-5(\bs{v}\cdot\bs{k})^4  \big)\\
&&+\:\:...
\end{eqnarray*}
Thus, the operator
\begin{equation}
\frac{1}{8\gamma^3m^3}\left(\bs{D}^4-3\left\{\bs{D}^2,\:(\bs{v}\cdot\bs{D})^2 \right\}
+5(\bs{v}\cdot\bs{D})^4\right) \label{eq:kin_term_m3}
\end{equation}
must be included into the moving NRQCD Lagrangian (\ref{eq:mNRQCD_1msqr}).
We ordered the terms with products of $(\bs{v}\cdot\bs{D})$ and $\bs{D}^2$
in the form of anticommutators, as the anticommutator-ordering is what one
would have obtained from field redefinitions.

For heavy-heavy mesons at $\bs{v}=0$, the power counting is different for
temporal- and spatial components of Lorentz vectors but they will mix in a
frame with $\bs{v} \neq 0$. The rules for both $\bs{v}=0$ and $\bs{v} \neq 0$
are summarized in Table~\ref{table:powercounting}.

Care has to be taken when dealing with quantities like $D\cdot D$ and $u\cdot D$;
their power counting cannot be derived by na\"{\i}vely multiplying the power
counting rules for each factor. For example, for $\bs{v} \rightarrow 1$ the
product $D\cdot D$ does not scale like $(\gamma m \vnr)^2$ but as $m^2\vnr^2$
instead. The correct values are shown in the last two rows of Table~\ref{table:powercounting}.

\subsection{Euclidean mNRQCD}

The Euclidean action $S_E=\int\mathrm{d}^4x_E\:\mathcal{L}_E(x_E)$ can be
obtained from the Minkowski space action $S=\int\mathrm{d}^4x\:\mathcal{L}(x)$
in the usual way by making the formal replacements
\begin{eqnarray*}
\Psi_v(x) &\rightarrow& \Psi_v(x_E),\\
\overline{\Psi}_v(x) &\rightarrow& \overline{\Psi}_v(x_E),\\
\bs{A}(x) &\rightarrow& \bs{A}(x_E),\\
A_0(x) &\rightarrow& iA_4(x_E),\\
x^0&\rightarrow&-i x_E^4\equiv-i\tau,
\end{eqnarray*}
so that the integration measure and derivatives become
$\mathrm{d}^4x\rightarrow(-i)\mathrm{d}^4x_E,$ $\partial_0\rightarrow i\partial_4$.
Finally, the result must be multiplied by $(-i)$. In the following, we drop the
subscript $E$ (``Euclidean''). Note that we do not introduce Euclidean
gamma matrices in this paper; the same definition as in Minkowski space
is used (see Appendix~\ref{app:notation}).

It is also convenient to define the relation between the chromoelectric
field $\bs{E}$ and the 4-dimensional $F_{\mu\nu}$ with a different sign
in Euclidean space, \textit{i.e.}\ $E_j=F_{j4}$, while the definition of the
chromomagnetic field is unchanged, $B_j=-\frac12\epsilon_{jkl}F_{kl}.$

With this definition, (\ref{eq:Btransf-mink}) turns into
the symmetric form
\begin{align}
\bs{E'} & = \gamma\biggl(\bs{E}+i\bs{v}\times\bs{B} - 
\frac{\gamma}{\gamma+1}\bs{v}\bigl(\bs{v}\cdot\bs{E}\bigr)\biggr)\,, 
\nonumber \\ 
\bs{B'} & = \gamma\biggl(\bs{B}+i\bs{v}\times\bs{E}-
\frac{\gamma}{\gamma+1}\bs{v}\bigl(\bs{v}\cdot\bs{B}\bigr)\biggr) \,.
\label{eq:Btransf_eucl}
\end{align}
The Euclidean Lagrangian, in which we now include the relativistic
correction term (\ref{eq:kin_term_m3}), becomes
\begin{eqnarray}
\nonumber\mathcal{L}&=&\overline{\Psi}_v\bigg[\dgamma^0D_4
- i\dgamma^0\bs{v}\!\cdot\!\bs{D}-\frac{\bs{D}^2
-(\bs{v}\!\cdot\!\bs{D})^2}{2\gamma m}\\
\nonumber&&\hspace{5ex}-\frac{g}{2\gamma m}\bs{\Sigma}\!\cdot\!{\bs{B'}}\\
\nonumber&&\hspace{5ex}-\frac{i}{4\gamma^2 m^2}\dgamma^0
\left(\left\{\bs{v}\!\cdot\!{\bs{D}},\:\bs{D}^2\right\}
-2(\bs{v}\!\cdot\!\bs{D})^3\right)\\
\nonumber&&\hspace{5ex}+\frac{g}{8m^2}\dgamma^0\left(i\bs{D}^{\mathrm{ad}}
\cdot\bs{E}+\bs{v}\cdot(\bs{D}^{\mathrm{ad}}\times\bs{B})\right)\\
\nonumber&&\hspace{5ex}-\frac{g}{8\gamma m^2}\dgamma^0\:\bs{\Sigma}
\cdot\left(\bs{D}\times{\bs{E'}}-{\bs{E'}}\times\bs{D} \right)\\
\nonumber&&\hspace{5ex}+\frac{g}{8(\gamma+1)m^2}\dgamma^0
\left\{\bs{v}\cdot\bs{D},\:\:\bs{\Sigma}\cdot(\bs{v}\times{\bs{E'}}) \right\}\\
\nonumber&&\hspace{5ex}-\frac{(2-\bs{v}^2)g}{16 m^2}\dgamma^0
\left(D^{\mathrm{ad}}_4+i\bs{v}\cdot\bs{D}^{\mathrm{ad}}\right)
\left(\bs{v}\cdot\bs{E}\right)\\
\nonumber&&\hspace{5ex}-\frac{ig}{4\gamma^2 m^2}\dgamma^0
\left\{\bs{v}\cdot\bs{D},\:\:\bs{\Sigma}\cdot{\bs{B'}} \right\}\\
\nonumber&&\hspace{5ex}-\frac{1}{8\gamma^3m^3}\bigg(\bs{D}^4
-3\left\{\bs{D}^2,\:(\bs{v}\cdot\bs{D})^2 \right\}\\
&&\hspace{16ex}+5(\bs{v}\cdot\bs{D})^4\bigg)\bigg]\Psi_v.
\label{eq:Euclid_mNRQCD_1msqr_v4}
\end{eqnarray}
As in (\ref{eq:mNRQCD_1m_2comp}) one can introduce two-component fields
for quark and antiquark. It turns out that in Euclidean space, the antiquark
action can be obtained from the quark action by replacing
$\psi_v \rightarrow \left(\xi_v^\dag\right)^T$,
$\psi_v^\dag \rightarrow \left(\xi_v\right)^T$,
$\bs{v} \rightarrow (-\bs{v})$ and taking the complex conjugate of the whole 
action kernel. This is an important result, because it implies that the Euclidean
antiquark Green function can be obtained from the Euclidean quark
Green function in a frame with the \textit{opposite} boost velocity, $-\bs{v}$. 
We define $G_{\xi_v}^{(+\bs{v})}(x,\:\:x') = \langle \xi_v(x)\xi_v^\dagger(x') \rangle$.
Writing out color, spin and position indices explicitly, one then has
\begin{eqnarray}
\nonumber\left[G_{\xi_v}^{(+\bs{v})}\right]_{cs\:c's'}\!\!(x,\:\:x')
&=&-\left[G_{\psi_v}^{(-\bs{v})}\right]_{c's'\:cs}^*\!\!(x',\:\:x) \\
\nonumber&=&-\left[G_{\psi_v}^{(-\bs{v})}\right]_{cs\:c's'}^\dag\!\!(x',\:\:x).
\\ \label{eqn:q_aq_rel}
\end{eqnarray}
%

%
\section{Lattice \lowercase{m}NRQCD}
\label{sec:lattice-mnrqcd}

\subsection{Construction of the Hamiltonian}
\label{sec:lattice-mnrqcd-construct}

We construct the lattice moving NRQCD action such that for $v=0$ it reduces
to the previously used lattice NRQCD action with conventions as in
\cite{Wingate:2002fh}. Thus, the quark action has the form
\begin{equation}
S_{\psi_v}=\sum_{\bs{x},\tau}\psi_v^\dagger(\bs{x},\tau)\big[{\psi_v}(\bs{x},\tau)
-K(\tau){\psi_v}(\bs{x},\tau-1) \big] \label{eq:latact}
\end{equation}
with the kernel
\begin{eqnarray}
\nonumber K(\tau)&=&\left(1-\frac{\delta H|_\tau}{2}\right)
\left(1-\frac{H_0|_\tau}{2n} \right)^nU_4^\dag(\tau-1)\\
&&\times\left(1-\frac{H_0|_{\tau-1}}{2n} \right)^n
\!\left(1-\frac{\delta H|_{\tau-1}}{2}\right) \,.
\label{eq:mNRQCD_action_kernel}
\end{eqnarray}
Note that the heavy-quark Green function for the action (\ref{eq:latact})
satisfies the evolution equation
\begin{equation}
G_{\psi_v}(\bs{x},\tau,\bs{x'},\tau')=K(\tau) \:
G_{\psi_v}(\bs{x},\tau-1,\bs{x'},\tau'). \label{eq:hq_ev_eq}
\end{equation}
For this, it is crucial that the Hamiltonian does not contain time derivatives
(other than the adjoint time derivative of the chromoelectric field).

This split into leading-order kinetic terms $H_0$ and higher-order
corrections $\delta H$ which satisfies time-reversal symmetry was introduced
in \cite{Lepage:1992tx}. Other than consistency with previous work, there are
no strong arguments (such as computational load, numerical stability or size
of discretization errors) for the relative ordering of $H_0$ and $\delta H$ in
the action. The time derivative in (\ref{eq:latact}) is implemented as a
backward (rather than forward) difference operator as this prevents mean-field
corrections to the wavefunction renormalization \cite{Lepage:1992tx}.

The leading evolution due to $H_0$ from one lattice time slice to the
next is effectively divided into $2n$ smaller steps to avoid the well-known
instability in the discretization of parabolic differential equations
(see, for instance, Sec.~19.2 of Ref.~\cite{press:numrec}). In this
way, one can allow the highest momentum modes in the theory to come into
equilibrium, while avoiding the need for a very small lattice spacing which
would render the theory too expensive to simulate. For NRQCD, where
$H_0$ is always positive, the integer-valued stability parameter $n$ has
to be chosen such that
\begin{equation}
  \max \left\{\Big|1-\frac{H_0}{2n}\Big|\right\} < 1.
\end{equation}
In the free field case this condition can be satisfied by choosing $n>3/(2am)$,
and gluons are known to reduce the factor of $3/2$ slightly \cite{Lepage:1992tx}.

In moving NRQCD, $H_0$ can be negative for values of $\bs{k}$ pointing
opposite to the frame velocity.  In this case the two-point function will grow
exponentially, but this is physical as we find the same behavior in the
continuum. In our numerical simulations, which included boost velocities up to
$v=0.6$, we did not encounter any instabilities with $n=2$, $am=2.8$.


The lattice $H_0$ and $\delta H$ are defined as
\begin{eqnarray}
H_0&=& - i \bs{v}\cdot\bs{\Delta}^\pm-\frac{\Delta^{(2)}-
\Delta_v^{(2)}}{2\gamma m} \,,\label{eq:H0}
 \label{eq:kinetic_energy_operator}\\
\nonumber\delta H&=&-\frac{g}{2\gamma m}\bs{\sigma}\!\cdot\!\bs{\tilde{B}'}\\
\nonumber&&-\frac{i}{4\gamma^2 m^2}
\left(\left\{\Delta^{(2)},\:\:\bs{v}\cdot\bs{\Delta}^\pm\right\}
-2\Delta^{(3)}_v\right)\\
\nonumber&&+\frac{g}{8 m^2}\Big(i(\bs{\Delta}^\pm\cdot\bs{\tilde{E}}
-\bs{\tilde{E}}\cdot\bs{\Delta}^\pm)+\bs{v}
\cdot(\bs{\Delta}^{\mathrm{ad}}\times\bs{\tilde{B}})\Big)
\label{eq:adder1}\\
\nonumber&&-\frac{g}{8\gamma m^2}\:\:\bs{\sigma}\cdot
\left(\bs{\tilde{\Delta}}^\pm\times\bs{\tilde{E}'}
-\bs{\tilde{E}'}\times\bs{\tilde{\Delta}}^\pm \right)\\
\nonumber&&+\frac{g}{8(\gamma+1)m^2}\left\{\bs{v}\cdot
\bs{\tilde{\Delta}}^\pm,\:\:\bs{\sigma}\cdot(\bs{v}
\times\bs{\tilde{E}'}) \right\}\\
\nonumber&&-\frac{(2-\bs{v}^2)g}{16 m^2}
\left(\Delta^{\mathrm{ad}}_4+i\bs{v}\cdot\bs{\Delta}^{\mathrm{ad}}\right)
\left(\bs{v}\cdot\bs{\tilde{E}}\right) \label{eq:adder2} \\
\nonumber&&-\frac{ig}{4\gamma^2 m^2}\left\{\bs{v}\cdot
\bs{\tilde{\Delta}}^\pm,\:\:\bs{\sigma}\cdot\bs{\tilde{B}'} \right\}\\
\nonumber&&-\frac{1}{8\gamma^3m^3}\Big(\left(\Delta^{(2)}\right)^2
-3\left\{\Delta^{(2)},\:\Delta_v^{(2)}\right\}+5\Delta_v^{(4)}\Big)\\
\nonumber&&\\
&&+\:\delta H_\mathrm{corr} \,.
\label{eq:dH_full}
\end{eqnarray}
The lattice derivative operators and field strength are defined in Appendix
\ref{app:latt_deriv}. Note that in the continuum the Leibniz rule
$\bs{D}^{\mathrm{ad}}\cdot\bs{E}=\bs{D}\cdot\bs{E}-\bs{E}\cdot\bs{D}$ holds.
For consistency with previous work we discretize the right hand side of this
expression on the lattice.  However, the other adjoint derivatives in the action, 
which enter only at $v\neq0$, are discretized as lattice adjoint derivatives. This is more
efficient and for the term $D^{\mathrm{ad}}_4\left(\bs{v}\cdot\bs{E}\right)$
it is crucial since it avoids a time derivative acting on the quark field.

Note that in the static limit ($m\rightarrow\infty$) one has $H_0=- i
\bs{v}\cdot\bs{\Delta}^\pm$. The symmetric derivative ${\Delta}^\pm$ leads to
zero-energy modes at the corners of the Brillouin zone (``doublers'').  With a
finite mass, these doublers are shifted to higher energy due to the
second-order derivatives in $H_0$. However, the second-order derivatives are
suppressed by a factor of $1/(2\gamma m)$ and hence $\gamma m$ must not be too
large.

The terms in $\delta H_\mathrm {corr}$ provide the spatial and temporal
lattice spacing improvement. We perform tree-level Symanzik improvement
to order $\mathcal{O}(a^4)$, as explained in the next section. This means
that the we expect the leading errors to be of order $\mathcal{O}(\alpha_s a^2)$.

\subsubsection{Improvement corrections}
An $\mathcal{O}(a^4)$-improved version of $H_0$ is given by
\begin{eqnarray}
\tilde{H}_0&=& - i \bs{v}\cdot\bs{\tilde{\Delta}}^\pm-\frac{\tilde{\Delta}^{(2)}
-\tilde{\Delta}_v^{(2)}}{2\gamma m} \label{eq:H0_imp}
\end{eqnarray}
with the improved derivatives given in Appendix \ref{app:latt_deriv}. However,
we do not simply replace $H_0$ by $\tilde{H}_0$.  Let us first consider the
time derivative in the lattice action. Improving it in the standard way would
introduce next-to-nearest neighbor couplings, preventing the use of an
evolution equation like (\ref{eq:hq_ev_eq}). Instead, we try to find an
operator $\tilde{H}_0^*$ such that (explicitly re-introducing the lattice
spacing $a$)
\begin{equation}
\left(1-\frac{a\tilde{H}_0^*}{2n} \right)^{n}=
\:\:\:\exp\left(-\frac{a}{2}\tilde{H}_0\right),
\end{equation}
which yields a more continuum-like behavior~\cite{Lepage:1992tx}.  We obtain
\begin{eqnarray}
a\tilde{H}_0^*&=&2n\left[1-\exp\left(-\frac{a\tilde{H}_0}{2n}\right)\right].
 \label{eq:temp_imp}
\end{eqnarray}
One could now replace $H_0\rightarrow\tilde{H}_0^*$ in the lattice action.
However, for performance reasons and consistency with previous work, we choose
to put all correction terms into $\delta H$. We consider the operator on the
right-hand side of the temporal link in the lattice action
(\ref{eq:mNRQCD_action_kernel}); the operator acting in the timeslice at time
$\tau-1$. Then $\delta H_\mathrm {corr}$, the lattice spacing improvement term
in (\ref{eq:dH_full}) is defined by
\begin{equation}
\left(1-\frac{a\tilde{H}_0^*}{2n} \right)^{n}=\left(1-\frac{aH_0}{2n} \right)^{n}
\left(1-\frac{a\:\delta H_\mathrm {corr}}{2}\right)
\end{equation}
for $\delta H_\mathrm {corr}$. This gives
\begin{eqnarray*}
a\:\delta H_\mathrm {corr}&=&2\left[1-\left(1-\frac{aH_0}{2n} \right)^{-n}
\left(1-\frac{a\tilde{H}_0^*}{2n} \right)^{n}\:\:\right]\\
&=&2\left[1-\left(1-\frac{aH_0}{2n} \right)^{-n}
\exp\left(-\frac{a\tilde{H}_0}{2}\right)\:\:\right],
\label{eq:corr_expansion}
\end{eqnarray*}
and, expanding in powers of $a$,
\begin{eqnarray*}
a\:\delta H_\mathrm {corr}&=&a(\tilde{H}_0-H_0)\\
&&+\,\frac{a^2}{4n}\left(-(1+n)H_0^2-n\tilde{H}_0^2+2n H_0\tilde{H}_0 \right)\\
&&+\,\frac{a^3}{24n^2}\bigg(-(2+3n+n^2)H_0^3\\
&&\hspace{9.5ex}+(3n+3n^2)H_0^2 \tilde{H}_0\\
&&\hspace{9.5ex}-3n^2 H_0 \tilde{H}_0^2+n^2\tilde{H}_0^3 \bigg)\\
&&+\,\frac{a^4}{192n^3}\bigg(-(6+11n+6n^2+n^3)H_0^4\\
&&\hspace{10.5ex}+(8n+12n^2+4n^3)H_0^3\tilde{H}_0 \\
&&\hspace{10.5ex}-(6n^2+6n^3)H_0^2\tilde{H}_0^2\\
&&\hspace{10.5ex}+4n^3H_0\tilde{H}_0^3-n^3\tilde{H}_0^4\bigg)\\
&&+\,\mathcal{O}(a^5) \,.
\end{eqnarray*}
The term $C \equiv \tilde{H_0}-H_0$ is of third order, while $H_0$ is
of first order. Neglecting all operators of order 5 and higher, we
obtain
\begin{eqnarray}
\nonumber a\:\delta H_\mathrm {corr}&=&a\:C-
\frac{a^2}{4n}\left(H_0^2+n[C,\:H_0]  \right)\\
&&-\frac{a^3H_0^3}{12n^2}-\frac{(2+n)a^4 H_0^4}{64n^3}.
\label{eq:tempcorrright}
\end{eqnarray}
Had we considered the operators on the left-hand side of 
the temporal link in the lattice action (\ref{eq:mNRQCD_action_kernel})
instead, the ordering of $H_0$ and $\tilde{H}_0$ would be interchanged, and this would
change the sign of the commutator $[C,\:H_0]$ in (\ref{eq:tempcorrright}),
thereby cancelling the term in the lattice action up to operators of order 5
and higher. We therefore remove this term on both sides.
%
%

Let us go back to lattice units now. Writing $H_0=A+B$ with
\begin{eqnarray*}
A&=&-i \bs{v}\cdot\bs{\Delta}^\pm,\\
B&=&-\frac{\Delta^{(2)}-\Delta_v^{(2)}}{2\gamma m},
\end{eqnarray*}
we obtain
\begin{eqnarray}
\nonumber\delta H_\mathrm {corr}&=&\tilde{H}_0-H_0-\frac{1}{4n}
\left(A^2+\left\{A\:,\: B\right\}+B^2\right)\\
\nonumber&&-\frac{1}{12n^2}\left(A^3+\left\{A^2\:,\:B\right\}+ABA\right)\\
&&-\frac{(2+n)}{64n^3}A^4.\label{eq:tempcorr}
\end{eqnarray}
For performance reasons, we replace some 3rd- and 4th-order derivatives
in (\ref{eq:tempcorr}) by more local expressions (the resulting change is of
order 5 or higher):
\begin{eqnarray*}
(\bs{v}\cdot\bs{\Delta}^\pm)^3&\rightarrow&\Delta_v^{(3)},\\
\big\{\bs{v}\cdot\bs{\Delta}^\pm,\:\Delta_v^{(2)}\big\}
&\rightarrow&2\Delta_v^{(3)},\\
(\bs{v}\cdot\bs{\Delta}^\pm)^4&\rightarrow&\Delta_v^{(4)},\\
(\Delta_v^{(2)})^2&\rightarrow&\Delta_v^{(4)},\\
\big\{(\bs{v}\cdot\bs{\Delta}^\pm)^2\:,\:\Delta^{(2)}\big\}
&\rightarrow&\big\{\Delta_v^{(2)}\:,\:\Delta^{(2)}\big\},\\
\big\{(\bs{v}\cdot\bs{\Delta}^\pm)^2\:,\:\Delta_v^{(2)}\big\}
&\rightarrow&2\Delta_v^{(4)},\\
(\bs{v}\cdot\bs{\Delta}^\pm)\Delta_v^{(2)}(\bs{v}\cdot\bs{\Delta}^\pm)
&\rightarrow&\Delta_v^{(4)},\\
(\bs{v}\cdot\bs{\Delta}^\pm)\Delta^{(2)}(\bs{v}\cdot\bs{\Delta}^\pm)
&\rightarrow&\sfrac12(\bs{v}\cdot\bs{\Delta}^-)
\Delta^{(2)}(\bs{v}\cdot\bs{\Delta}^+)\\
&+&\sfrac12(\bs{v}\cdot\bs{\Delta}^+)\Delta^{(2)}(\bs{v}\cdot\bs{\Delta}^-)\,.
\end{eqnarray*}
This finally gives
\begin{widetext}
\begin{eqnarray}
\nonumber\delta H_\mathrm {corr}&=&\tilde{H}_0-H_0\\
\nonumber&&-\frac{1}{4n}\bigg(-(\bs{v}\!\cdot\!\bs{\Delta}^\pm)^2
+\frac{\big\{i \bs{v}\!\cdot\!\bs{\Delta}^\pm,\:\Delta^{(2)}\big\}
-2i\Delta_v^{(3)}}{2\gamma m}+\frac{(\Delta^{(2)})^2-\big\{\Delta^{(2)},
\:\Delta_v^{(2)}\big\}+\Delta_v^{(4)}}{4\gamma^2 m^2} \bigg)\\
&&\nonumber-\frac{1}{12n^2}\bigg(i\Delta_v^{(3)}
+\frac{\big\{\Delta^{(2)},\:\Delta_v^{(2)}\big\}
-3\Delta_v^{(4)}+\frac12\big((\bs{v}\!\cdot\!\bs{\Delta}^{\!-})
\Delta^{(2)}(\bs{v}\!\cdot\!\bs{\Delta}^{\!+})+(\bs{v}
\!\cdot\!\bs{\Delta}^{\!+})\Delta^{(2)}(\bs{v}\!\cdot\!\bs{\Delta}^{\!-}) \big)}
{2\gamma m}  \bigg)\\
&&-\frac{(2+n)}{64n^3}\:\Delta_v^{(4)}.
\label{eq:final_corr}
\end{eqnarray}
\end{widetext}
The result (\ref{eq:final_corr}) can be simplified further since
most operators are already in the Hamiltonian. 

\subsubsection{Radiative corrections}
In principle, all operators in the Hamiltonian are multiplied by coefficients
$c_i$ which contain radiative corrections that correct for lattice artifacts
appearing beyond tree-level, including the missing contributions of UV modes with 
momenta greater than the lattice cut-off: $|k_\mu| > \pi/a$. They can be expanded 
as a power series in $\alpha_s$:
\begin{equation*}
c_i = c_i^{(0)} + \alpha_s  c_i^{(1)} + \ldots +  (\alpha_s)^n c_i^{(n)} +
\ldots\;,
\end{equation*}
where the tree level $c_i^{(0)} = 1$ and the radiative corrections $c^{(n)}_i$
depend on the bare quark mass and the frame velocity.  These radiative
corrections are calculated using lattice perturbation theory by matching
standard on-shell processes computed in mNRQCD with the continuum counterpart.
Four-quark operators can only arise at $\mathcal{O}(\alpha_s^2)$ and for this
reason will not be considered in our analysis.

For the calculations in this paper, we use the tree level values of the
couplings $c_i$. However, we account for a large amount of the expected
renormalizations via tadpole improvement.

\subsection{Tadpole improvement of the Hamiltonian}
\label{sec:tad_imp}

It is well-known that the perturbative expansion in the bare lattice coupling
is poorly behaved.  Tadpole diagrams, which do not contribute in continuum
schemes, give large contributions to coefficients multiplying powers of the
bare coupling.  Tadpole improvement (also known as mean-field improvement)
fixes this problem by resumming diagrams containing tadpoles
\cite{Lepage:1992xa}.  As tadpole improvement reduces the size of perturbative
corrections, even the tree-level couplings in the action will give accurate
results. Gauge links $U_\mu$ and $U_\mu^\dagger$ in the action and operators
are divided by a factor $u_0$ which is designed to correct for the fact that
the expectation value of the mean link (using some gauge-fixed or
gauge-independent definition) is much less than unity. We choose $u_0$ to be
the mean link in Landau gauge.  The fourth root of the mean plaquette
is another frequently used definition of $u_0$.

Care has to be taken when replacing $U_\mu \mapsto U_\mu/u_0$ and
$U_\mu^\dagger \mapsto U_\mu^\dagger/u_0$ in the action. The action is
composed of Wilson lines or ``paths''. If, due to application of a lattice
derivative for example, the product $U_\mu(x) U_\mu^\dagger(x)$ appears, one
should not multiply by a factor of $1/u_0^2$ since the product is trivial and
does not contribute to tadpole contamination.  Some paths are not explicit in
our simulation code, where we evolve the heavy quark green function by
subsequently applying the individual blocks of the action kernel
(\ref{eq:mNRQCD_action_kernel}) rather than expanding it in terms of paths
first.  Explicitly coding (\ref{eq:mNRQCD_action_kernel}) in terms of products
of link variables would be forbiddingly time-consuming.  Therefore we only take
into account link-pair cancellation separately within $H_0$ and $\delta H$.
Also, no extra cancellations are made when derivative operators act on field
strengths in $\delta H$.

For perturbative studies the tadpole counter-term must be computed to
the appropriate order in $\mathcal{O}(\alpha_s)$. The
tadpole improvement of perturbation theory is discussed in
subsection~\ref{sec:tad_imp_PT}
%
%
%
\section{Renormalization of \lowercase{m}NRQCD}
\label{sec:renorm}

In the previous sections we derived the tree-level continuum mNRQCD Lagrangian
and its lattice version. The radiative corrections to the couplings $c_i$
include a renormalization of the external momentum whose origins are discussed
below. The momentum renormalization is important because it is the coupling of
the $\bs{v}\cdot\bs{D}$ term (= $\bs{P}_0\cdot\bs{k}/\gamma m$,
$\bs{P}_0=\gamma m\bs{v}$) in the action which is leading order in the $1/m$
expansion. The momentum renormalization must be well-determined for accurate
results. Fortunately, as described below, approximate reparametrization
invariance ensures that this renormalization is small; the renormalization 
constant is close to unity.

\subsection{Derivation of the \lowercase{m}NRQCD renormalization parameters}
\label{sec:ren-param}

The low-momentum properties of the moving heavy quark inverse
propagator can be expressed as a general power series in the energy
$p_4$ and the three-momentum $\bs{p}$. The coefficients of this power
series determine the renormalization of the wavefunction $Z_\psi$,
the quark mass $Z_m$, the shift in the origin of energy $E_0$ and of
the frame velocity $Z_v$.

\subsubsection{Wavefunction renormalization}
The wavefunction renormalization $Z_\psi$ can be computed using the
following simple arguments.  The tree-level quark propagator is given
by:
\begin{equation}
G_0(z)=\frac{z}{z-z_0}\,,
\label{eq:treelevel_propagator}
\end{equation}
where $z=e^{ip_4}$ and
\begin{equation}
z_0 \equiv
\left(1-\nobreak\frac{H_0(\bs{p})}{2n}\right)^{\!2n}\!
\left(1-\nobreak\frac{\delta H(\bs{p})}{2}\right)^{\!2}\, .
\label{eq:tree_pole_z}
\end{equation}
Then $z = z_0$ is the on-shell (tree-level) value. At one loop
\begin{equation*}
G^{-1}(z)=G^{-1}_0(z)-\alpha_s\Sigma(z) = Z_\psi^{-1}\frac{z-z_1}{z}
\end{equation*}
where $\alpha_s\Sigma(z)$ is the self-energy (to order $\alpha_s$), containing
both rainbow and tadpole diagrams. Let the new ``one-loop'' on-shell
value be $z_1$, which is the solution of
\begin{equation}
G^{-1}(z_1)=G^{-1}_0(z_1) - \alpha_s\Sigma(z_1)=0\,.
\label{eq:new_on_shell_value}
\end{equation}
Expanding $\Sigma(z)$ around the new on-shell value we have:
\begin{equation}
\Sigma(z)=\Sigma(z_1)+(z-z_1)\left.\frac{\partial\Sigma}{\partial z}
\right|_{z=z_1}+\cdots\;.
\end{equation}
Therefore
\begin{eqnarray*}
G^{-1}(z)&=&\frac1z(z-z_0)-\alpha_s\Big[\Sigma(z_1) \\
&&\quad+\;\;(z-z_1) \left. \frac{\partial\Sigma}{\partial z}
\right|_{z=z_1} + \ldots\Big] \; .
\end{eqnarray*}
Eliminating $z_0$ in this expression in favor of $z_1$ using
(\ref{eq:new_on_shell_value}), we obtain
\begin{equation}
G^{-1}(z)=\frac1z(z-z_1)\Biggl(1-\alpha_s\Big[\Sigma(z_1)
+z\left.\frac{\partial\Sigma}{\partial z}\right|_{z=z_1}\Big]\Biggr)+\cdots\,
\end{equation}
Thus, as $z_1-z_0 = \mathcal{O}(\alpha_s)$, the wavefunction
renormalization is, at one loop,
\begin{eqnarray}
Z_\psi & = & 1+\alpha_s\Big[\Sigma(z_0)
+z\,\left.\frac{\partial\Sigma}{\partial z}\right|_{z=z_0}\Big] \nonumber \\
& = & 1+\alpha_s\Big[\Sigma-i\left.\frac{\partial\Sigma}{\partial p_4}
\right|_{\rm on\,shell}\Big]\,.
\label{eq:wavefunction_renormalization}
\end{eqnarray}

\subsubsection{Other renormalization parameters}
To derive the other renormalization parameters, we use the following
argument which can easily be extended to higher order kinetic terms 
\cite{Morningstar:1994qe}.
At tree level we have in momentum space (up to $\mathcal{O}(p^2)$):
\begin{eqnarray}
  H_0(\bs{p}) &=& \bs{v}\cdot\bs{p}+\frac{\bs{p}^2-(\bs{v}\cdot\bs{p})^2}
		{2\gamma m} + \dots\\
	\notag
	\delta H(\bs{p}) &=& -\frac{1}{4n}(\bs{v}\cdot\bs{p})^2 + \dots \,
\end{eqnarray}
By combining this with (\ref{eq:tree_pole_z}) and
expanding in $\bs{p}$ we find that the pole in the tree level
propagator (\ref{eq:treelevel_propagator}) is given by
\begin{equation}
\omega=\omega_0(\bs{p}) = \bs{v}\cdot\bs{p}+\frac{\bs{p}^2-(\bs{v}\cdot\bs{p})^2}{2\gamma m}
\end{equation}
where $\omega=-ip_4$ is the energy in Minkowski space. At one loop the inverse propagator is
\begin{eqnarray*}
	G(\bs{p},\omega)^{-1} &=& 1 - e^{\omega-\omega_0(\bs{\scriptstyle p})} -
		\alpha_s \Sigma(\bs{p},\omega_0(\bs{p}))
\end{eqnarray*}
so that 
\begin{eqnarray}
	\omega(\bs{p}) &=& \omega_0(\bs{p}) 
		- \alpha_s \Sigma(\bs{p},\omega_0(\bs{p}))
	\label{eq:ren_dispersion_relation}
	\\
	&\equiv&\bs{v}_R\cdot\bs{p} + \frac{\bs{p}^2-(\bs{v}_R\cdot\bs{p})^2}{2\gamma_R m_R} 
		- \alpha_s \delta \omega(\bs{p})\notag
\end{eqnarray}
with $\bs{v}_R = Z_v \bs{v}$, $\gamma_R=(1-\bs{v}_R^2)^{-1/2}$,
$m_R = Z_m m$ and $\alpha_s \delta \omega(\bs{p}) = E_0 + \dots$.
Here and in the following we assume that the boost velocity points in one of the
lattice directions, which guarantees that only the magnitude of $\bs{v}$ is
renormalized. The self energy can now be expanded in small momenta
\begin{equation*}
	\Sigma(\bs{p},\omega) = \Sigma_0(\omega) + \Sigma_v(\omega)\;\bs{v}\cdot\bs{p}
		+ \Sigma_1(\omega) \frac{\bs{p}^2}{2\gamma m} + \dots
\end{equation*} and the renormalization constants can be expressed in 
terms of the coefficients $\Sigma_j^{(\ell)}$ in the expansion
\begin{equation*}
	\Sigma_j(\omega) = \sum_{\ell=0}^\infty \Sigma_j^{(\ell)} \omega^\ell.
\end{equation*}
We find
\begin{eqnarray}
	E_0 &=& \alpha_s\Sigma_0^{(0)},\\
	Z_v &=& 1 -\alpha_s (\Sigma_0^{(1)} + \Sigma_v^{(0)}), \notag\\
	Z_m &=& 1 + \alpha_s((\Sigma_0^{(1)}+\Sigma_1^{(0)}) 
	        + \gamma^2 \bs{v}^2 (\Sigma_v^{(0)}+\Sigma_0^{(1)})),\notag
\end{eqnarray}
and have for the renormalization of the external momentum
$\bs{P} = \gamma_R m_R \bs{v}_R \equiv
Z_p \bs{P}_0$ with
\begin{eqnarray}
	Z_p &=& 1 + \alpha_s(\Sigma_1^{(0)} - \Sigma_v^{(0)}).
\end{eqnarray}
In actual calculations we consider the real parts of parameters 
$\Sigma_j^{(\ell)}$. It is convenient to define
\begin{align}
\Omega_0 & = \mathrm{Re}\:\Sigma_0^{(0)} = \Sigma(0)
\nonumber \,,\\
\Omega_1 & = -\mathrm{Re}\:\Sigma_0^{(1)}
= \mathrm{Im}\:\left.\frac{\partial\Sigma}{\partial p_4}\right|_{p=0}
\nonumber \,,\\
\Omega_2 & = \mathrm{Re}\: \Sigma_1^{(0)}
= \gamma m\mathrm{Re}\:\left.\frac{\partial^2\Sigma}{\partial p_z^2}\right|_{p=0}
\nonumber \,,\\
\Omega_v & = \mathrm{Re}\: \Sigma_v^{(0)}
= \frac1v\mathrm{Re}\:\left.\frac{\partial\Sigma}{\partial p_x}\right|_{p=0}\,,
\end{align}
taking the frame velocity $\bs{v}$ to lie in the $x$-direction. The
renormalization parameters are then expressed as
\begin{align}
Z_\psi & = 1+\alpha_s(\Omega_0+\Omega_1),
 \nonumber\\
E_0 & = \alpha_s\Omega_0,
\nonumber \\
Z_v & = 1-\alpha_s(\Omega_v-\Omega_1)\,.
\nonumber \\
Z_m & = 1+\alpha_s(\Omega_2-\Omega_1) + 
\alpha_s(\Omega_v-\Omega_1)\bs{v}^2\gamma^2\,,
\nonumber \\
Z_p & = 1-\alpha_s(\Omega_v-\Omega_2)\,.
\label{eq:ren_par2}
\end{align}
%
\subsection{Dispersion relation and energy shift}
\label{sec:dispersion_relation}
The renormalized dispersion relation in 
(\ref{eq:ren_dispersion_relation})
has to be compared to the corresponding expression in QCD
\begin{eqnarray}
  \omega^{\rm (QCD)}(\bs{p}) &=& 
	\sqrt{(\gamma_R m_R \bs{v}_R+\bs{p})^2+m_R^2} \\
		&=& \gamma_R m_R + \bs{v}_R\cdot\bs{p}\notag\\&&\qquad+\;\;
		\frac{\bs{p}^2-(\bs{v}_R\cdot\bs{p})^2}{2\gamma_R m_R} + \dots
\notag
\end{eqnarray}
from which one obtains a shift in the zero point energy of a heavy quark 
of 
\begin{eqnarray}
	C_v &=& \omega^{\rm (QCD)}(\bs{p}=0) - \omega(\bs{p}=0) \\
	&=& \gamma_R m_R + E_0 \,. \notag
\end{eqnarray}
We write $C_v = \gamma m(1+\alpha_s \delta C_v+\dots)$ and the one-loop 
correction is given by
\begin{eqnarray}
	\delta C_v &=& \Omega_2-\Omega_1 + \frac{\Omega_0}{\gamma m}\,.
\end{eqnarray}

The shift $C_v$ and the renormalization of the external momentum can
be obtained nonperturbatively by computing the
energy $E_v(\bs{p})$ of a heavy-heavy system which is up to lattice artifacts
given by
\begin{equation*}
	E_v(\bs{p}) + 2C_v = \sqrt{(2Z_p \gamma m\bs{v}
	+\bs{p})^2+M_{\operatorname{kin}}^2} \,.
\end{equation*}
A corresponding dispersion relation with $2C_v \mapsto C_v$ and $2Z_p \mapsto
Z_p$ holds for heavy-light mesons containing only one heavy quark.  We will
compare values for the energy shift and the renormalization of the external
momentum calculated in perturbation theory and nonperturbatively using the
dispersion relation of heavy-heavy and heavy-light mesons in section
\ref{sec:comparison_NP_PT}.
%
%
%
%
%
\subsection{Reparametrization invariance}
\label{sec:reparam-invar}

One thing we expect from our results is that, because of lattice
reparametrization invariance
\cite{Foley:2004rp},
the deviation of the momentum renormalization parameter~$Z_p$ from its
tree level result is much smaller than for other renormalization
parameters. Reparametrization invariance is a symmetry that has been
studied in the context of heavy quark effective theories
\cite{Neubert:1993mb,Luke:1992cs,Finkemeier:1997re}.
This symmetry arises from the fact that the division of the full
momentum $p$ into a ``fixed'' external part $m\: u$ and a ``dynamic''
residual part $k$ is not unique. We can always write $p=m\:
u+k=m\: u'+k'$ where $k'=k - m\epsilon$, $u'=u+\epsilon$. The
4-velocities $u$ and $u'$ have unit norm which implies the constraint
on $\epsilon$ that $2\epsilon\cdot u + \epsilon\cdot\epsilon = 0$. It
can be shown
\cite{Luke:1992cs,Neubert:1993mb}
that this reparametrization of the full momentum is a symmetry of the
effective heavy quark Lagrangian in the continuum.

Because mNRQCD is a non-relativistic formulation Lorentz symmetry is
not manifest in the action. This is apparent from: the form of the FWT
transformation (\ref{eq:TFWT}); the field redefinition (section
\ref{h_order}) required to remove time derivatives in the Hamiltonian;
the truncation of the action to a given order in $1/m$; and the
non-relativistic field normalization (\ref{eq:rescale}). To adapt the
discussion of reparametrization invariance to mNRQCD we study
the ambiguity in division of the total
3-momentum, $\bs{p} = \gamma(\bs{v})m\:\bs{v} + \bs{k}$, keeping
$|\bs{v}|$ fixed since it is a parameter in the Hamiltonian. We first
consider a simple action with Hamiltonian
\begin{equation}
  H_0 = 
    -i\bs{v}\cdot \bs{D} - \frac{\bs{D}^2}{2\gamma m}
  \label{eq:invariant_action}
\end{equation}
omitting the term $(\bs{v}\cdot\bs{D})^2/(2\gamma m)$ for the moment.
This action is invariant under the transformation
\begin{xalignat}{2}
  v_j &\mapsto v_j+\epsilon_j, &
  \psi &\mapsto e^{-i\gamma m {\bs{\epsilon}\cdot\bs{x}}}\psi
  \label{eq:RPI_transformation}
\end{xalignat}
with $2\bs{v}\cdot\bs{\epsilon}+\bs{\epsilon}\cdot\bs{\epsilon} = 0$.
This constraint ensures that $|\bs{v}'| = |\bs{v}|$. 
This is an exact
symmetry which implies that the external momentum $\bs{P}_0 = \gamma m
\bs{v}$ is not renormalized as the relative coefficients of the two
terms in (\ref{eq:invariant_action}) are fixed even after
renormalization.

On the lattice, where we use the discretized Hamiltonian \vspace{-2ex}
\begin{eqnarray}
  H_0^{(\operatorname{lat})} &=& 
  -i\bs{v}\cdot\bs{\Delta}^\pm - \frac{\Delta^{(2)}}{2\gamma a m},
\end{eqnarray}
this symmetry is broken. Under (\ref{eq:RPI_transformation})
$H_0^{(\operatorname{lat})}$ transforms according to
\begin{eqnarray*}
  H_0^{(\operatorname{lat})} &\mapsto& H_0^{(\operatorname{lat})}
  + \frac{1}{2}\gamma m a^2 \sum_j \psi^\dagger v_j \epsilon_j
     \Delta_j^+ \Delta_j^- \psi 
  + {\mathcal O}(\epsilon^2).
\end{eqnarray*}
If $\bs{v}$ is chosen along a lattice axis $v_j = v\delta_{j1}$, say, then
using the constraint on $\epsilon$ the factor $v_j\epsilon_j$ can be replaced
by $-\frac{1}{2}|\bs{\epsilon}|^2\delta_{j1}$ which is small for small
$\epsilon$. We might therefore expect the breaking of reparametrization
invariance by lattice artifacts in this case to be small. In the corresponding
derivation with improved derivatives in $H_0^{(\operatorname{lat})}$, we find
that the lattice artifacts which break reparametrization invariance are of
$\mathcal{O}(a^4)$.

Reparametrization invariance is broken even for the continuum theory unless
the FWT transformation and the truncation of the action as a series in $1/m$
respect it.  The field redefinition, designed to remove time derivatives in
the Hamiltonian, must also be invariant under the reparametrization
transformation. This will be satisfied only if the velocity and the covariant
derivative appear in the combination
\cite{Luke:1992cs,Neubert:1993mb}
\begin{equation}
\bs{v} - \frac{i\bs{D}}{2\gamma m}\;.
\label{eq:RPI}
\end{equation}
This implies that terms of different order in $1/m$ are mixed by the
reparametrization transformation and so any truncation of the action as a
series in $1/m$ will break this invariance. It would be possible to include
selected higher-order terms in $1/m$ by rewriting the action in terms of the
combination (\ref{eq:RPI}) but in practice this is unnecessary since the
approximate reparametrization invariance of the action is sufficient to
restrict the renormalization $Z_p$ of the total quark momentum $P_0$ to be
close to unity. It would also introduce extra terms of little significance in
the non-relativistic expansion but which are expensive to evaluate
computationally for the lattice theory. In any case discretization breaks the
invariance as already discussed.  We shall compute $Z_p$ both perturbatively
and nonperturbatively.

The mixing is evident in our simple example above. It is easy to see that
adding the term $(\bs{v}\cdot\bs{D})^2/(2\gamma m)$ will break the invariance
for non-zero frame velocities even in the continuum. This breaking is
proportional to $v^2/(2\gamma m)$, so it increases to reach a maximum at
$v=\sqrt{2/3}\approx 0.8$ and then drops to zero due to the suppression by
$1/\gamma$. Numerically we find this behavior in our perturbative results for
the simple action we discuss in Appendix
\ref{sec:mnrqcd_simple}.
The one-loop contribution to the external momentum renormalization vanishes
for small $v$, rises to a maximum at $v\approx 0.75$ and then drops again.
At this velocity we also computed $\delta Z_p$ with the action
(\ref{eq:invariant_action})
both with na\"{\i}ve and improved derivatives. We find that the use of
improved derivatives reduces $\delta Z_p$ by roughly a factor 2.

Our numerical results (see Table~\ref{tab:ren_parm_full_action}, to be
discussed in Sec.~\ref{sec:results}) do indeed show
that on the lattice $Z_p$ is very close to 1 for small frame velocities.  For
larger frame velocities the perturbative results show a deviation of $Z_p$
from the tree level value of at most 10\% for practical choices of
frame-velocity $\bs{v}$.
%
%
%
%
%
\subsection{Current construction} \label{sec:current_construction}
For calculations of hadronic matrix elements of weak interaction operators
involving the heavy quark, the continuum QCD currents must be replaced by appropriate
lattice currents. Let us, for example, consider
the vector current \vspace{-1ex}
\begin{equation*}
J^\mu(x)=\overline{q}(x)\dgamma^\mu\Psi(x)
\end{equation*}
where $q(x)$ is the Dirac field of the light quark and $\Psi(x)$ is the Dirac
field of the heavy quark. At tree-level, it suffices to express $\Psi(x)$ via
the Euclidean version of the field redefinition (\ref{eq:MNRQCD_field_redef}).

Recall that Eq. (\ref{eq:TFWT0}) contains a factor of $e^{-im{x'}^0\dgamma^0}$
which removes the mass term from the Lagrangian. For a heavy quark, the lower two components
of the non-relativistic field $\tilde{\Psi}'(x')$ in Eq. (\ref{eq:TFWT0}) are zero,
so that $\dgamma^0 \tilde{\Psi}'(x') = \tilde{\Psi}'(x')$
and hence $e^{-im{x'}^0\dgamma^0} \tilde{\Psi}'(x') = e^{-im{x'}^0} \tilde{\Psi}'(x')$.
Since the FWT transformation in this frame does not contain time derivatives,
the factor $e^{-im{x'}^0}$ can be moved to the left of $T'_{\scriptscriptstyle\mathrm{FWT}}$.
(In the antiquark case, where the upper two components of $\tilde{\Psi}'(x')$
are zero, one has $e^{+im{x'}^0}$.)

Performing the other steps of the derivation in Section \ref{sec:mnrqcd_derivation} again,
it then follows that also the factor of $e^{-im\:u\cdot x\:\dgamma^0}$ in Eq. (\ref{eq:MNRQCD_field_redef})
can be moved to the left of $T_{\scriptscriptstyle\mathrm{FWT}}$ in the case
where ${\xi_v}(x)=0$. Thus, in correlation functions the
factor $e^{-im\:u\cdot x\:\dgamma^0}$ trivially shifts energy and momentum
and can be removed. We obtain
\begin{equation*}
J^\mu(x)=\overline{q}(x)\dgamma^\mu\frac{1}{\sqrt{\gamma}}S(\Lambda)
\:T_{\scriptscriptstyle \mathrm{FWT}}\:\:A_{\scriptscriptstyle D_t}\left(\begin{array}{c}{\psi_v}(x) \\
0 \end{array}\right).
\end{equation*}
For on-shell quantities, time derivatives in $T_{\scriptscriptstyle
\mathrm{FWT}}$ and $A_{\scriptscriptstyle D_t}$ can be eliminated using the
equations of motion, $D_4=i\bs{v}\cdot\bs{D}+\mathcal{O}(1/m)$.  The continuum
derivatives are then replaced by lattice derivatives.

Beyond tree-level, additional lattice operators are required and matching
coefficients must be introduced to correct for the different ultraviolet
behavior of QCD and lattice mNRQCD. These matching coefficients can be
computed perturbatively by comparing matrix elements between on-shell states
in the continuum and lattice theories.

Note that the renormalization of the boost velocity also affects the spinorial
boost matrix $S(\Lambda)$. We have for bare quantities \vspace{-1ex}
\begin{eqnarray}
  S(\Lambda(\bs{v})) &=& \frac{1}{\sqrt{2(1+\gamma)}}\left(\begin{array}{cc} 1
  +\gamma & \gamma\:\bs{\sigma}\cdot\bs{v} \\ \gamma\:\bs{\sigma}\cdot\bs{v} &
  1+\gamma \end{array} \right)\nonumber
\end{eqnarray}
and the renormalized matrix is obtained from this by an additional Lorentz
boost,
\begin{eqnarray}
  S(\Lambda(\bs{v}_R))
  &=& S(\Lambda(\delta \bs{v}))S(\Lambda(\bs{v}))\nonumber.
\end{eqnarray}
(No Wigner rotation is needed here as only the magnitude of $\bs{v}$ is renormalized
for $\bs{v}$ pointing in one of the lattice directions.) We find
\begin{eqnarray}
  S(\Lambda(\delta \bs{v})) &=& \left(
  \begin{array}{cc}
    1 & \frac12 \delta \bs{v}\cdot \bs{\sigma} \\
    \frac12 \delta \bs{v}\cdot \bs{\sigma} & 1
  \end{array}
  \right)\nonumber
\end{eqnarray}
with
\begin{eqnarray}
  \delta \bs{v} = \frac{\bs{v}_R-\bs{v}}{1-\bs{v}\cdot\bs{v}_R}
  &=& \alpha_s \delta Z_v \gamma^2 \bs{v}\nonumber.
\end{eqnarray}
We will not consider the current matching any further here; this
will be discussed in another paper.

\subsection{Lattice Perturbation Theory}
\label{sec:PTh}

Feynman rules for lattice actions are complicated and for all but the
simplest cases an automated procedure is needed to obtain them. The
formalism for this is due to L\"uscher and Weisz
\cite{Luscher:1985wf}.
This was extended by Nobes and Trottier
\cite{Nobes:2003nc}
and Hart et al. 
\cite{Hart:2004bd,Hart:2008zi,Hart:2009nr}
to include both relativistic and non-relativistic fermion actions
such as HISQ 
\cite{Follana:2006rc} 
and, as used here, mNRQCD. In this paper we use the implementation of
Hart et al.
\cite{Hart:2004bd,Hart:2008zi,Hart:2009nr}
to compute the one-loop self-energy $\Sigma(z)$ for various choices of
mNRQCD Hamiltonian. The Feynman rules, vertices and propagators, are
generated in machine-readable form using the Python program \hippy\
and then used in the Fortran 95 code \hpsrc\ to construct the diagrams
and carry out the loop momentum integrations. The latter are done
using \vegas\
\cite{Lepage:1977sw,Lepage:vegas2}
or, in the case of small lattices, by mode-summation. All perturbative
results presented in this paper are obtained on an infinite lattice.

As more correction terms are added to the action, the number of terms
in the perturbative expansion grows very fast, and so does computation
time.  We have used a version of \vegas\ that has been adapted
to parallel computing using MPI (Message Passing Interface).

\begin{figure}
\centering
\includegraphics[width=\linewidth]{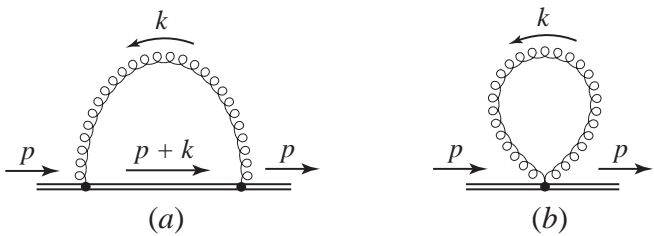}
\caption[Diagrams to be evaluated for the renormalization parameters 
calculation]{Diagrams to be evaluated: ($a$)~rainbow diagram and
($b$)~tadpole diagram. Numerical calculations show that contributions
to the heavy quark self-energy from diagram ($b$) are approximately 
an order of magnitude bigger than those of diagram ($a$),
demonstrating the crucial importance of tadpole improvements for
any lattice-based perturbation theory calculation.}
\label{fig:self_energy}
\end{figure}

The diagrams we evaluate to obtain the heavy quark self energy at one loop 
are shown in Fig.~\ref{fig:self_energy}.
The renormalization parameters require derivatives of the self
energy. The derivatives of the Feynman rules were calculated exactly
(rather than from small finite differences due to their associated
errors and instabilities) and then automatically combined to form
diagram derivatives using code based on the \taylur\ package
\cite{vonHippel:2005dh,vonHippel:2007xd}
(which overloads arithmetic operations so as to respect Leibniz's rule 
and the chain rule).

As an alternative to perturbation theory based on loop integrals,
renormalized quantities may be measured by simulation in the weak
coupling regime of the theory (\textit{i.e.}\ at high $\beta$)
\cite{Trottier:2001vj,Hart:2004jn}
on small lattices using \mbox{'t Hooft} twisted boundary conditions
\cite{tHooft:1979uj,Luscher:1985wf}.
While not the subject of this paper, knowledge from analytic calculation of
the one-loop corrections allows accurate fitting to extract the two-loop
contributions.  To implement twisted boundary conditions is straight-forward;
it requires the spectrum of the momenta used to be appropriately modified and
the vertices to carry a momentum-dependent phase rather than the usual color
factor. We will discuss such calculations for mNRQCD in more detail in a
forthcoming publication.

\subsection{Contour shift}

\begin{figure}
 \begin{center}
   \includegraphics[width=0.7\linewidth]{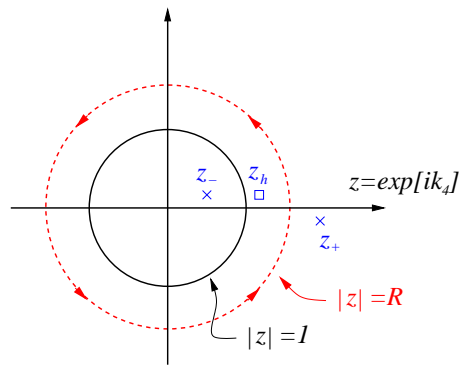}
 \end{center}
 \caption[Integration contour in the $z$ plane]{%
 Position of poles in the complex $z$ plane and integration contour 
 (dashed circle).
 The two poles in the Wilson gluon action are $z_\pm$ with
$z_+z_- = 1$ whereas the heavy quark pole can be found at $z_h$. If $z_h > z_-$
we shift the contour according to $z\mapsto Rz$ with 
$R=\sqrt{z_hz_+} > 1$.
 }\label{fig:poles_in_z_plane}
\end{figure}

For a Euclidean lattice field theory the energy integral is nominally over the
unit circle $|(z=e^{i k_4})|=1$. However, the positions of the poles in the
integrand are functions of the loop three-momentum and care must be taken that
no pole crosses the contour: the contour must be distorted to avoid this
happening. In particular, the heavy quark pole $z_h$ must remain inside the
contour of integration in order to represent a forward-propagating heavy
quark.  This can be done by choosing $|z| = R, R > 1$ where $R$ is chosen so
that the contour is large enough to enclose $z_h$ and as distant from any pole
as is possible to improve convergence of the integration. In
Fig.~\ref{fig:poles_in_z_plane} we show the position of the poles in the $z$
plane.

This contour shift applies to the case of the rainbow diagram
Fig.~\ref{fig:self_energy}a but is not necessary for the tadpole graph in
Fig.~\ref{fig:self_energy}b as the poles in the gluon propagator corresponding
to solutions moving forward/backward in time always come in pairs with $z_+
z_- =1$.

Finding the pole of the heavy quark propagator is straightforward as
the Lagrangian only contains first order time derivatives
\cite{Hart:2006ij}.
Exact expressions for the position of the poles of the Wilson gluon
action can also be derived.  These and the extension to more
complicated gauge actions are discussed in
Appendix~\ref{app:improved_gluon_poles}. There we show that
$|z_-^{(\operatorname{imp})}| < z_- < 1 < z_+ <
|z_+^{(\operatorname{imp})}|$ so that the contour shift derived for
the Wilson action remains valid.

\begin{figure}
 \begin{center}
   \includegraphics[width=0.7\linewidth]{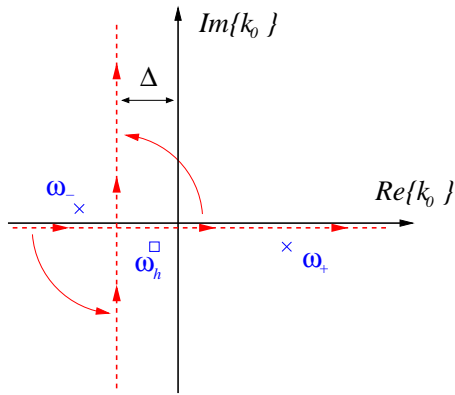}
 \end{center}
 \caption[Wick rotation for HQET in the continuum]{%
 Wick rotation to Euclidean space for continuum HQET; the integration contour is
 shown as a dashed line.
 If the heavy quark pole at $\omega_h = \bs{v}\cdot\bs{k}$
 lies to the left of the imaginary axis the contour has to be rotated around
 $-\Delta=\bs{v}\cdot\bs{k}-\delta$.
 The gluon poles are denoted by $\omega_\pm$.
 }\label{fig:wick_rotation}
\end{figure}

The additional contour shift which is necessary when formulating the
theory in Euclidean space has been discussed in the literature
\cite{Aglietti:1992in,Aglietti:1993hf}. In
Ref.~\cite{Aglietti:1992in}, Aglietti \textit{et al.}\ conclude that deriving
Feynman rules for HQET in the Euclidean theory is problematic as a
simple Wick rotation will generate unphysical solutions propagating
backwards in time. However, in a subsequent paper
\cite{Aglietti:1993hf}, Aglietti extends the analysis and realizes
that this is due to an incorrect rotation of the integration contour
to Euclidean time. To avoid crossing the heavy quark pole at
$\bs{v}\cdot\bs{k}$ it is necessary to rotate the contour around
$-\Delta = \bs{v}\cdot\bs{k}-\delta$ instead of the origin of the
$k_0$ plane (see Fig.~\ref{fig:wick_rotation}).

%
%
%
%
\subsection{Treatment of infrared divergences}
\label{sec:IR-divergences}

To deal with infrared divergences, we note that any lattice theory has
the same infrared behavior as the equivalent continuum theory. Therefore we
consider the diagrams of Fig.~\ref{fig:self_energy} where
lattice Feynman rules have been replaced by equivalent continuum ones
(noting that the two-gluon vertex is still present in continuum
(m)NRQCD).

To analyze the infrared behavior of these diagrams we first
perform the integration over the temporal component of the loop
momentum as a contour integration, then look at the behavior of the
remaining three-dimensional spatial integral for small loop
momentum. 

In the non-moving case ($v=0$) this can conveniently be done
in spherical polar coordinates; for the moving case we need to take
into account the fact that the external velocity introduces a
preferred direction.  It is convenient to take the velocity to lie
along the $x$-axis, for instance. 

After performing these calculations we see that the rainbow diagram
Fig.~\ref{fig:self_energy}a, as well as the tadpole diagram
Fig.~\ref{fig:self_energy}b and all derivatives of the tadpole diagram
are infrared-finite; however the derivatives of the rainbow diagram
behave for low momentum as $\sim\int\frac{dk}{k}$ and thus are
logarithmically divergent.
To regulate this divergence we introduce a small gluon mass
$\lambda$, which we may do because the rainbow diagram has Abelian color
structure. 

To find the infrared behavior of $\Omega_1$, $\Omega_2$, and $\Omega_v$ we
perform the analytic calculations as detailed above, keeping track of all
prefactors in the integration.  After doing this, we obtain the
infrared-divergent part of the derivative of the rainbow diagram:
\begin{equation}
-\frac{2}{3\pi}\log\lambda^2\; 
\label{eq:IR_divergence}
\end{equation}
which is the same as the IR divergence in continuum QCD, using the same
regulator in both theories. In the matching coefficients between lattice
mNRQCD and QCD the logarithmic dependence on the gluon mass will cancel out
and we can set $\lambda = 0$ at the end of the calculation.

We discuss three approaches to verify that this same divergence is present
in the full lattice Feynman integrals.

\subsubsection{Infrared subtraction function}
The first approach is to construct a suitable subtraction function which can
be integrated analytically and has the same infrared behavior as the lattice
integrand. The subtracted lattice integral is then infrared-finite and the
full result can be obtained by adding the analytical expression for the
integral over the subtraction function.  This method was also used in the
current matching in Ref.~\cite{Hart:2006ij}.

Only the wavefunction renormalization (in Feynman gauge) is infrared
divergent. All other renormalization constants are IR finite and can
be computed directly. To construct a suitable subtraction function
$f^{(\operatorname{sub})}$ for $\delta Z_\psi$ we start from the
continuum integral in heavy quark effective theory. (Note that in
principle $f^{(\operatorname{sub})}$ is arbitrary as long as it:
agrees with the lattice integrand for small loop momenta $k$; is
ultraviolet-finite in $d=4$ dimensions; and can be integrated
analytically.)

The logarithmic UV divergence can be regulated without changing the infrared
behavior by replacing
\begin{equation}
  \frac{-i}{k_0-i\bs{v}\cdot \bs{k}} \mapsto
  \frac{2\gamma m}{(k+m\: u)^2+m^2}
\end{equation}
in the (Euclidean) heavy quark propagator. The resulting integral (which is 
\textit{not} restricted to the Brillouin zone) 
is readily evaluated and gives
\begin{equation}
  \delta Z_\psi^{(\operatorname{sub})} = -\frac{2}{3\pi} \log \lambda^2 
  + \mathcal{O}(\lambda/m).
\end{equation}
This is exactly the logarithmic divergence found in
(\ref{eq:IR_divergence}). The subtracted integral
$\delta\overline{Z}_\psi$ is evaluated numerically, defined through
\begin{eqnarray}
  \delta Z_\psi &=& \int\frac{d^4k}{(2\pi)^4} 
  \left(\theta_{\operatorname{BZ}}(k)f^{(\operatorname{lat})}(k) - 
  f^{(\operatorname{sub})}(k) \right)\notag\\&&\qquad+\;\;
\delta Z_{\psi}^{(\operatorname{sub})} \nonumber \\
  &\equiv & \delta \overline{Z}_\psi - \frac{2}{3\pi}\log \lambda^2
\label{eq:defdeltaZbarpsi}
\end{eqnarray}
where $\theta_{\operatorname{BZ}}(k)$ is equal to 1 inside the Brillouin zone 
and vanishes for any $|k_\mu| > \pi/a$.

While this method is easy to carry out for the case of the
self-energy and vertex correction calculations, it becomes
increasingly complicated when considering other calculations.

\subsubsection{Direct calculation for different $\lambda$}

\begin{figure}
\centering
\ifpdf
\includegraphics[width=\linewidth]{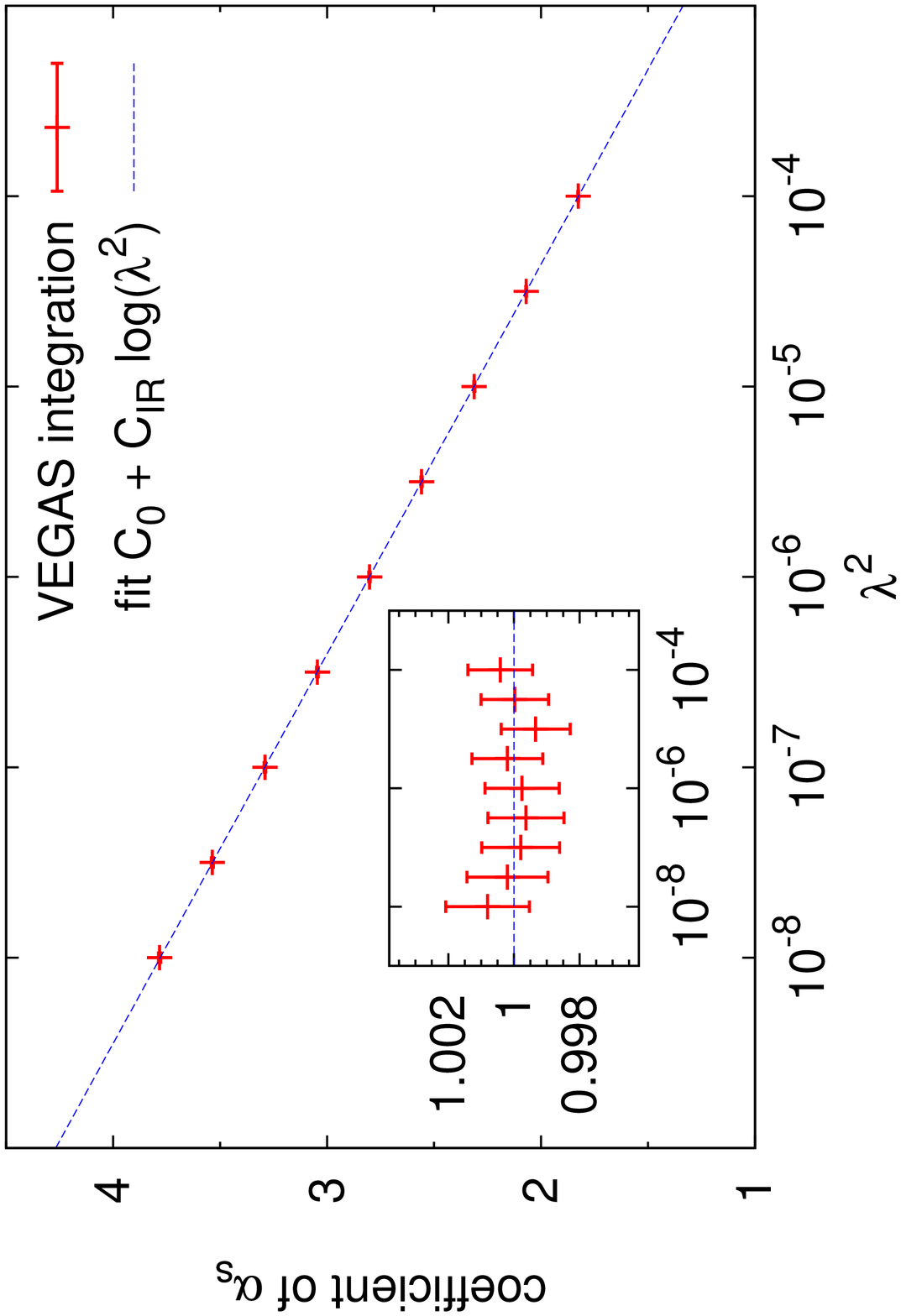}
\else
\includegraphics[height=\linewidth,angle=270]{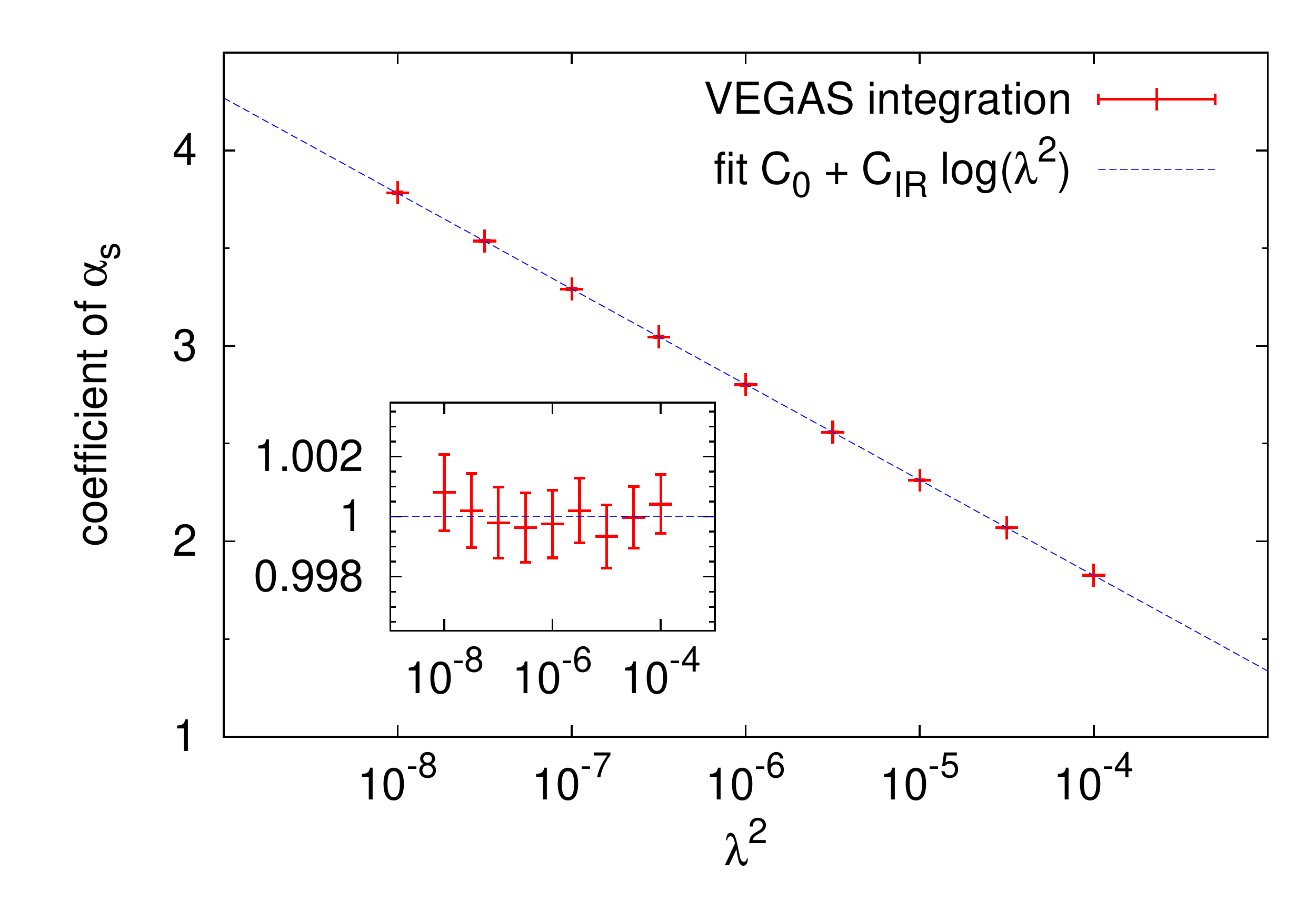}
\fi

\caption[Plot of the wavefunction renormalization for different values of the
infrared regulator]{Plot of the wavefunction renormalization for different
values of the infrared regulator $\lambda$.  The results exhibit a very clear
dependence on $\log\lambda^2$, and by fitting a straight line through the
points one obtains
$C_{\scriptscriptstyle\mathrm{IR}}=-0.21220(14)\simeq-\frac{2}{3\pi}$ and
$C_0=-0.1291(18)$ with a reduced $\chi^2$ of 0.17.  The value of $C_0$ agrees
well with $-0.13124(52)$, the value of $\delta \overline{Z}_\psi$ obtained
directly with $\lambda^2=10^{-6}$ (a less precise value is obtained by adding
$\Omega_0$ and $\Omega_1$ in Table~\ref{tab:Omega_onlyH0_action}).  The inset
shows the data divided by the fit. We use the simple action setup described in
Appendix \ref{sec:mnrqcd_simple}. The frame velocity is $v=0.3$ in this
example.}
\label{fig:deltaZpsi_IRlog}
\end{figure}

The alternative, more generic way of isolating the IR divergent behavior is to
run our integration for different values of $\lambda$ and then obtain the
desired $\log\lambda^2$ behavior by numerically fitting a line through the
points. For example, in Fig.~\ref{fig:deltaZpsi_IRlog} we show the
wavefunction renormalization for $\lambda^2$ varying from $10^{-8}$ to
$10^{-4}$. Using a logarithmic scale on the horizontal axis we see a very
clear linear behavior, which demonstrates the desired dependence on
$\log\lambda^2$. The fit to $C_0 +
C_{\scriptscriptstyle\mathrm{IR}}\log\lambda^2$ yields, with a $\chi^2$ per
degree-of-freedom of $0.17$,
$C_{\scriptscriptstyle\mathrm{IR}}=-0.21220(14)\log\lambda^2$,
which agrees well with the analytic result $-2/(3\pi) = -0.2122\ldots$.

The latter method can be applied to all kinds of calculations such as current
matching calculations in mNRQCD. It can also be used when the expressions for
the diagrams are so complex that obtaining the infrared counterterms
analytically is not feasible.  For the integrands considered here this method
is not very resource- or time-intensive; even a preliminary investigation,
with short integration runs and a small number of sampling points, can yield a
plot with a very good fit, demonstrating clear $\log\lambda^2$-dependence.
For more complicated integrands, subtraction functions may still be necessary:
the computer time required for \vegas\ to sufficiently reduce the statistical
errors as we lower $\lambda^2$ may well be prohibitive and, in addition,
strong IR divergences can confuse the importance sampling used by \vegas.
%
\subsubsection{Twisted boundary conditions}
%
Alternatively infrared divergences can be regulated by working on a
lattice of finite size and using twisted periodic boundary conditions
\cite{tHooft:1979uj,Luscher:1985wf} 
which provide a lower momentum cutoff. We have successfully
implemented and tested this method but will not discuss it further
here. More details will appear in a forthcoming publication.
%
%
%
\subsection{Tadpole improvement}
\label{sec:tad_imp_PT}

The tadpole improvement of the action was described in
Section~\ref{sec:tad_imp}.  We define $u_0$ to be
the mean-link in Landau gauge.  In perturbation theory 
$u_0 = 1 - \alpha_s u_0^{(2)}+ \ldots$, with $u_0^{(2)} = 0.750$ for
the Symanzik-improved gluon action
\cite{Nobes:2001tf}.
Mean-field corrections are then included as counterterms in the action. This
leads to
\begin{equation}
\Omega_j\rightarrow\Omega_j+\Omega_j^{(\mathrm{tadpole})}
\end{equation}
where $\Omega_j^{(\mathrm{tadpole})}$ are the resulting tadpole factors
which we give explicitly below.

We choose the form of the time derivative in
(\ref{eq:latact}) so that the wavefunction
renormalization is immune from mean-field corrections
\cite{Lepage:1992tx}.
Thus we expect (and, indeed, find) that the tadpole improvement
contributions to $\Omega_0$ and $\Omega_1$ are exactly equal and
opposite. The approximate reparametrization invariance implies that
the radiative corrections to $Z_p$ should be small, which suggests the
tadpole corrections to $\Omega_2$ and to $\Omega_v$ should be very
similar. Again, we find this to be the case.

The computation of the tadpole factors was checked in two separate
calculations. We find
\begin{eqnarray*}
\Omega^{(\mathrm{tadpole})}_0 &=& -\Omega^{(\mathrm{tadpole})}_1 \\
&=& u_0^{(2)} \bigg[1+7\frac{3-v^2}{6\gamma m}-\frac{3-6v^2+5v^4}{2\gamma^3m^3}\\
&&\hspace{6ex}+\frac{1}{4n}\left(-v^2+\frac{-3+2v^2-v^4}{\gamma^2m^2}\right)\\
&&\hspace{6ex}+\frac{1}{6n^2}\frac{-5v^2+3v^4}{\gamma m}-\frac{n+2}{16n^3}v^4\bigg],\\[2mm]
%
\Omega^{(\mathrm{tadpole})}_2 &=& -u_0^{(2)} \bigg[\frac{5}{3} +
7\frac{3-v^2}{6\gamma m}+\frac{3-3v^2}{\gamma^2 m^2}\\
&&\hspace{7ex}-\frac{3-6v^2+5v^4}{2\gamma^3 m^3}\\
&&\hspace{7ex}+\;\;\frac{1}{4n}
\left(-v^2+\frac{-3+2v^2-v^4}{\gamma^2m^2}\right)\\
&&\hspace{7ex}+\frac{1}{6n^2}\left(2v^2+\frac{-5v^2+3v^4}{\gamma m}\right)\\
&&\hspace{7ex}-\frac{n+2}{16n^3}v^4\bigg],\\[2mm]
%
\Omega^{(\mathrm{tadpole})}_v &=& \Omega^{(\mathrm{tadpole})}_2
\;-\; u_0^{(2)}\bigg[\frac{2v^2}{\gamma^2 m^2} - \frac{v^2}{6n^2}\bigg] \,.
\end{eqnarray*}
For $v=0$ these expressions reduce to the ones obtained in
\cite{Dalgic:2003uf}.  Numerical values are given below in
Table~\ref{tab:OmegaMF_full_action}.

We give the corresponding expression for an alternative treatment of tadpole
cancellation in Appendix~\ref{app:tadpoles} and list the tadpole improvement
factors for other, simpler actions in
Appendix~\ref{app:results-simple-actions}.  
%
%
%
\subsection{Perturbative results}
\label{sec:results}

In this section we present one-loop perturbative results for the
renormalization of the mNRQCD propagator. Further results for a variety of simpler
mNRQCD actions are given in Appendix~\ref{app:results-simple-actions}.

To obtain agreement with our numerical simulations, it is important that we
use the L\"uscher-Weisz gauge action \cite{Luscher:1984xn,Luscher:1985zq}
which is used for the generation of MILC lattices \cite{Bernard:2001av}. For
the heavy quark self energy at one-loop level, this action is equivalent to
the tree-level Symanzik-improved gauge action
\begin{multline}
\nonumber S_G = 
-\beta \sum_{\genfrac{}{}{0pt}{2}{x}{\mu<\nu}} \left( 
\frac{5}{3} P_{\mu \nu}(x) - \frac{1}{12} R_{\mu\mu\nu}(x)-\frac{1}{12}
R_{\mu\nu\nu}(x) \right)\\
+ \mathcal{O}(\alpha_s)\; ,
\end{multline}
where $P$, $R$ are $1 \times 1$ and $2 \times 1$ Wilson loops
respectively. $\mathcal{O}(\alpha_s)$ denotes possible radiative corrections and
tadpole improvements of the action that only contribute at higher loop orders in the 
perturbative calculation of the heavy quark self energy.

For the squared gluon mass we choose a value of $\lambda^2 = 10^{-6}$.  The
infrared-finite part of the wavefunction renormalization was extracted using a
suitable subtraction function and we also checked that our results are indeed
infrared-finite by varying $\lambda$. The stability parameter is $n=2$ and for
the heavy quark mass we use $m=2.8$. 

In Table \ref{tab:Omega_full_action} we list numerical results for $\Omega_j$
for a range of frame velocities before including mean-field corrections. We
only give the finite parts of the $\Omega_j$, the infrared divergence
$-2/(3\pi) \log \lambda^2$ is not included in the results for $\Omega_1$,
$\Omega_2$ and $\Omega_v$.

\begin{figure}
\centering
\ifpdf
\includegraphics[width=\linewidth]{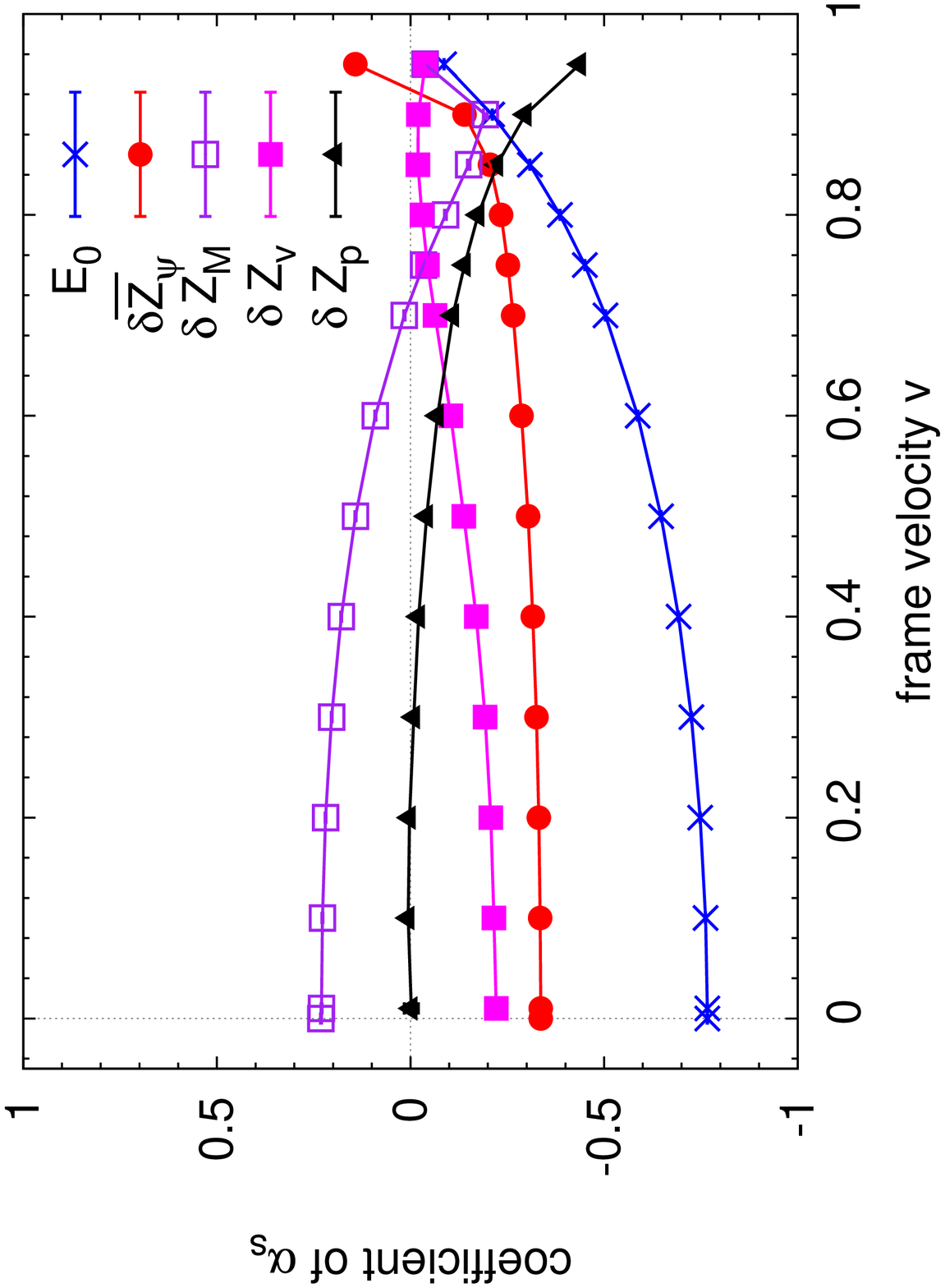}
\else
\includegraphics[height=\linewidth,angle=270]{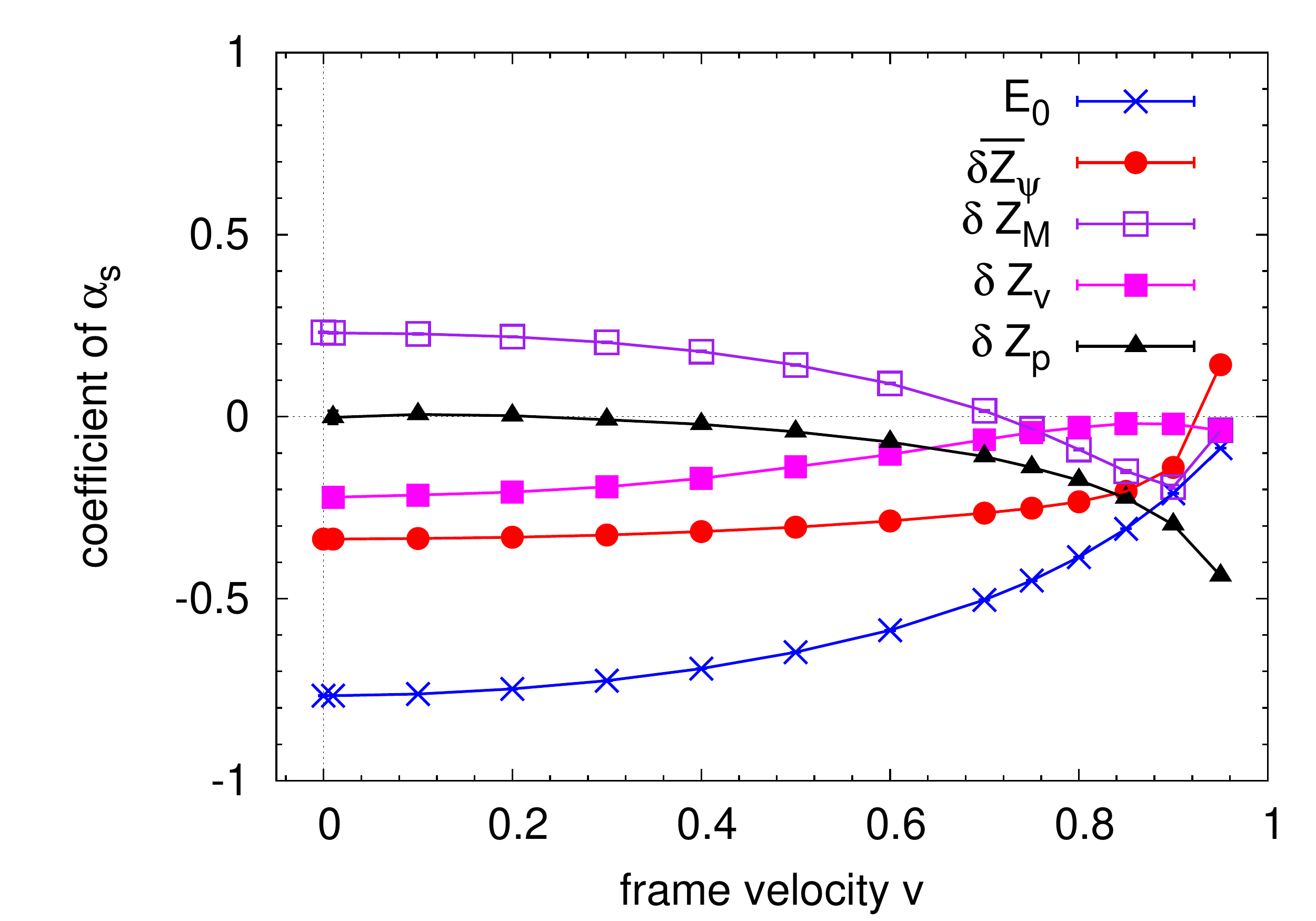}
\fi
\caption[Renormalization parameters for the full
$\mathcal{O}(1/m^2,\vnr^4)$ action]
{Heavy quark renormalization parameters for the full
$\mathcal{O}(1/m^2,\vnr^4)$ action.
The gluon action is Symanzik-improved with
$\lambda^2 = 10^{-6}$ and we use $m=2.8$, $n=2$.
$\delta \overline{Z}_\psi$
is the infrared-finite part of the wavefunction renormalization as defined
in (\ref{eq:defdeltaZbarpsi}).
Violation of reparametrization invariance is very small, 
as indicated by the smallness of $\delta Z_p$.} 
\label{fig:ren_parm_full_action}
\end{figure}

We give results for the tadpole improvement coefficients
$\Omega^{(\mathrm{tadpole})}_j$ in Table~\ref{tab:OmegaMF_full_action}
(see Table~\ref{tab:OmegaMF_full_action_complete} in
Appendix~\ref{app:tadpoles} for an alternative prescription).
 Finally we show the infrared-finite renormalization
parameters, including mean-field corrections, in
Table~\ref{tab:ren_parm_full_action} and Fig.~\ref{fig:ren_parm_full_action}.
In particular, note that the one-loop coefficient renormalizing the
momentum is indeed small, as expected from the arguments presented in
Sec.~\ref{sec:reparam-invar}.

\begin{table}
\begin{tabular}{ccccccccc}
\hline\hline
 \\[-2ex]
  $v$ && $\Omega_0$ && $\Omega_1$ && $\Omega_2$ && $\Omega_v$ \\
 \\[-2ex]
  \hline
$0.00$ && $-2.36685(40)$ && $2.03045(62)$ && $3.0487(13)$ &&  --- \\
$0.01$ && $-2.36672(39)$ && $2.03042(62)$ && $3.0470(13)$ && $3.039(18)$ \\
$0.10$ && $-2.35534(40)$ && $2.02033(62)$ && $3.0276(13)$ && $3.0192(24)$ \\
$0.20$ && $-2.32049(39)$ && $1.98900(62)$ && $2.9668(13)$ && $2.9695(16)$ \\
$0.30$ && $-2.26205(38)$ && $1.93675(62)$ && $2.8646(14)$ && $2.8857(14)$ \\
$0.40$ && $-2.17678(37)$ && $1.86081(61)$ && $2.7199(14)$ && $2.7636(13)$ \\
$0.50$ && $-2.06318(35)$ && $1.75964(61)$ && $2.5330(15)$ && $2.6023(12)$ \\
$0.60$ && $-1.91598(33)$ && $1.62928(62)$ && $2.3020(17)$ && $2.4059(12)$ \\
$0.70$ && $-1.72666(31)$ && $1.46150(63)$ && $2.0220(20)$ && $2.1623(11)$ \\
$0.75$ && $-1.61272(30)$ && $1.36128(65)$ && $1.8614(24)$ && $2.0247(11)$ \\
$0.80$ && $-1.48224(28)$ && $1.24847(69)$ && $1.6828(29)$ && $1.8794(11)$ \\
$0.85$ && $-1.33083(27)$ && $1.12528(82)$ && $1.4925(41)$ && $1.7275(12)$ \\
$0.90$ && $-1.15125(25)$ && $1.0118(11)$ && $1.2930(68)$ && $1.5972(15)$ \\
$0.95$ && $-0.92738(24)$ && $1.0698(21)$ && $1.236(19)$ && $1.6559(25)$ \\
\hline\hline
\end{tabular}
\caption[$\Omega_j$ for the full $\mathcal{O}(1/m^2)$ action]
{Infrared-finite part of $\Omega_j$ for the full
$\mathcal{O}(1/m^2,\vnr^4)$ action. The gluon action
is Symanzik-improved with $\lambda^2 = 10^{-6}$ and we use $m=2.8$, $n=2$.
Mean-field corrections are not included and the errors shown are purely
statistical from the \vegas\ integration.}
\label{tab:Omega_full_action}
\end{table}

\begin{table}
\begin{tabular}{cccc}
\hline\hline
 \\[-2ex]
$v$ & $\Omega^{(\mathrm{tadpole})}_0/u_0^{(2)}$
& $\Omega^{(\mathrm{tadpole})}_2/u_0^{(2)}$
& $\Omega^{(\mathrm{tadpole})}_v/u_0^{(2)}$ \\
 \\[-2ex]\hline
$0.00$ & $2.13384$ & $-3.18316$ &  --- \\
$0.01$ & $2.13375$ & $-3.18300$ & $-3.18302$\\
$0.10$ & $2.12459$ & $-3.16713$ & $-3.16923$\\
$0.20$ & $2.09650$ & $-3.11915$ & $-3.12728$\\
$0.30$ & $2.04863$ & $-3.03967$ & $-3.05682$\\
$0.40$ & $1.97963$ & $-2.92963$ & $-2.95725$\\
$0.50$ & $1.88797$ & $-2.79071$ & $-2.82813$\\
$0.60$ & $1.77221$ & $-2.62561$ & $-2.66939$\\
$0.70$ & $1.63091$ & $-2.43793$ & $-2.48127$\\
$0.75$ & $1.54999$ & $-2.33677$ & $-2.37612$\\
$0.80$ & $1.46143$ & $-2.23103$ & $-2.26313$\\
$0.85$ & $1.36379$ & $-2.12013$ & $-2.14118$\\
$0.90$ & $1.25365$ & $-2.00163$ & $-2.00714$\\
$0.95$ & $1.12074$ & $-1.86625$ & $-1.85110$\\
\hline\hline
\end{tabular}
\caption[Tadpole improvement corrections for the full
$\mathcal{O}(1/m^2,\vnr^4)$ action.]
{Tadpole improvement corrections $\Omega^{(\mathrm{tadpole})}_j$ for
the full $\mathcal{O}(1/m^2,\vnr^4)$ action. The heavy quark mass is $m=2.8$
and the stability parameter $n=2$. Note that $\Omega^{(\mathrm{tadpole})}_1
= - \Omega^{(\mathrm{tadpole})}_0$.}
\label{tab:OmegaMF_full_action}
\end{table}

\begin{table*}
\begin{center}
\begin{tabular}{ccccccccccccc}
\hline\hline
 \\[-2ex]
  $v$ && $E_0$ && $\delta \overline{Z}_\psi$ && $\delta Z_m$ && $\delta Z_v$ && $\delta Z_p$ && $\delta C_v$\\
 \\[-2ex]\hline
$0.00$ && $-0.76647(40)$ && $-0.33639(48)$ && $0.2313(12)$ &&  --- &&  --- && $-0.0425(12)$ \\
$0.01$ && $-0.76641(39)$ && $-0.33630(47)$ && $0.2297(12)$ && $-0.221(18)$ && $-0.002(18)$ && $-0.0441(12)$\\
$0.10$ && $-0.76190(40)$ && $-0.33501(47)$ && $0.2275(12)$ && $-0.2154(23)$ && $0.0061(20)$ && $-0.0454(12)$\\
$0.20$ && $-0.74812(39)$ && $-0.33149(48)$ && $0.2194(12)$ && $-0.2074(15)$ && $0.0025(12)$ && $-0.0510(12)$\\
$0.30$ && $-0.72558(38)$ && $-0.32530(48)$ && $0.2037(12)$ && $-0.1928(12)$ && $-0.0087(10)$ && $-0.0626(12)$\\
$0.40$ && $-0.69206(37)$ && $-0.31597(49)$ && $0.1789(13)$ && $-0.1696(11)$ && $-0.02131(89)$ && $-0.0799(13)$\\
$0.50$ && $-0.64720(35)$ && $-0.30354(50)$ && $0.1421(13)$ && $-0.1376(10)$ && $-0.04175(86)$ && $-0.1039(14)$\\
$0.60$ && $-0.58682(33)$ && $-0.28670(52)$ && $0.0910(15)$ && $-0.1037(10)$ && $-0.06974(87)$ && $-0.1350(16)$\\
$0.70$ && $-0.50349(31)$ && $-0.26516(55)$ && $0.0158(17)$ && $-0.06305(91)$ && $-0.10943(92)$ && $-0.1731(19)$\\
$0.75$ && $-0.45023(30)$ && $-0.25144(58)$ && $-0.0337(19)$ && $-0.04380(89)$ && $-0.1394(10)$ && $-0.1964(23)$\\
$0.80$ && $-0.38616(28)$ && $-0.23377(63)$ && $-0.0901(24)$ && $-0.02967(90)$ && $-0.1746(11)$ && $-0.2256(29)$\\
$0.85$ && $-0.30798(27)$ && $-0.20554(77)$ && $-0.1502(32)$ && $-0.01915(93)$ && $-0.2235(11)$ && $-0.2580(40)$\\
$0.90$ && $-0.21101(25)$ && $-0.1395(11)$ && $-0.1933(51)$ && $-0.0203(10)$ && $-0.2966(13)$ && $-0.3127(67)$\\
$0.95$ && $-0.08682(24)$ && $0.1425(21)$ && $-0.038(14)$ && $-0.0383(13)$ && $-0.4374(16)$ && $-0.402(18)$\\

\hline\hline
\end{tabular}
\caption[Renormalization parameters for the full $\mathcal{O}(1/m^2,\vnr^4)$ action]
{Heavy quark renormalization parameters for the full
$\mathcal{O}(1/m^2,\vnr^4)$ action. The gluon action is Symanzik
improved with $\lambda^2 = 10^{-6}$ and we use $m=2.8$, $n=2$. All mean-field
corrections are
included and the results are infrared-finite. The errors shown are purely
statistical from the \vegas\ integration.}
\label{tab:ren_parm_full_action}
\end{center}
\end{table*}

%
%
%
\section{Numerical simulation results}
\label{sec:simulation_results}

In addition to the perturbative calculations described in the previous
sections, we have performed a wide range of nonperturbative computations with
the full mNRQCD action on unquenched gluon configurations.  We have computed
two-point correlation functions for various heavy-heavy and heavy-light mesons
at different momenta and boost velocities. These allow the extraction of both
energies and amplitudes. From the combination of simulation energies at
different momenta, we have obtained nonperturbative results for the external
momentum renormalization, the energy shift and the kinetic masses of the
mesons. We have also examined the dependence of several energy splittings on
the boost velocity. In addition to these spectral properties, we studied the
behavior of decay constants.

The next section describes the simulations with heavy-heavy mesons and is
followed by a section on heavy-light mesons. All results are given in lattice
units.  

\begin{table}[ht]
\begin{center}
\begin{tabular}{lllllll@{~~~~}l}
\hline\hline
 \\[-2ex]
Name & $n$ & $L$ & $S$ & $J$ & $P$ & $C$ & $\Gamma(\bs{r})$ \\
 \\[-2ex]
\hline
 \\[-2ex]
$\eta_b(1S)$ & 1 & 0 & 0 & 0 & $-$ & $+$ & $\displaystyle
\exp[-|\bs{r}|/r_s]\:\dgamma^5$ \\
 \\[-2ex]
$\eta_b(2S)$ & 2 & 0 & 0 & 0 & $-$ & $+$ & $\displaystyle
\left[1-|\bs{r}|/(2r_s)\right]\:\exp[-|\bs{r}|/(2r_s)]\:\dgamma^5$ \\
 \\[-2ex]
$\Upsilon(1S)$ & 1 & 0 & 1 & 1 & $-$ & $-$ & $\displaystyle
\exp[-|\bs{r}|/r_s]\:\dgamma^j$ \\
 \\[-2ex]
$\Upsilon(2S)$ & 2 & 0 & 1 & 1 & $-$ & $-$ & $\displaystyle
\left[1-|\bs{r}|/(2r_s)\right]\:\exp[-|\bs{r}|/(2r_s)]\:\dgamma^j$ \\
 \\[-2ex]
$\chi_{b1}(1P)$ & 1 & 1 & 1 & 1 & $+$ & $+$ & $\displaystyle
\:\exp[-|\bs{r}|/(2r_s)]\:(\bs{r}\times\bs{\dgamma})^j/r_s$ \\
 \\[-2ex]
\hline\hline
\end{tabular}
\end{center}
\caption{Some (continuum) quantum numbers and smearing functions for the
bottomonium system.}
\label{tab:bott_ops}
\end{table}

\subsection{Heavy-heavy mesons}

\subsubsection{Methods}

We begin by constructing ``smeared'' interpolating fields for
quarkonium. To demonstrate the effect of the moving NRQCD field
redefinition, we start the construction with the QCD fields
$\overline{\Psi}$, $\Psi$. A meson with momentum $\bs{p}$ can be
obtained from
\begin{equation}
\nonumber O_\Gamma(\bs{p}, \bs{\tau})=\sum_{\bs{x}_1,\bs{x}_2}\overline{\Psi}
(\bs{x}_1,\tau)\Gamma(\bs{x}_1-\bs{x}_2)\Psi(\bs{x}_2,\tau)e^{-i\bs{p}
\frac{\bs{x}_1+\bs{x}_2}{2}}
\end{equation}
where $\Gamma(\bs{r})$ is a Dirac-matrix-valued smearing function. We do not
include gauge links in $\Gamma(\bs{r})$; instead we fix the gauge
configurations to Coulomb gauge. The (continuum) quantum numbers and
corresponding functions $\Gamma(\bs{r})$ used in the simulations are listed in
Table \ref{tab:bott_ops}.

We now express $\overline{\Psi}$ and $\Psi$ through the tree-level
moving NRQCD field redefinition. To lowest order one has
\begin{eqnarray*}
\Psi(x)&=&\frac{1}{\sqrt{\gamma}}S(\Lambda)e^{-i\gamma m( -i\tau-\bs{v}
\cdot\bs{x})\dgamma^0}\Psi_v(x),\\
\overline{\Psi}(x)&=&\frac{1}{\sqrt{\gamma}}\:\overline{\Psi_v}(x)
\: e^{i\gamma m( -i\tau-\bs{v}\cdot\bs{x})\dgamma^0}\:\:\overline{S(\Lambda)}.
\end{eqnarray*}
Let us, for example, consider the $\Upsilon$ states with polarization
$j=1,2,3$.  We allow different smearing at source and sink, so that
$\Gamma_{\rm sc}(\bs{r})=\dgamma^j\:f_{\rm sc}(\bs{r})$ and $\Gamma_{\rm
sk}(\bs{r})=\dgamma^j\:f_{\rm sk}(\bs{r})$. Using
\begin{equation}
\overline{S(\Lambda)} \: \dgamma^j \: 
S(\Lambda)=\Lambda^j_{\:\:\:\mu}\dgamma^\mu,
\end{equation}
we obtain
\begin{widetext}
\begin{eqnarray}
\nonumber O_{\Gamma_{\rm sk}}(\bs{p},\tau)O_{\Gamma_{\rm sc}}^\dag(\bs{p},\tau')
&=&\frac{1}{\gamma^2}e^{-2\gamma m(\tau-\tau')}\sum_{\bs{x}_1, \bs{x}_2, 
\bs{x}'_1, \bs{x}'_2 }e^{-i\bs{k}\frac{\bs{x}_1+\bs{x}_2}{2}}f_{\rm sk}
(\bs{x}_1-\bs{x}_2)e^{i\bs{k}\frac{\bs{x}'_1+\bs{x}'_2}{2}}
f_{\rm sc}(\bs{x}'_1-\bs{x}'_2)\\
&&\times\:\:\Lambda^j_{\:\:\:l}\Lambda^j_{\:\:\:m}
\xi_v^\dag(\bs{x}_1,\tau)\sigma^l\psi_v(\bs{x}_2,\tau)
\psi_v^\dag(\bs{x}'_2,\tau')\sigma^m\xi_v(\bs{x}'_1,\tau')+\:\:...\label{eqn:elipsis}
\end{eqnarray}
(no summation over $j$ here) where 
\begin{equation}
\bs{k}\equiv\bs{p}-2\gamma m\bs{v}. \label{eqn:momentumrel}
\end{equation}
The ellipsis in (\ref{eqn:elipsis}) denotes terms that do not
contribute to the connected meson correlator for $\tau>\tau'$.
The correlator is then given by
\begin{eqnarray}
\nonumber\langle O_{\Gamma_{\rm sk}}(\bs{p},\tau)O_{\Gamma_{\rm sc}}^\dag(\bs{p},\tau')
\rangle&=&\frac1N\sum_U\frac{1}{\gamma^2}e^{-2\gamma m(\tau-\tau')}
\sum_{\bs{x}_1, \bs{x}_2, \bs{x}'_1, \bs{x}'_2 }e^{-i\bs{k}\frac{\bs{x}_1
+\bs{x}_2}{2}}f_{\rm sk}(\bs{x}_1-\bs{x}_2)e^{i\bs{k}\frac{\bs{x}'_1+\bs{x}'_2}{2}}
f_{\rm sc}(\bs{x}'_1-\bs{x}'_2)\nonumber\\
&&\times\:\:\Lambda^j_{\:\:\:l}\Lambda^j_{\:\:\:m}\mathrm{Tr}\left(\sigma^l
\left[ G_{\psi_v}^{U,\bs{v}}\left((\bs{x}_2,\tau),(\bs{x}'_2,\tau')\right)\right]
\sigma^m\left[ G_{\psi_v}^{U,-\bs{v}}\left((\bs{x}_1,\tau),(\bs{x}'_1,\tau')
\right)\right]^\dag\right),
\end{eqnarray}
\end{widetext}
where we average over $N$ gauge configurations $U$. The trace is over color
and spin indices. We have also used equation (\ref{eqn:q_aq_rel}) to express
the antiquark green function $G_{\xi_v}^{U,\bs{v}}$ in terms of the quark green
function $G_{\psi_v}^{U,-\bs{v}}$ with the opposite boost velocity.

The summations over all quark and antiquark source locations would render the
lattice computation too expensive. Therefore, using translation invariance, we
remove the summation over the antiquark source location
$\bs{x}_1'$. Furthermore, we remove the factor of $e^{-2\gamma m(\tau-\tau')}$
which corresponds to the tree-level energy shift.  Hence, the quantity
\begin{widetext}
\begin{eqnarray*}
C(\Gamma_{\rm sk}, \Gamma_{\rm sc}, \bs{k}, \tau, \tau')&=&\frac1N\sum_U
\frac{1}{\gamma^2}\sum_{\bs{x}_1,\bs{x}_2}e^{-i\bs{k}\frac{\bs{x}_1
+\bs{x}_2}{2}}f_{\rm sk}(\bs{x}_1-\bs{x}_2)\\
&&\times\Lambda^j_{\:\:\:l}\Lambda^j_{\:\:\:m}\mathrm{Tr}
\left(\sigma^l\left[ \tilde{G}_{\psi_v}^{U,\bs{v}}\left((\bs{x}_2,\tau),
(\bs{x}'_1,\tau')\right)\right]\sigma^m\left[ G_{\psi_v}^{U,-\bs{v}}
\left((\bs{x}_1,\tau),(\bs{x}'_1,\tau')\right)\right]^\dag\right)
\label{eqn:lat_corr}
\end{eqnarray*}
with
\begin{equation}
\tilde{G}_{\psi_v}^{U,\bs{v}}\left((\bs{x}_2,\tau),(\bs{x}'_1,\tau')\right)
=\sum_{\bs{x}'_2}e^{i\bs{k}\frac{\bs{x}'_1+\bs{x}'_2}{2}}
f_{\rm sc}(\bs{x}'_1-\bs{x}'_2)G_{\psi_v}^{U,\bs{v}}\left((\bs{x}_2,\tau),
(\bs{x}'_2,\tau')\right)
\label{eqn:initial_green}
\end{equation}
\end{widetext}
is computed on the lattice. The correlator (\ref{eqn:initial_green}) can be
computed by using the function
\begin{equation}
e^{i\bs{k}\frac{\bs{x}'_1+\bs{x}'_2}{2}}f_{\rm sc}(\bs{x}'_1-\bs{x}'_2)
\label{eqn:src_momentum}
\end{equation}
as the initial condition in the mNRQCD evolution equation (\ref{eq:hq_ev_eq}).
The momentum-dependent phase factor $\exp(i\bs{k}(\bs{x}'_1+\bs{x}'_2)/2)$ at
the source improves the overlap with the momentum considered. However, since
there is no sum over $\bs{x}'_1$, one may omit this factor to allow the
calculation of correlators with different momenta from the same source. 

In order to maintain the periodic boundary conditions,
we set $f(\bs{r})$ to zero for $|\bs{r}|>R_s$ with some cut-off radius $R_s$
smaller than half the length of the lattice.

On the finite volume lattice with periodic boundary conditions, the momentum
$\bs{k}$ takes on discrete values, $k_j={2\pi\:n_j}/{L_j}$ where $L_j$ are the
spatial extents of the lattice. However, the physical meson momentum $\bs{p}$
is expected to deviate from the tree-level relation (\ref{eqn:momentumrel}),
since mass and velocity are renormalized. One has
\begin{equation}
\bs{p}=2Z_p\bs{P}_0+\bs{k}\hspace{3ex}\mathrm{with}\hspace{3ex}\bs{P}_0
=\gamma m\bs{v}.  \label{eqn:momshift}
\end{equation}
We fit a matrix of correlators with different smearings at source
and sink with the functional form
\begin{eqnarray}
C(\Gamma_{\rm sk}, \Gamma_{\rm sc}, \bs{k}, \tau, \tau')\rightarrow
A^{\rm sk}(A^{\rm sc})^* \bigg[e^{-E(\tau-\tau')}\hspace{4ex}\nonumber \\
\nonumber \\ \hspace{4ex}+\sum_{n=1}^{n_{\rm exp}-1}B^{\rm sk}_n(B^{\rm sc}_n)^*
e^{-(E+\Delta E_1+...+\Delta E_n)(\tau-\tau')}\bigg]
\label{eqn:fitfunction}
\end{eqnarray}
where $E$ is the energy of the meson ground state, $A^{\rm sc}$ and
$A^{\rm sk}$ are the (real) ground state amplitudes of the operators at
source and sink and $B^{\rm sc}_n$, $B^{\rm sk}_n$ are (real) amplitudes
for the $n$-th excited state, relative to the ground state amplitude.
We use the constrained fitting method described in \cite{Lepage:2001ym},
and increase the number of exponentials until the fit results and error
estimates become independent of $n_{\rm exp}$.

The full (physical) energy differs from the energy $E=E_v(\bs{k})$
obtained from the fit by twice the mNRQCD energy shift,
\begin{equation}
E_{\rm phys}=E_v(\bs{k})+2C_v. 
\label{eqn:Ephys}
\end{equation}
In perturbation theory, one has
\begin{equation}
C_v=Z_m Z_\gamma \gamma m + E_0.
\end{equation}
Given expression (\ref{eqn:momshift}) for the full (physical) momentum,
we expect that, up to lattice artifacts,
\begin{eqnarray}
\nonumber E_{\rm phys}&=&\sqrt{\bs{p}^2+M_\mathrm{kin}^2}\\
&=&\sqrt{(2Z_p\bs{P}_0+\bs{k})^2+M_\mathrm{kin}^2}
\label{eqn:phys_energy}
\end{eqnarray}
where $M_\mathrm{kin}$ is the kinetic mass of the meson.

Using (\ref{eqn:phys_energy}), we can obtain nonperturbative results
for $C_v$, $Z_p$ and $M_{\rm kin}$ from the energies at various non-zero lattice
momenta in combination with the energy at $\bs{k}=0$:
\begin{eqnarray}
C_v&\!=\!&\frac12\frac{\bs{k}^2_\perp-\left(E^2_v(\bs{k}_\perp)
-E^2_v(0)\right)}{2(E_v(\bs{k}_\perp)-E_v(0))},
\label{eq:C_v}\\
\nonumber Z_p&\!=\!&\frac{E^2_v(\bs{k}_\parallel)\!-
\!E^2_v(-\bs{k}_\parallel)\!+\!4C_v(E_v(\bs{k}_\parallel)
\!-\!E_v(-\bs{k}_\parallel))}{4\bs{k}_\parallel\cdot2\bs{P_0}},\\
 \label{eq:Z_p}\\
M_\mathrm{kin}&\!=\!&\sqrt{(E_v(\bs{k})+2C_v)^2
-(2Z_p\bs{P_0}+\bs{k})^2}.
\label{eq:M_kin}
\end{eqnarray}
Here, $\bs{k}_\parallel$ is parallel to $\bs{v}$, and $\bs{k}_\perp$ is
perpendicular to $\bs{v}$. In order to fully take into account correlations in
the energies at different momenta, we use the bootstrap method, performing
fits on 500 bootstrap ensembles and computing the final quantity 500
times. The errors are then estimated as the 68\% width of the resulting
distribution.

Ultimately we will be interested in semileptonic $B$ decay matrix elements.
As a simpler test we first study the decay of the $\eta_b(1S)$ meson via a
fictitious axial vector current.  The corresponding decay constant is
defined by
\begin{equation}
\langle 0 | \mathsf{J}_5^\mu(0) |\eta_b(1S),\bs{p}\rangle = if\:p^\mu.
\label{eqn:eta_decay_const}
\end{equation}
Here, $\mathsf{J}_5^\mu$ is the mNRQCD field operator associated with the
axial current
\begin{equation}
J_5^\mu(x)=\overline{\Psi}(x)\:\dgamma^5\dgamma^\mu\:\Psi(x).
\label{eqn:axialcurrent}
\end{equation}
For simplicity, we have only considered the temporal component
and, as above, used only the leading-order tree-level mNRQCD field
redefinition to construct the lattice current. To extract the amplitude,
we compute $2\times2$ matrix correlators with the local smearing function
\begin{equation}
\Gamma(\bs{r})=\delta(\bs{r})\:\dgamma^5\dgamma^0
\end{equation}
for the temporal axial current, and the $\eta_b(1S)$ smearing function from Table
\ref{tab:bott_ops}. The product of the ground state amplitudes in
(\ref{eqn:fitfunction}) is given by
\begin{eqnarray}
\nonumber A^{\rm sk}(A^{\rm sc})^*&=&\frac{1}{2E_{\rm phys}}
\langle \eta_b(1S),\bs{p}|\mathsf{O}_{\Gamma_{\rm sk}}(0)|0\rangle \\
&&\hspace{3.2ex}\times\:\:\:\:\:\:\langle 0|\mathsf{O}_{\Gamma_{\rm sc}}(0)
|\eta_b(1S),\bs{p}\rangle,
\label{eqn:eqn:eta_decay_const_ampl}
\end{eqnarray}
as can be seen from the spectral decomposition of the two-point correlator.
Using (\ref{eqn:eta_decay_const}) with $p^0=E_{\rm phys}$, (\ref{eqn:Ephys})
and (\ref{eqn:eqn:eta_decay_const_ampl}), we obtain
\begin{equation}
f=A\:\sqrt{\frac{2}{E_v(\bs{k})+2C_v}}
\label{eq:f_Cv}
\end{equation}
where $A=A^{\rm sk/sc}$ is the amplitude from the fit corresponding to
$\Gamma_{\rm sk/sc}=\delta(\bs{r})\:\dgamma^5\dgamma^0$.

\subsubsection{Lattice parameters}

The computations were performed using 400 MILC gauge configurations (fixed to
Coulomb gauge) of size $20^3\times64$ with 2+1 flavors of rooted staggered
light quarks, at $\beta=6.76$ \cite{Bernard:2001av}.  The light quark masses
were $m_u=m_d=0.007$ and $m_s=0.05$ (in the MILC convention for lattice
masses).  The Landau gauge mean link, used in the mNRQCD action, was
$u_0=0.836$.  The inverse lattice spacing of these ``coarse'' MILC
configurations is known to be approximately 1.6 GeV \cite{Gray:2005ur}.

Heavy quark propagators were computed using full mNRQCD lattice action
described in section \ref{sec:lattice-mnrqcd} and used in the perturbative
calculation.  The bare heavy quark mass was set to $m=2.8$, which gave the
correct $\Upsilon$ kinetic masses using non-moving NRQCD
\cite{Gray:2005ur}. The boost velocity was always pointing in the
$x$-direction, $\bs{v}=(v,0,0)$.  The stability parameter was set to $n=2$.

In order to increase statistics, between 16 and 120 correlators with different
origins $(\bs{x}'_1,\tau')$ spread over the lattice were calculated and
averaged over on each gauge configuration. These origins were also shifted
randomly to reduce autocorrelations. The smearing parameter $r_s$ was set to 1
for the S wave states and 0.5 for the P wave states.

\subsubsection{Results}
\label{sec:heavy_heavy_results}

\begin{table*}
\begin{center}
\begin{tabular}{lllllllll}
\hline \hline \\[-2ex]
& & \multicolumn{3}{c}{ $|\bs{k}_\perp|=|\bs{k}_\parallel|=2\pi/L$ }
& & \multicolumn{3}{c}{ $|\bs{k}_\perp|=|\bs{k}_\parallel|=4\pi/L$ } \\
 \\[-2ex]
\hline
 \\[-2ex]
$|\bs{v}|$ & \hspace{8ex} & $Z_p$ & $M_\textrm{kin}$ & $C_v/(\gamma m)$ & \hspace{8ex} & $Z_p$
& $M_\textrm{kin}$ & $C_v/(\gamma m)$ \\
 \\[-2ex]
\hline
 \\[-2ex]
$0$   & &  ---          & $5.974(48)$  &  $1.0182(86)$ &  & ---          & $5.979(37)$ & $1.0190(65)$ \\
$0.2$ & & $1.008(19)$   & $5.95(10)$   &  $1.015(18)$  &  & $1.009(12)$  & $5.969(62)$ & $1.017(11)$  \\
$0.4$ & & $0.9953(78)$  & $5.931(44)$  &  $1.0084(77)$ &  & $0.9830(65)$ & $5.954(40)$ & $1.0101(70)$ \\
$0.6$ & & $0.898(27)$   & $6.22(18)$   &  $1.010(28)$  &  & $0.843(27)$  & $6.37(15)$  & $1.011(21)$ \\
\hline  \hline
\end{tabular}
\caption{Nonperturbative results (using the $\eta_b(1S)$) for
$M_\mathrm{kin}$, $Z_p$, $C_v$.}
\label{tab:eta1S_disp_rel_results}
\end{center}
\end{table*}

Results for the $\eta_b(1S)$ kinetic mass $M_\mathrm{kin}$ and the
renormalization parameters $Z_p$, $C_v$ are shown in Table
\ref{tab:eta1S_disp_rel_results}. The energies were obtained from
6-exponential fits to $2\times2$ matrix correlators with the $\eta_b(1S)$
smearing and the local axial current. For the calculation of $C_v$
using (\ref{eq:C_v}), we averaged the results over the 4 different
perpendicular lattice momenta 
\begin{equation}
\bs{k}_\perp\in\left\{\frac{2\pi}{L}(0,\pm1,0),\frac{2\pi}{L}(0,0,\pm1)\right\}.
\end{equation}
The momentum parallel to the boost velocity in (\ref{eq:Z_p}) was
chosen to be $\bs{k}_\parallel=\frac{2\pi}{L}(1,0,0)$, and in (\ref{eq:M_kin}),
for the measurement of $M_\mathrm{kin}$, we use $\bs{k}=0$.

Because the lattice is of finite extent, $L=20$ in our test case,
the estimates for $C_v$ and $Z_p$ will be affected by the choice  
of momenta in (\ref{eq:C_v}) and (\ref{eq:Z_p}) since the formulae
are accurate only in the limit that the momenta are infinitesimal.
Note that the uncertainty due to using non-infinitesimal momenta will decrease
for larger lattices for which smaller momenta are available.

To estimate the size of the resulting systematic error we also performed the
calculations with the larger momenta
\begin{equation}
\bs{k}_\perp\in
\left\{\frac{2\pi}{L}(0,\pm2,0),\frac{2\pi}{L}(0,0,\pm2)\right\},
\bs{k}_\parallel=\frac{2\pi}{L}(2,0,0).
\end{equation}
For $C_v$, the results from $|\bs{k}_\perp|=2\pi/L$ agree with those
obtained from $|\bs{k}_\perp|=4\pi/L$ within statistical errors,
indicating that the systematic error is small and does not increase
significantly when increasing the momentum perpendicular to $\bs{v}$
in the measurement.
For the measurement of $Z_p$ at $|\bs{v}|=0.6$ we find a 6\% ($2\sigma$)
change in $Z_p$ when going from $|\bs{k}_\parallel|=2\pi/L$ to
$|\bs{k}_\parallel|=4\pi/L$. At $|\bs{v}|=0.4$ and smaller boost velocities
the results are equal within statistical errors.
For the kinetic mass, which depends on both $C_v$ and $Z_p$, we again find
agreement within statistical errors between the results from the two different
momenta for all boost velocities considered.
At small velocities, we find that both $Z_p$ and $C_v/(\gamma m)$ are close to
their tree-level value of 1, demonstrating that renormalizations are indeed
small.

We also obtained the amplitude for the axial current and extracted the
pseudoscalar decay constant from the same $2\times2$ matrix fits using
(\ref{eq:f_Cv}). For the energy shift $C_v$ in (\ref{eq:f_Cv}) we used the
result from $|\bs{k}_\perp|=2\pi/L$. The meson momentum is given by
$\bs{p}=2Z_p\gamma m\bs{v}+\bs{k}$. In the following we compare two methods of
reaching large $|\bs{p}|$. First, at $\bs{v}=0$, \textit{i.e.}\ with standard
NRQCD, we computed the decay constant at large non-zero lattice momentum
$\bs{k}$; the results are shown in Table \ref{tab:f_NRQCD}. Second, we
computed the decay constant with $\bs{k}=0$ and three different boost
velocities $\bs{v}$; the results are shown in Table \ref{tab:f_mNRQCD}. In
this case the uncertainty in $Z_p$ leads to an uncertainty in the meson
momentum.

\begin{table}
\begin{tabular}{cccc}
\hline\hline
 \\[-2ex]
 $|\bs{p}| L/(2\pi)$ & $|\bs{p}|$ & \hspace{2ex} & 
$f$ \\
 \\[-2ex]
\hline
$0$  & $0$       &  & $0.4724(23)$ \\
$1$  & $0.31416$ &  & $0.4731(23)$ \\
$2$  & $0.62832$ &  & $0.4755(24)$ \\
$3$  & $0.94248$ &  & $0.4772(43)$ \\
$4$  & $1.25664$ &  & $0.4835(77)$ \\
$5$  & $1.57080$ &  & $0.4971(78)$ \\
$6$  & $1.88496$ &  & $0.5209(46)$ \\
$7$  & $2.19911$ &  & $0.5527(44)$ \\
$8$  & $2.51327$ &  & $0.6006(45)$ \\
$9$  & $2.82743$ &  & $0.6740(49)$ \\
$10$ & $3.14159$ &  & $0.715(29)$ \\
\hline\hline
\end{tabular}
\caption{$\eta_b(1S)$ decay constant with standard NRQCD
(i.e. $\bs{v}=0$) computed with several values of meson momentum
$|\bs{p}|$ by varying $|\bs{k}|$.}
\label{tab:f_NRQCD}
\end{table}

\begin{figure}
\begin{center}
\includegraphics[width=\linewidth]{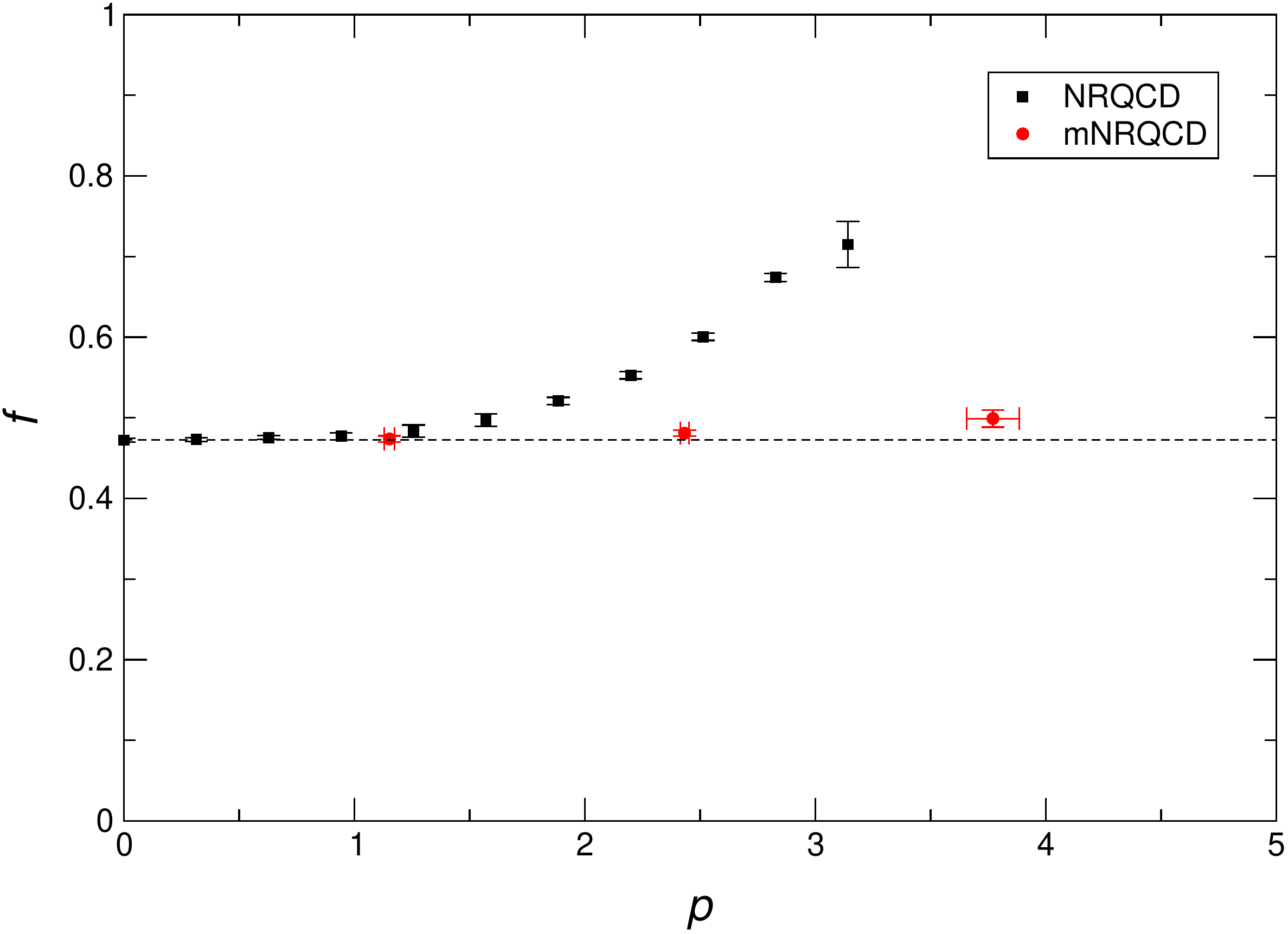}
\caption{Heavy-heavy decay constant in NRQCD and mNRQCD for different
values of the meson's momentum, $|\bs{p}|/(2\pi/L)=0\dots 10$ (NRQCD)
and $\bs{p}=Z_p\:2\gamma m \bs{v}$ for $|\bs{v}|= 0.2, 0.4, 0.6$
(mNRQCD). The horizontal line indicates the value at $\bs{p}=0$.}
\label{fig:f_eta_ex}
\end{center}
\end{figure}

\begin{table}
\begin{tabular}{ccccc}
\hline\hline
 \\[-2ex]
$|\bs{v}|$ & \hspace{2ex} & $|\bs{p}|$ & \hspace{2ex} & 
$f$ \\
 \\[-2ex]
\hline
$0$   && $0$         && $0.4724(23)$ \\
$0.2$ && $1.152(22)$ && $0.4739(38)$ \\
$0.4$ && $2.433(19)$ && $0.4810(36)$ \\
$0.6$ && $3.77(11)$  && $0.499(11)$ \\
\hline\hline
\end{tabular}
\caption{$\eta_b(1S)$ decay constant with mNRQCD at $\bs{k}=0$
computed with several values of meson momentum $|\bs{p}|$ by varying
$|\bs{v}|$.}
\label{tab:f_mNRQCD}
\end{table}
A plot of the decay constant against the total momentum (with $Z_p$ from
(\ref{eq:Z_p}) with $|\bs{k}_\parallel|=2\pi/L$) for the two
methods is shown in Fig.~\ref{fig:f_eta_ex}. The decay constant is a
Lorentz scalar and should be independent of the momentum. However, with
NRQCD we see large deviations due to both relativistic and discretization
errors. With moving NRQCD the deviation is very small, giving evidence
that the formalism works very well. Small deviations are still expected here,
since only the leading-order current was used; \textit{i.e.}\ 
$T_{\scriptscriptstyle \mathrm{FWT}}$ and $A_{\scriptscriptstyle D_t}$
were set to unity in (\ref{eq:MNRQCD_field_redef}) for this calculation.

Next, we studied the velocity-dependence of various energy splittings between
the bottomonium states listed in Table \ref{tab:bott_ops}.  For the $\Upsilon$
and $\eta_b$ states, we used 6-exponential $2\times2$ matrix fits with the
$1S$ and $2S$ smearings; for the $\chi_{b1}$ states a 6-exponential
single-correlator fit with the $1P$ smearing at both source and sink was
used. The results for the $\Upsilon(2S)-\Upsilon(1S)$,
$\chi_{b1}(1P)-\Upsilon(1S)$ and $\Upsilon(1S)-\eta_b(1S)$ splittings are
listed in Tables \ref{tab:Upsilon2S-Upsilon1S_splitting},
\ref{tab:chi1P1-Upsilon1S_splitting} and \ref{tab:Upsilon1S-eta1S_splitting},
respectively.

\begin{table}
\begin{tabular}{ccccc}
\hline\hline
 \\[-2ex]
$|\bs{v}|$ & \hspace{2ex} & $\Delta E_v(0)$ & \hspace{2ex} &
 $\displaystyle\frac{\Delta E_v(0)}{\Delta E_0(0)}$\\
 \\[-2ex]
\hline
$0.0$ && $0.3334(68)$ &&  1           \\
$0.2$ && $0.329(10)$ &&  $0.986(37)$ \\
$0.4$ && $0.320(15)$ &&  $0.958(48)$ \\
$0.6$ && $0.20(11)$  &&  $0.59(33)$\\
\hline\hline
\end{tabular}
\caption{$\Upsilon(2S)-\Upsilon(1S)$ energy splitting as a function
of the boost velocity.}
\label{tab:Upsilon2S-Upsilon1S_splitting}
\end{table}

\begin{table}
\begin{tabular}{ccccc}
\hline\hline
 \\[-2ex]
$|\bs{v}|$ & \hspace{2ex} & $\Delta E_v(0)$ & \hspace{2ex} & $\displaystyle
\frac{\Delta E_v(0)}{\Delta E_0(0)}$\\
 \\[-2ex]
\hline
$0.0$ && $0.2703(89)$  && 1           \\
$0.2$ && $0.264(12)$   && $0.976(56)$ \\
$0.4$ && $0.270(23)$   && $0.998(91)$  \\
$0.6$ && $0.227(57)$   && $0.84(21)$  \\
\hline\hline
\end{tabular}
\caption{$\chi_{b1}(1P)-\Upsilon(1S)$ energy splitting as a
function of the boost velocity.}
\label{tab:chi1P1-Upsilon1S_splitting}
\end{table}

\begin{table}
\begin{tabular}{ccccc}
\hline\hline
 \\[-2ex]
$|\bs{v}|$ & \hspace{2ex} & $\Delta E_v(0)$ & \hspace{2ex} & $\displaystyle
\frac{\Delta E_v(0)}{\Delta E_0(0)}$\\
 \\[-2ex]
\hline
$0.0$ &&  $0.031469(98)$ && 1            \\
$0.2$ &&  $0.03039(20)$  && $0.9656(71)$ \\
$0.4$ &&  $0.02837(85)$  && $0.901(27)$  \\
$0.6$ &&  $0.0281(28)$   && $0.894(88)$   \\
\hline\hline
\end{tabular}
\caption{$\Upsilon(1S)-\eta_b(1S)$ energy splitting as a
function of the boost velocity.}
\label{tab:Upsilon1S-eta1S_splitting}
\end{table}

\begin{figure}
\begin{center}
\includegraphics[width=\linewidth]{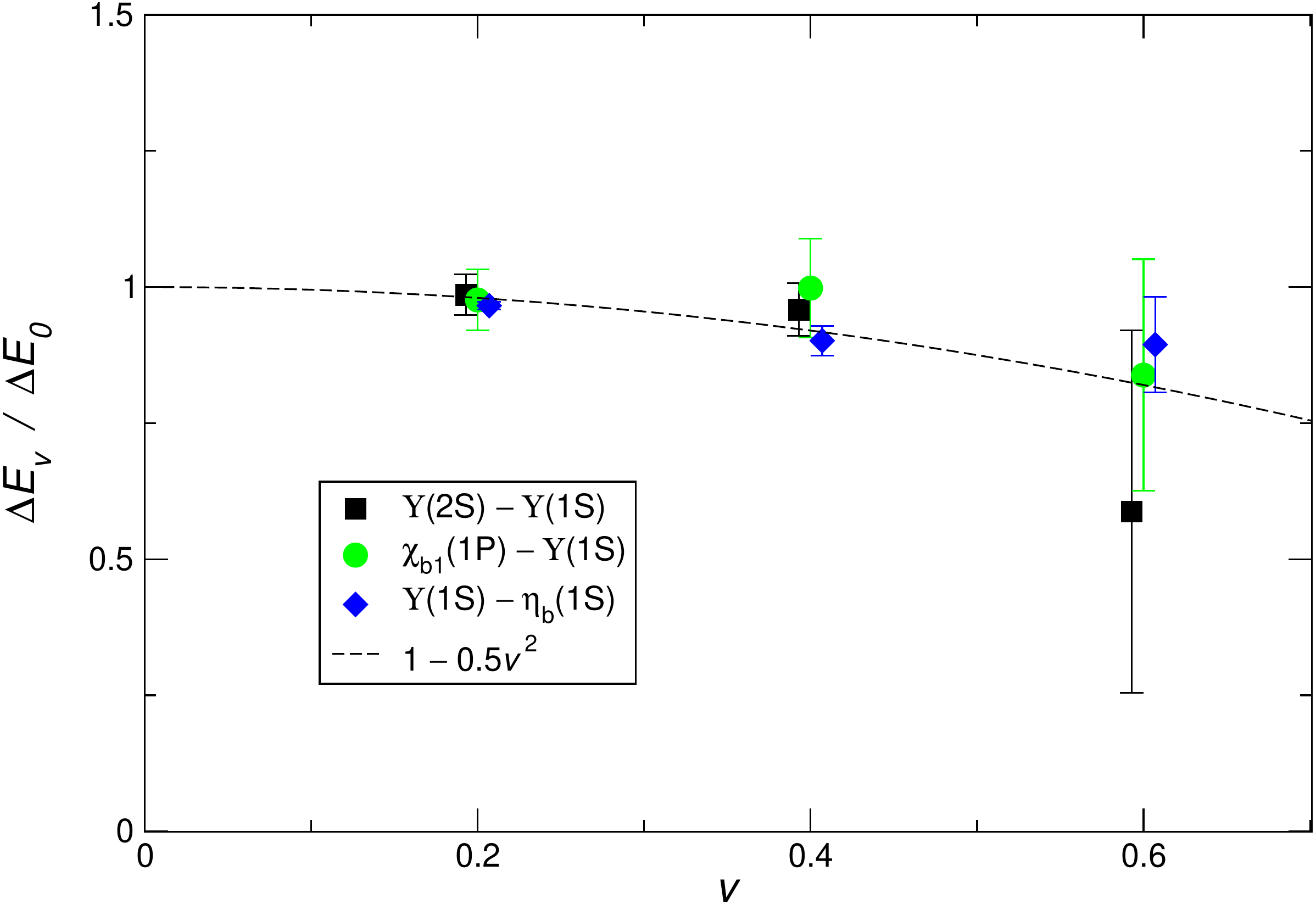}
\end{center}
\caption{Bottomonium energy splittings relative to $v=0$ as a
function of the boost velocity. Points are offset horizontally
for legibility.  The data agree with an estimate for the leading
$v^2$ dependence (see text).}
\label{fig:hh_splittings}
\end{figure}

\begin{table}
\begin{tabular}{ccccccc}
\hline\hline
 \\[-2ex]
$|\bs{v}|$ & \hspace{2ex} & $\Delta E_v(0)|_1$ & \hspace{2ex} & $\Delta E_v(0)|_2$ & \hspace{2ex} & $\Delta E_v(0)|_3$ \\
 \\[-2ex]
\hline
$0$    && $-0.000009(63)$  &&   $-0.000039(68)$  &&   $0.000053(73)$  \\
$0.2$  && $-0.00012(26)$   &&   $-0.00005(28)$   &&   $0.00017(30)$   \\
$0.4$  && $-0.00046(56)$   &&   $0.00055(62)$    &&   $-0.00010(57)$   \\
$0.6$  && $-0.0176(96)$    &&   $0.0107(62)$     &&   $0.0069(75)$   \\
\hline\hline
\end{tabular}
\caption{Dependence of the $\Upsilon(1S)$ energy on the polarization
direction. $\Delta E_v(0)|_j$ is the difference between $E_v(0)|_j$ and
the polarization-averaged energy.}
\label{tab:Upsilon1S_spin_dep}
\end{table}

Note that the energy splittings are not Lorentz scalars. Using (\ref{eqn:phys_energy}),
we expect that the splitting between two states $A$ and $B$ at zero lattice momentum
is given by
\begin{eqnarray*}
E^A_\bs{v}(0)-E^B_\bs{v}(0)&=&\sqrt{(2Z_p\gamma m \bs{v})^2
+(M^A_\mathrm{kin})^2}\\
&&-\sqrt{(2Z_p\gamma m \bs{v})^2+(M^B_\mathrm{kin})^2}.
\end{eqnarray*}
If we set $Z_p=1$ and expand the splitting at velocity $\bs{v}$ relative to
$\bs{v}=0$ in powers of the boost velocity, we obtain
\begin{equation*}
\frac{E^A_\bs{v}(0)-E^B_\bs{v}(0)}{E^A_0(0)-E^B_0(0)}
=1-\underbrace{\left(\frac{2m^2}{M^A_\mathrm{kin}\:M^B_\mathrm{kin}}\right)}
_{\approx0.5}\:\bs{v}^2+\mathcal{O}(\bs{v}^4),
\end{equation*}
that is, we expect a quadratic decrease like $1-0.5|\bs{v}|^2$. The numerical
results, shown in Fig.~\ref{fig:hh_splittings}, are consistent with this estimate
as desired.

Finally, for the $\Upsilon(1S)$ meson, we studied the dependence of the energy
on the polarization direction. If moving NRQCD works well, then there should
be no difference for polarizations parallel and perpendicular to the boost
velocity. In Table \ref{tab:Upsilon1S_spin_dep} we show the difference between
the energy with definite polarization direction, $E_v(0)|_j$ and the
polarization-direction-averaged energy
$\frac13(E_v(0)|_1+E_v(0)|_2+E_v(0)|_3)$.  No significant dependence on the
polarization direction can be seen (except maybe at $v=0.6$, where a $1.8\sigma$
deviation in the energies was found).

\subsection{Heavy-light mesons}

\subsubsection{Methods}

Starting with the standard Dirac fields, we construct interpolating fields for
the $B_s$ and $B_s^*$ mesons with momentum $\bs{p}$ from
\begin{equation}
O_\Gamma(\bs{p}, \bs{\tau})=\sum_{\bs{x},\bs{y}}\overline{\Psi}_l(\bs{x},\tau)
\Gamma(\bs{x}-\bs{y})\Psi_H(\bs{y},\tau)e^{-i\bs{p}\cdot\bs{y}} \; ,
\label{eqn:B_interpol_field}
\end{equation}
where $\Psi_l$ is the Dirac spinor for the valence strange quark and $\Psi_H$
is the Dirac spinor for the $b$ quark. We use $\Gamma(\bs{r})=\dgamma^5\:f(\bs{r})$
for the $B_s$ pseudoscalar meson, $\Gamma(\bs{r})=\dgamma^j\:f(\bs{r})$ with
$j=1, 2, 3$ for the $B^*_s$ vector meson and $\Gamma(\bs{r})=\dgamma^5\dgamma^0\:f(\bs{r})$
for the computation of the decay constant $f_{B_s}$. We compute $2\times2$ matrix
correlators with Gaussian and local smearing,
$f(\bs{r})=e^{-|\bs{r}|^2/r_s^2},\:\:\delta(\bs{r})$.

In terms of the standard Dirac propagators, the two-point function reads
\begin{eqnarray}
\nonumber&& \langle O_{\Gamma_{\rm sk}}(\bs{p},\tau)
O_{\Gamma_{\rm sc}}^\dag(\bs{p},\tau') \rangle
=\frac1N\sum_U\sum_{\bs{x},\bs{y},\bs{x'},\bs{y'}}\\
\nonumber&&\times\:\:\mathrm{Tr}\left[
\:\Gamma_{\rm sk}(\bs{x}-\bs{y})\:G_l\left(x',\:x\right)
\:\Gamma_{\rm sc}^\dag(\bs{x'}-\bs{y'})\: G_H\left(y,\:y'\right)\:\right]\\
&&\times \:\: e^{-i\bs{p}\cdot\bs{y}}e^{i\bs{p}\cdot\bs{y'}} \; ,
\end{eqnarray}
with $x=(\bs{x},\tau)$, $y=(\bs{y},\tau)$, $x'=(\bs{x'},\tau')$,
$y'=(\bs{y'},\tau')$. For $\tau>\tau'$, the tree-level leading-order
mNRQCD field redefinition (\ref{eq:MNRQCD_field_redef}) leads to the
following expression for the $b$ propagator:
\begin{eqnarray*}
G_H\left(y,\:y'\right)&=&\frac{1}{\gamma}e^{-\gamma m(\tau-\tau')
+i\gamma m\bs{v}\cdot(\bs{y}-\bs{y'})}\\
&&\times \:\: S(\Lambda)\left(\begin{array}{cc} G_{\psi_v}(y,\:y')
& 0 \\ 0 & 0\end{array}\right)\overline{S}(\Lambda).
\end{eqnarray*}
For the light quark, we use the ASQTAD staggered fermion action
\cite{Wingate:2002fh}. The 4-component \emph{na\"{\i}ve} light quark
propagator can be obtained from the 1-component staggered
propagator $G_\chi(x',x)$ via
\begin{equation}
G_l(x',x)=\:G_\chi(x',x)\otimes\Omega(x')\Omega^\dag(x)
\end{equation}
with
\begin{equation}
\Omega(x)=(\dgamma^0)^{x_4}(-i\dgamma^1)^{x_1}(-i\dgamma^2)^{x_2}(-i\dgamma^3)^{x_3}.
\end{equation}
(Recall our convention for the Dirac matrices is as given
in Appendix~\ref{app:notation}.)  We also employ $\dgamma^5$-hermiticity
\begin{equation}
G_l(x',x) = \dgamma^5 G_l^\dag(x,x') \dgamma^5,
\end{equation}
to interchange the points $x$ and $x'$ for the light quark propagator.
As before, we remove the factor of $e^{-\gamma m(\tau-\tau')}$ and the
summation over $\bs{x'}$.

In the case where $\Gamma_{\rm sk}$ and $\Gamma_{\rm sc}$ contain the same
Dirac matrix, we arrive at the following expression:
\begin{align}
\nonumber &C(\Gamma_{\rm sk}, \Gamma_{\rm sc}, \bs{k}, \tau, \tau')
=\frac1N\!\sum_U\!\frac{1}{\gamma}\sum_{\bs{x},\bs{y}}f_{\rm sk}
(\bs{x}\!-\!\bs{y})e^{-i\bs{k}\cdot\bs{y}}\eta(x, x')\\
\nonumber&\times\mathrm{Tr}\left[G^\dag_\chi(x,x')\:
\overline{S}(\Lambda)\Omega(x')\Omega^\dag(x)S(\Lambda)\!
\left(\!\!\begin{array}{cc} \tilde{G}_{\psi_v}(y,\:x') & 0 \\ 0 & 0\end{array}
\!\!\right)\right]\\
\label{eq:hl_corr}
\end{align}
with $\bs{k}\equiv\bs{p}-\gamma m \bs{v}$ and
\begin{equation*}
\tilde{G}_{\psi_v}(y,\:x')=\sum_{\bs{y'}}f(\bs{x'}
-\bs{y'})e^{i\bs{k}\cdot\bs{y'}}G_{\psi_v}(y,y').
\end{equation*}
The phase factor $\eta(x, x')$ in (\ref{eq:hl_corr}) depends on the
Dirac matrix in $\Gamma_{\rm sk}$ and $\Gamma_{\rm sc}$. It is given by
\begin{equation*}
\eta(x, x')=\left\{\begin{array}{ll} 1 & \mathrm{for}\:\:\dgamma^5,\\
(-1)^{x'_j-x_j} &\mathrm{for}\:\: \dgamma^j, \\
(-1)^{\sum_{j}(x_j+x'_j)} &\mathrm{for}\:\: \dgamma^5\dgamma^0.
\end{array}\right.
\end{equation*}
As before, we set $f(\bs{r})$ to zero for $|\bs{r}|>R_s$ with some cut-off
radius $R_s$ smaller than half the length of the lattice.

The staggered/na\"{\i}ve light quark action used here suffers from the
doubling problem. As shown in \cite{Wingate:2002fh}, the spatial doublers
do not contribute to the correlators. However, the temporal doubler leads
to a coupling to additional opposite parity states, which manifest themselves
as oscillating exponentials in the correlators. We therefore fit the
heavy-light correlators to
\begin{align*}
C(\Gamma_{\rm sk}, \Gamma_{\rm sc}, \bs{k}, \tau, \tau')\:\:\:\:
\rightarrow\:\:\:\: A^{\rm sk}(A^{\rm sc})^* \bigg[e^{-E(\tau-\tau')}\hspace{6ex}\\
\hspace{6ex}+\sum_{n=1}^{n_{\rm exp}-1}B^{\rm sk}_n(B^{\rm sc}_n)^*
e^{-(E+\Delta E_1+...+\Delta E_n)(\tau-\tau')}\bigg]\\
+\:\:\:\:(-1)^{\tau-\tau'+1}\tilde{A}^{\rm sk}(\tilde{A}^{\rm sc})^*
\bigg[e^{-\tilde{E}(\tau-\tau')}\hspace{15ex}\\
\hspace{8ex}+\sum_{m=1}^{m_{\rm exp}-1}\tilde{B}^{\rm sk}_m(\tilde{B}^{\rm sc}_m)^*
e^{-(\tilde{E}+\Delta \tilde{E}_1+...+\Delta \tilde{E}_m)(\tau-\tau')}\bigg].
\end{align*}
The quantities $C_v$, $Z_p$ ,$M_\mathrm{kin}$ and the decay constants
$f_B$, $f_{B_s}$ can be extracted in a completely analogous manner as for
the heavy-heavy-mesons, with the replacements
$2C_v\rightarrow C_v$ and $2\bs{P}_0\rightarrow\bs{P}_0$,
since now there is only one heavy quark.

\subsubsection{Lattice parameters}

The heavy-light simulations have been performed with the same gauge
configurations as the heavy-heavy simulations, and the same heavy-quark
action and parameters were used. Again, the boost velocity was always
pointing in $x$-direction, $\bs{v}=(v,0,0)$.
The valence strange quark mass for the $B_s$ and $B_s^*$ mesons
was set to 0.040.
Four staggered propagators with source times $\tau'=0,\: 16,\: 32,\: 48$
were used for each gauge configuration. Both forward- and
backward-propagating meson correlators were computed to increase statistics.
The smearing parameter $r_s$ was set to 2.5.

\subsubsection{Results}
\label{sec:heavy_light_results}

\begin{figure}
\begin{center}
\includegraphics[width=\linewidth]{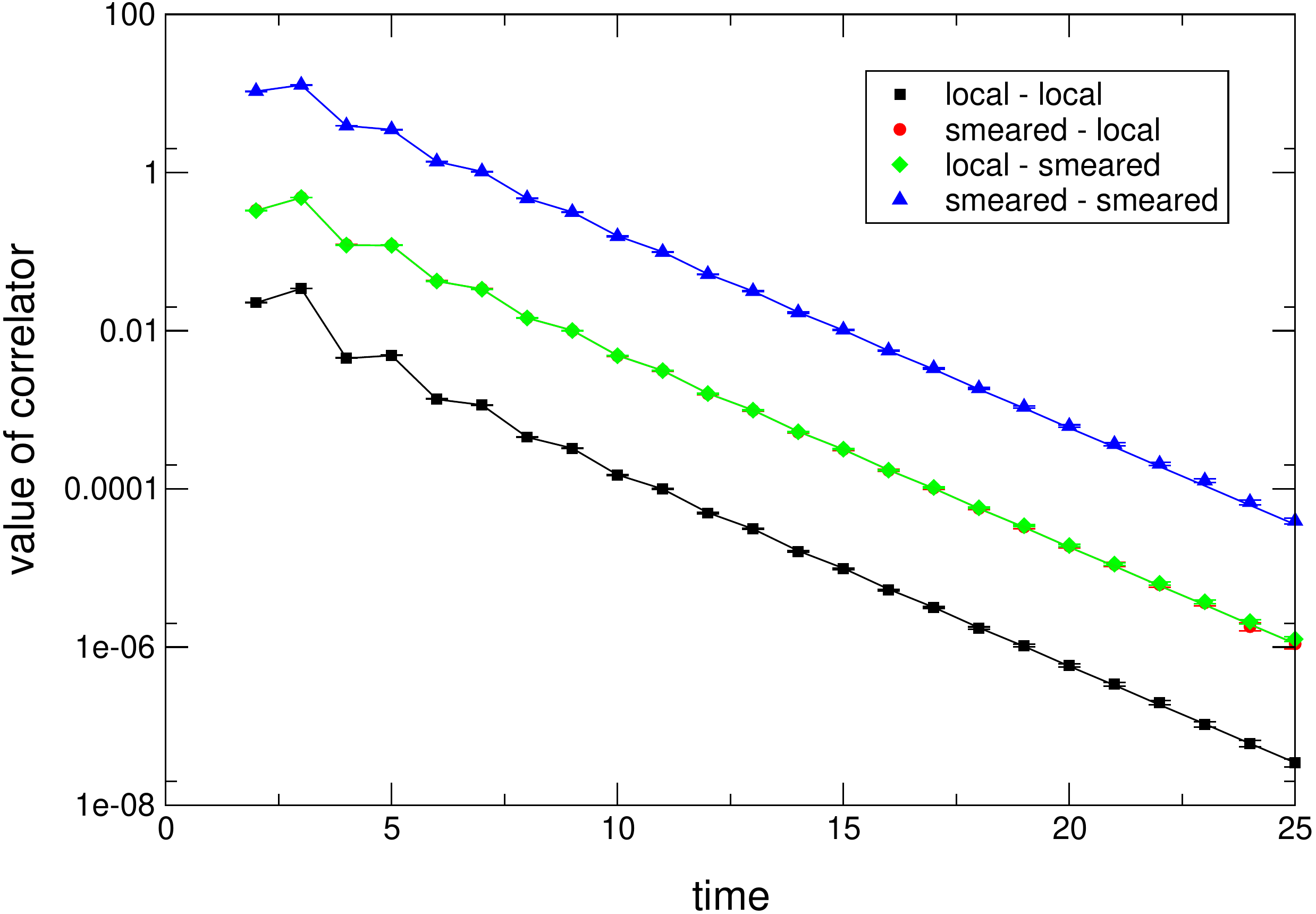}

\includegraphics[width=\linewidth]{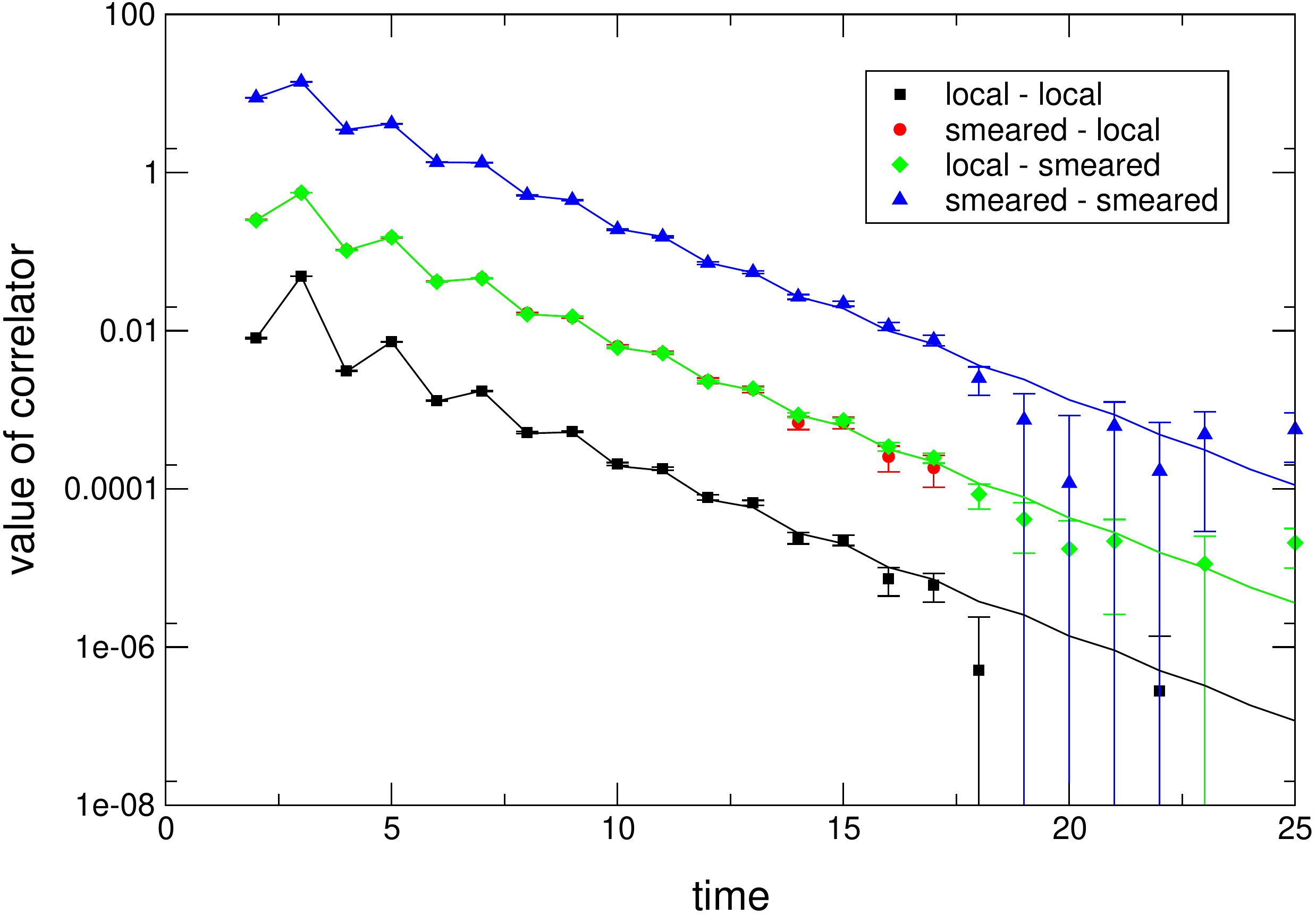}
\end{center}
\caption{$B_s$ matrix correlators at $\bs{k}=0$ and $\bs{v}=0$
(upper panel), $|\bs{v}|=0.4$ (lower panel).}
\label{fig:Bs_corr}
\end{figure}

Results for the $B_s$ kinetic mass $M_\mathrm{kin}$ and the renormalization
parameters $Z_p$, $C_v$ are shown in Table \ref{tab:Bs_disp_rel_results}.  The
energies and the amplitude required for the calculation of the decay constant
were obtained from 8-exponential (4 of which are oscillating) fits to
$2\times2$ matrix correlators with the Gaussian smearing and the local axial
current. Two sample plots of these correlators at $v=0$ and $v=0.4$ are shown
in Fig.~\ref{fig:Bs_corr}. This also demonstrates the worsening of the
signal-to-noise ratio as the boost velocity increases, in accordance with
(\ref{eq:noise2}).

\begin{table}
\begin{tabular}{ccccccc}
\hline\hline
 \\[-2ex]
$v$ & \hspace{2ex} & $Z_p$ & \hspace{2ex} & $M_\textrm{kin}$ & \hspace{2ex} & $C_v/(\gamma m)$ \\
 \\[-2ex]
\hline
$0$   &&              && $3.37(15)$  && $1.002(52)$ \\
$0.2$ && $1.05(15)$   && $3.72(47)$  && $1.13(16)$ \\
$0.4$ && $1.05(18)$   && $3.66(68)$  && $1.10(23)$  \\
\hline\hline
\end{tabular}
\caption{$B_s$ results for $M_\mathrm{kin}$, $Z_p$, $C_v$.}
\label{tab:Bs_disp_rel_results}
\end{table}

For the calculation of $C_v$, we again averaged the results over
the 4 different lattice momenta perpendicular to $\bs{v}$
\begin{equation}
\bs{k}_\perp=\frac{2\pi}{L}(0,\pm1,0),~~\frac{2\pi}{L}(0,0,\pm1),
\end{equation}
and the momentum parallel to the boost velocity required for the
determination of $Z_p$ was chosen to be
$\bs{k}_\parallel=\frac{2\pi}{L}(1,0,0)$.

As expected, the statistical errors are larger than for the heavy-heavy mesons,
partly due to a much smaller number of origins (four) per gauge configuration.
The results for $Z_p$ and $C_v$ agree with those obtained using heavy-heavy
mesons in section \ref{sec:heavy_heavy_results}.

\begin{figure}
\begin{center}
\includegraphics[width=\linewidth]{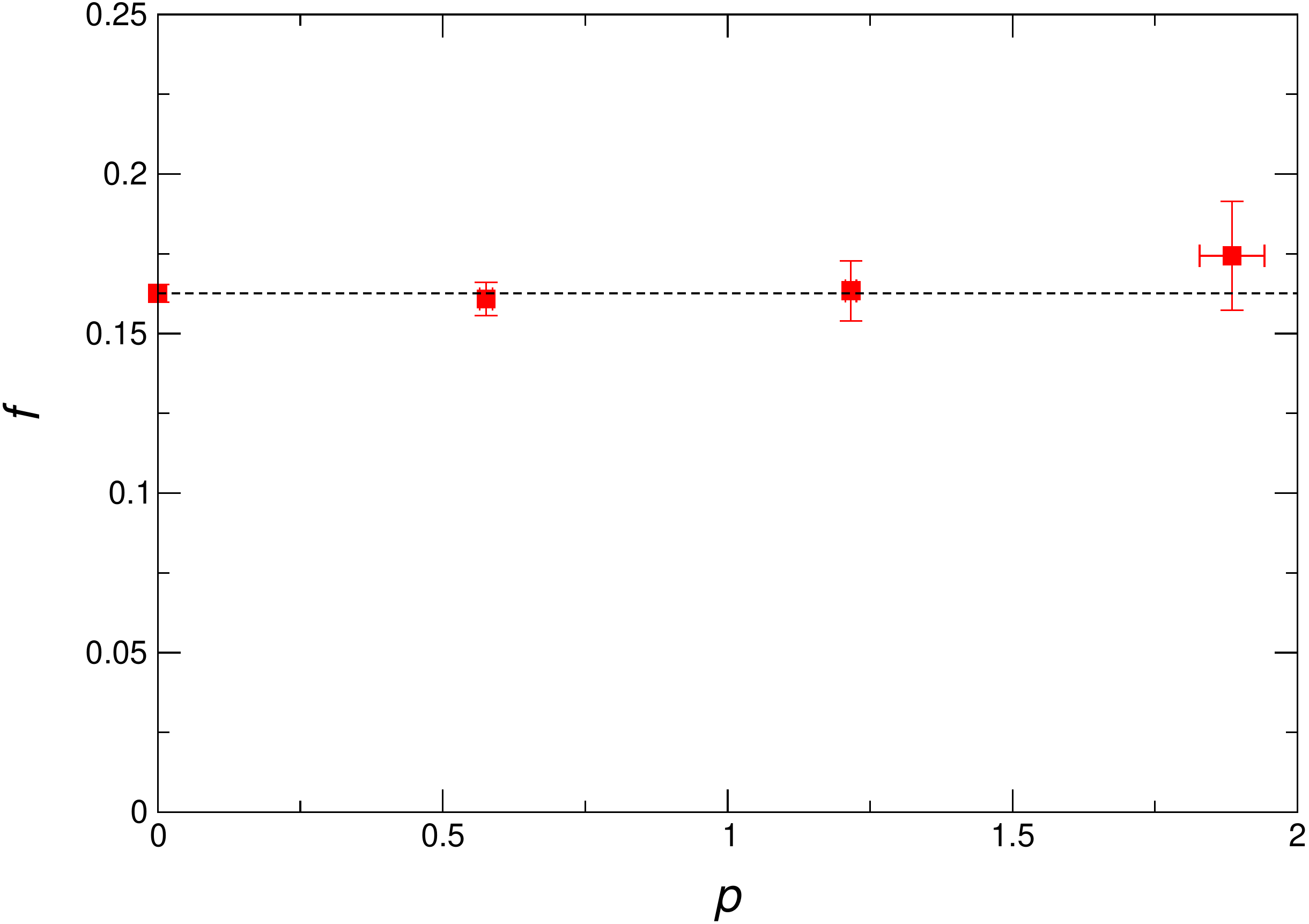}
\end{center}
\caption{The $B_s$ decay constant at $k=0$ and $v=0,\:0.2,\:0.4,\:0.6$
plotted against the total momentum $\bs{p}=Z_p\gamma m \bs{v}+\bs{k}$.
The horizontal line indicates the value at $v=0$.}
\label{fig:f_Bs}
\end{figure}

The results for the decay constant $f_{B_s}$ at $\bs{k}=0$ and
$v=0,0.2,0.4,0.6$ are listed in Table \ref{tab:f_Bs} and plotted against the
total momentum in Fig.~\ref{fig:f_Bs}. In the calculation of the decay
constant, we used $C_v$ and $Z_p$ determined from the $\eta_b(1S)$ dispersion
relation since this is more precise. We find that the decay constant is
independent of the boost velocity within statistical errors.  (Even when
working with non-moving NRQCD, the discretization errors in the heavy-light
decay constant do not appear to grow as severely with momentum
\cite{Collins:2001nn} as in the heavy-heavy decay constant
(Fig.~\ref{fig:f_eta_ex}).)

\begin{table}
\begin{tabular}{ccccc}
\hline\hline
 \\[-2ex]
$|\bs{v}|$ & \hspace{2ex} & $|\bs{p}|$ & \hspace{2ex} & $f(\bs{k}=0)$ \\
 \\[-2ex]
\hline
$0$   && $0$          && $0.1626(27)$ \\
$0.2$ && $0.576(11)$  && $0.1608(52)$ \\
$0.4$ && $1.2163(96)$ && $0.1634(94)$ \\
$0.6$ && $1.885(57)$  && $0.174(17)$ \\
\hline\hline
\end{tabular}
\caption{$B_s$ decay constant (unrenormalized, and in lattice units) with
mNRQCD at $\bs{k}=0$.}
\label{tab:f_Bs}
\end{table}

We also computed the $B_s^*-B_s$ energy splitting as a function of $v$; the
results are shown in Table \ref{tab:Bsstar-Bs_splitting}. The statistical
errors are so large that no definite statement can be made about the
velocity dependence.

\begin{table}
\begin{center}
\begin{tabular}{ccccc}
\hline\hline
 \\[-2ex]
$v$ & \hspace{2ex} & $\Delta E_v(0)$ & \hspace{2ex} & $\displaystyle
\frac{\Delta E_v(0)}{\Delta E_0(0)}$\\
 \\[-2ex]
\hline
$0.0$ && $0.0261(35)$  && 1 \\
$0.2$ && $0.0262(65)$  && $1.00(28)$ \\
$0.4$ && $0.0310(80)$  && $1.18(34)$ \\
\hline\hline
\end{tabular}
\end{center}
\caption{$B_s^*-B_s$ energy splitting as a function of $v$.}
\label{tab:Bsstar-Bs_splitting}
\end{table}

\section{Comparison of perturbative  and nonperturbative results}
\label{sec:comparison_NP_PT}

In the following we compare our perturbative results given in
Section \ref{sec:results} to the nonperturbative numbers obtained in
Sections \ref{sec:heavy_heavy_results} and
\ref{sec:heavy_light_results}.

We use the strong coupling constant defined in the potential scheme
\cite{Lepage:1992xa} and choose $q^\star$ (for each quantity and each value of
$v$) using the Brodsky-Lepage-Mackenzie procedure \cite{Brodsky:1982gc}.  
The $q^*$ values range approximately between $0.5/a$ and $3/a$. As a reference,
$2/a=3.2$ GeV on the coarse MILC configurations \cite{Gray:2005ur}. Using
the running of the strong coupling constant $\alpha_V(q)$ \cite{Mason:2005zx}
this gives $\alpha_V(2/a) \approx 0.3$.

\begin{figure}
  \centering

\ifpdf
\includegraphics[width=\linewidth]{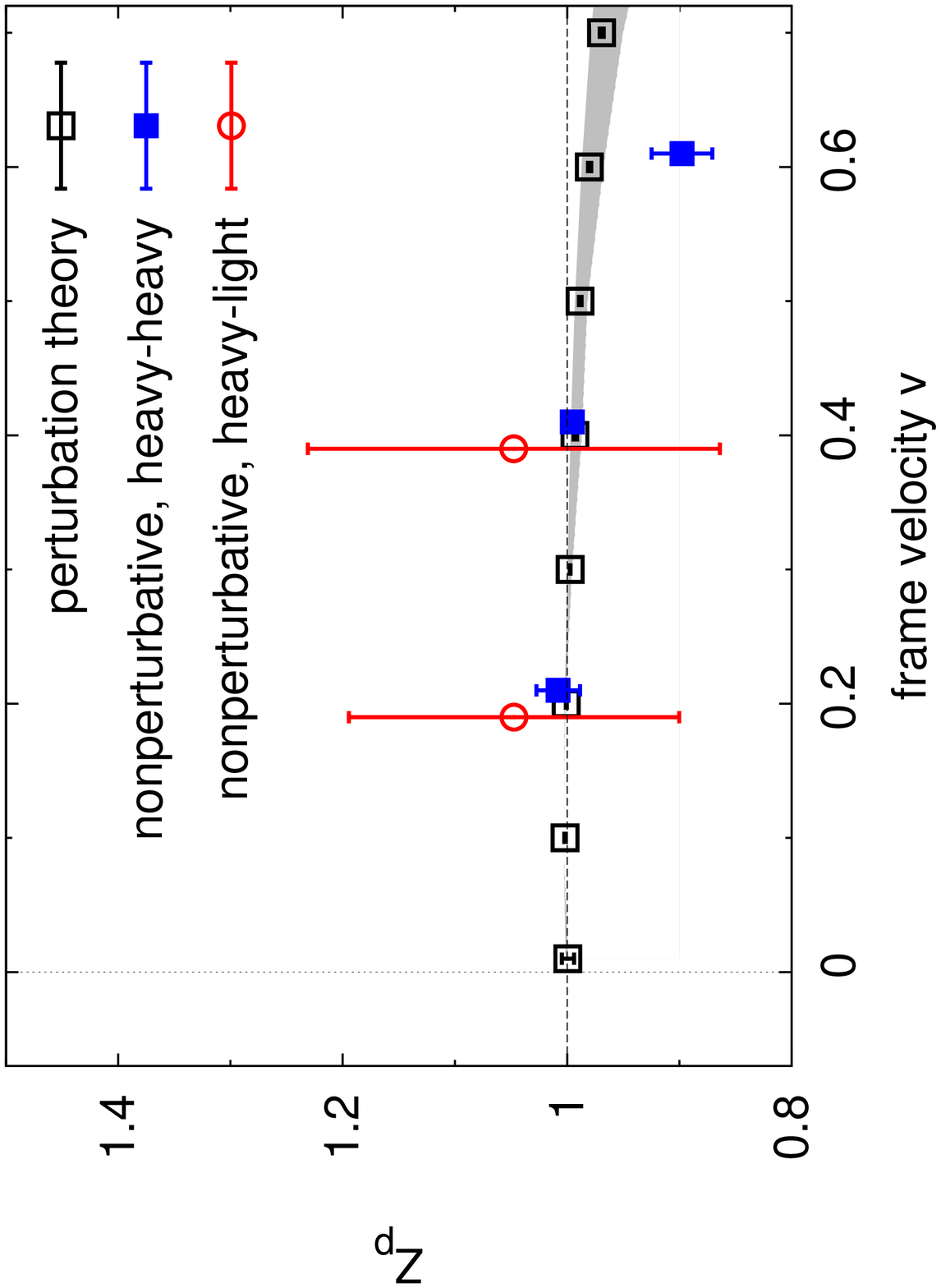}
\else
\includegraphics[height=\linewidth,angle=270]{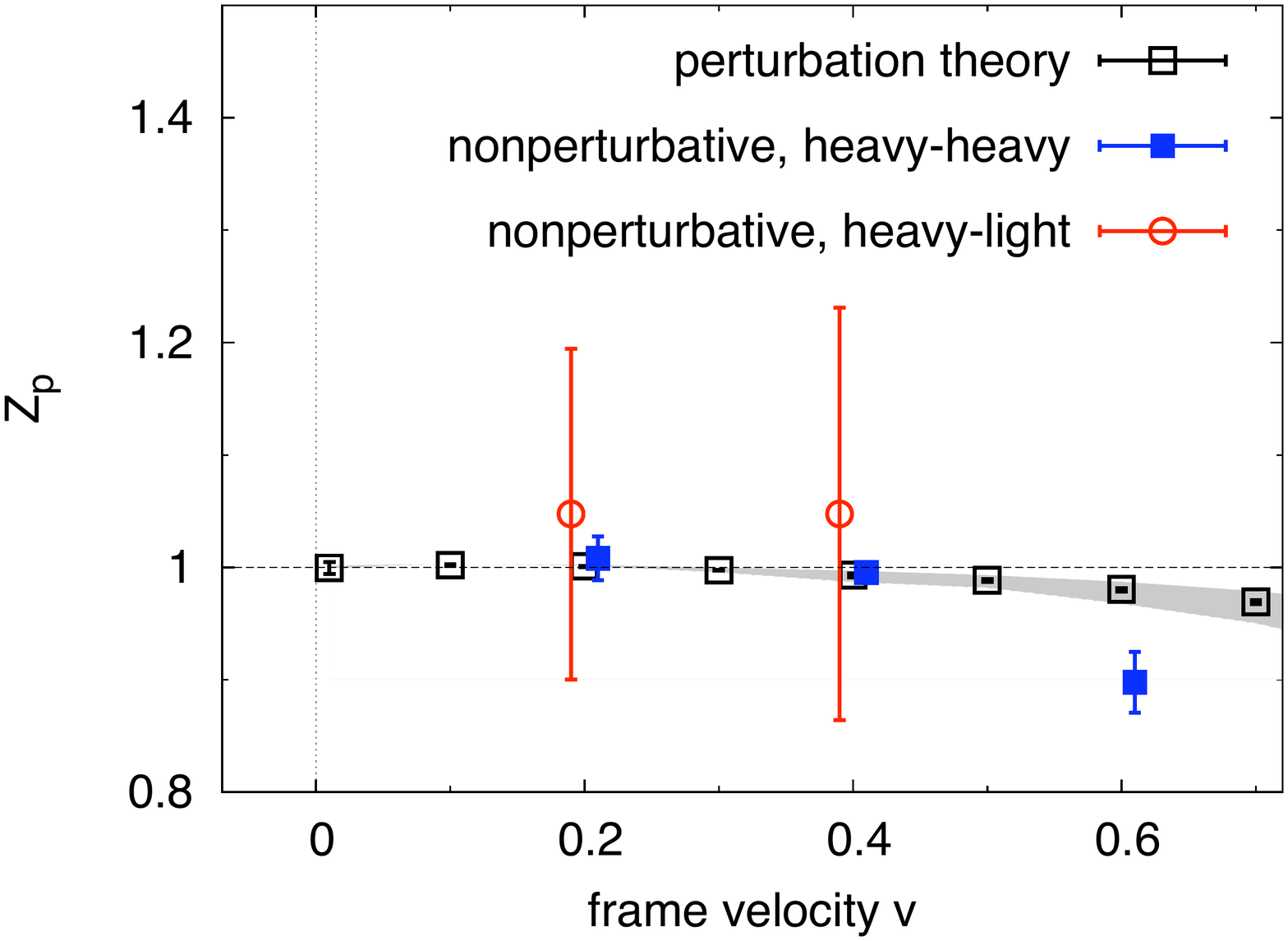}
\fi

    \caption[Renormalization of the external momentum.]
    {Renormalization $Z_p$ of the external momentum $\bs{P}_0=\gamma m
      \bs{v}$.  We show perturbative and nonperturbative
      results for heavy-heavy and heavy-light mesons, with
      a slight horizontal offset for legibility. The uncertainties
      shown on the data points for the perturbative results are purely
      statistical due to the \vegas\ integration. The strong coupling constant
      is taken to be $\alpha_s = \alpha_V(q^*)$ and the error band is obtained
      by varying the matching point in the range $[q^*/2,2q^*]$.}
  \label{selfenergy:fig:comparison_NP_PT_deltaZp}
\end{figure}

\begin{figure}
  \centering

\ifpdf
\includegraphics[width=\linewidth]{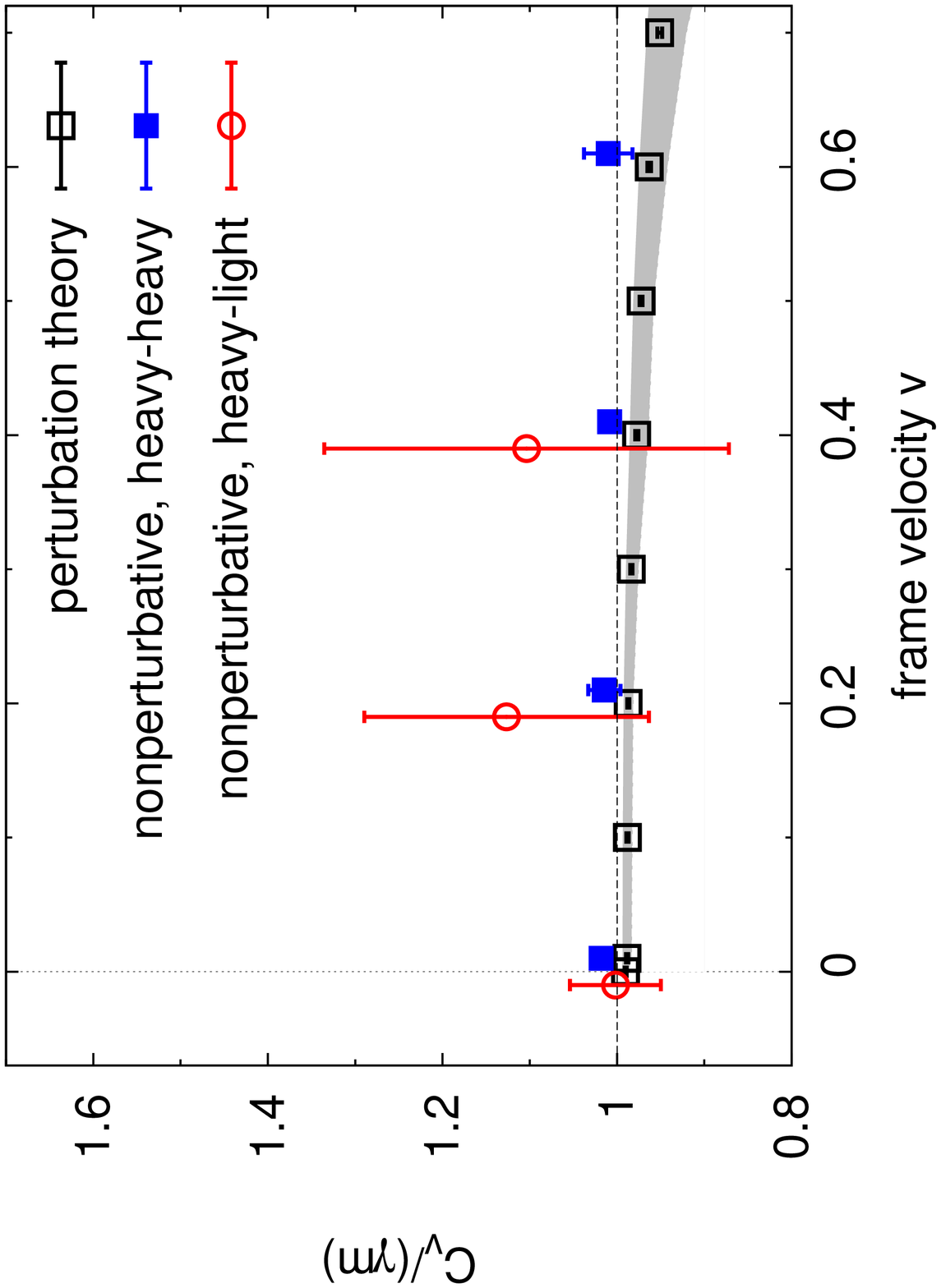}
\else
\includegraphics[height=\linewidth,angle=270]{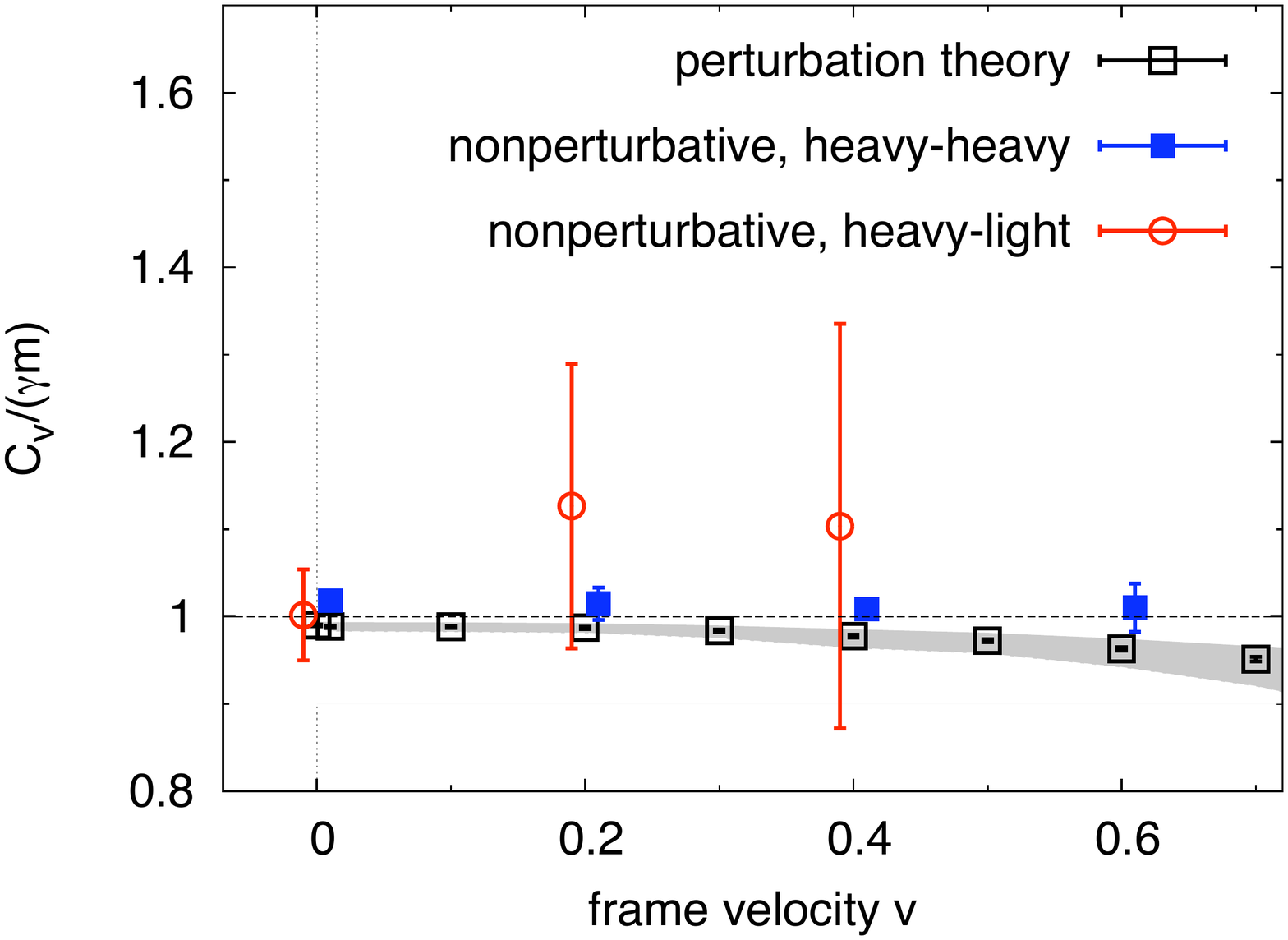}
\fi

    \caption[Renormalization of the energy shift $C_v$.]
    {Renormalization of the energy shift $C_v$ compared to the tree level
      value $\gamma m$. We show perturbative and
      nonperturbative results for heavy-heavy and heavy-light mesons, with
      a slight horizontal offset for legibility.
      Uncertainties are presented as in
      Fig.~\ref{selfenergy:fig:comparison_NP_PT_deltaZp}.}
    \label{selfenergy:fig:comparison_NP_PT_deltaCv}

\end{figure}

In Figs.~\ref{selfenergy:fig:comparison_NP_PT_deltaZp} and
\ref{selfenergy:fig:comparison_NP_PT_deltaCv} we show both
perturbative and nonperturbative results for the renormalization of
the external momentum and the energy shift between QCD and mNRQCD (see
Section \ref{sec:dispersion_relation}). The discrepancies we find at 
$v=0.6$ indicate sizable higher order loop contributions as $v$ grows. 
High-$\beta$ simulations verify the one-loop
perturbative calculation as described earlier,
and preliminary estimates of the gluonic (i.e., quenched) two-loop
contribution using high-$\beta$ simulations show that higher-order
loop corrections reduce this discrepancy; further work is in progress
and will be presented in a forthcoming publication.

%
%
%
%
\section{Conclusion}
\label{sec:conclusions}

We have derived the mNRQCD action through
$\mathcal{O}(1/m^2,v^4_{\mathrm{rel}})$ and discretized it with errors
starting at $\mathcal{O}(\alpha_s a^2)$ (tree-level errors begin at
$\mathcal{O}(a^5)$). The one-loop renormalizations of the
wavefunction, the external momentum, the frame velocity, and the
energy shift $E_0$ have been computed and presented here. In the cases
of the external momentum and the energy shift, we compared 
perturbative and nonperturbative results. Nonperturbative
calculations of heavy-heavy meson and heavy-light meson properties
were undertaken, with the aim of testing the specific action and the
general method. Fig.~\ref{fig:f_eta_ex} is particularly instructive;
it shows the reduction in discretization errors obtained by using
mNRQCD compared to non-moving NRQCD to compute the fictitious $\eta_b$
decay constant. Whether mNRQCD will prove indispensable in
determinations of heavy-to-light form factors is still to be seen.
Nevertheless, lattice calculations of these form factors are a pressing
need, and the more tools we have at our disposal, the more quickly can
we understand and reduce the errors in our calculations. In
particular, these methods will enable us to explore the $q^2 \to
0$ limit needed for the rare decay $B \to K^*\gamma$ while
maintaining control over lattice discretization errors for the light
vector meson. In future work we will employ mNRQCD, and other tools,
to move towards this goal.

%
%
%
%
%
\section*{Acknowledgments}

We thank L.C.~Storoni for useful
conversations. This work has made use of the resources provided by:
the Darwin Supercomputer of the University of Cambridge High
Performance Computing Service (\url{http://www.hpc.cam.ac.uk}),
provided by Dell Inc.\ using Strategic Research Infrastructure Funding
from the Higher Education Funding Council for England; the Edinburgh
Compute and Data Facility (\url{http://www.ecdf.ed.ac.uk}), which is
partially supported by the eDIKT initiative
(\url{http://www.edikt.org.uk}); and the Fermilab Lattice Gauge Theory
Computational Facility (\url{http://www.usqcd.org/fnal}).
We thank the DEISA Consortium (\url{http://www.deisa.eu}), co-funded
through the EU FP6 project RI-031513 and the FP7 project RI-222919,
for support within the DEISA Extreme Computing Initiative.
We thank the U.K.\ Royal Society (C.T.H.D. and A.H.) and the Leverhume Trust
(C.T.H.D.)  for financial support.  G.M.v.H.\ was supported by the Deutsche
Forschungsgemeinschaft in the SFB/TR 09. This work was supported in part by
the Sciences and Technology Facilities Council.  The Universities of Edinburgh
and Glasgow are supported in part by the Scottish Universities Physics
Alliance (SUPA).
\appendix

\section{Notation}

\label{app:notation}

In this Appendix we summarize for convenience our choices of notation and
convention detailed throughout the main text.

\begin{itemize}
\item Lorentz boost:
\begin{equation*}
\Lambda=\left(\begin{array}{cc} \gamma & \gamma\:v^k \\[2mm] 
\gamma\:v^j & ~~~\delta^{jk}+\frac{\gamma^2}{1+\gamma}\:v^j v^k ~~~
\end{array}\right)
\label{eqn:Lambda_explicit}
\end{equation*}
with $\gamma=(1-\bs{v}^2)^{-1/2}$
\item gamma matrices:
\begin{align*}
\dgamma^0&=\left(\begin{array}{cc} \sigma^0 & 0 \\ 0 & -\sigma^0
\end{array}\right),\hspace{3ex}\dgamma^j=\left(\begin{array}{cc} 0
& \sigma^j \\ -\sigma^j & 0 \end{array}\right),\\
\dgamma^5&=i\dgamma^0\dgamma^1\dgamma^2\dgamma^3
=\left(\begin{array}{cc} 0 & \sigma^0 \\ \sigma^0 & 0 \end{array}\right)
\end{align*}
with the Pauli matrices $\sigma^j$. We define $\sigma^0=\mathbb{1}_{2\times2}$.
\item spinorial Lorentz boost:
\begin{equation*}
 S(\Lambda)=\frac{1}{\sqrt{2(1+\gamma)}}\left(\begin{array}{cc}
1+\gamma & \gamma\:\bs{\sigma}\cdot\bs{v} \\ \gamma\:\bs{\sigma}\cdot\bs{v}
& 1+\gamma \end{array} \right)
\end{equation*}
\item covariant derivatives and field strength tensor:
\begin{equation*}
{D}_\mu=\frac{\partial}{\partial {x}^\mu}+ig{A}_\mu
\end{equation*}
\begin{equation*}
[D_\mu,D_\nu]=igF_{\mu\nu}
\end{equation*}
\item chromoelectric and chromomagnetic fields in Minkowski space:
\begin{equation*}
{E}_k={F}_{0k}, \hspace{4ex}{B}_j=-\frac12\epsilon_{jkl}{F}_{kl}
\end{equation*}
\item chromoelectric and chromomagnetic fields in Euclidean space:
\begin{equation*}
{E}_k=-{F}_{4k}, \hspace{4ex}{B}_j=-\frac12\epsilon_{jkl}{F}_{kl}
\end{equation*}
\end{itemize}

\section{Removing time derivatives in $\boldsymbol{H}$ at order $\boldsymbol{1/m^2}$}
\label{app:2nd_time_der_removal}
In this section we show in detail how additional time derivatives can be
removed from the mNRQCD Lagrangian at $\mathcal{O}(1/m^2)$. In particular we
give an explicit expression for the operator $V$ in (\ref{eqn:td_trans_2}).

The field redefinition (\ref{eqn:td_trans_2}) results in
\begin{align*}
\mathcal{L}&=\gamma\:\overline{\tilde{\Psi}}_{(2)}\left[O_0
+\frac{1}{\gamma m}O_{(2)1}+\frac{1}{(\gamma m)^2}O_{(2)2}\right]
\tilde{\Psi}_{(2)}\\
&+\mathcal{O}(1/m^3)
\end{align*}
with
\begin{eqnarray*}
O_{(2)1}&=&O_{(1)1},\\
O_{(2)2}&=&O_{(1)2}+\left\{V,\:O_0\right\}, \label{eq:O22}
\end{eqnarray*}
and we need to write $O_{(1)2}=O_{(2)2}-\left\{V,\:O_0\right\}$
with some operator $V$ such that $O_{(2)2}$ does not contain time derivatives.
We will treat the different terms in $O_{(1)2}$ (see (\ref{eq:O12}))
individually. Note that the last term, $-\left\{-\frac{1}{2}U^2,\:O_0\right\}$,
is already in the desired form. The time-derivative in the original $O_2$,
defined after (\ref{eq:Lagrangian-O}), can be treated as follows:
\begin{align}
\nonumber\frac{ig}{8}\gamma\:\dgamma^0&\epsilon_{jkl}\Sigma^j
\Lambda^0_{\:\:\:k}\left\{D_0,\: E'_l\right\}\\
\nonumber=&-\frac{ig}{8}\gamma\:\dgamma^0\epsilon_{jkl}\Sigma^j
\Lambda^0_{\:\:\:k}\left\{\bs{v} \cdot \bs{D},\: E'_l\right\}\\
&-\left\{-\frac{g}{8}\gamma\:\epsilon_{jkl}\Sigma^j\Lambda^0_{\:\:\:k}
E'_l,\:\:O_0 \right\}. \label{eq:V_1}
\end{align}
Next, using
\begin{equation}
U=\frac14(\gamma^2-1)O_0+\frac i2\dgamma^0\:\bs{v}\!\cdot\!\bs{D}
\label{eq:U_O0}
\end{equation}
we obtain 
\begin{eqnarray}
 \nonumber UO_0U&=&\frac12\left\{U^2,\:O_0\right\}
+\frac12\left[U,\:[O_0,\: U] \right]\\
 \nonumber&=&\frac12\left\{U^2,\:O_0\right\}\\
\nonumber&&\!\!+\frac12\left[U,\:\left[O_0,
\: \frac14(\gamma^2-1)O_0+\frac i2\dgamma^0\:\bs{v}\!\cdot\!\bs{D}\right] \right]\\
 \nonumber&=&\frac12\left\{U^2,\:O_0\right\}\\
\nonumber&&\!\!+\frac12\left[U,\:\left[i\dgamma^0(D_0
+ \bs{v}\!\cdot\!\bs{D}),\: \frac i2\dgamma^0\:\bs{v}\!\cdot\!\bs{D}\right] \right]\\
 \nonumber&=&\frac12\left\{U^2,\:O_0\right\}
-\frac 14\left[U,\:[D_0, \bs{v}\!\cdot\!\bs{D}] \right]\\
 \nonumber&=&\frac12\left\{U^2,\:O_0\right\}\\
\nonumber&&\!\!-\frac{i}{16}\left[\dgamma^0\left((\gamma^2\!-\!1) D_0
+(\gamma^2\!+\!1)\bs{v}\!\cdot\!\bs{D}\right),\:\:ig\bs{v}\!\cdot\!\bs{E} \right]\\
\nonumber&=&\!\!-\left\{-\frac12U^2,\:O_0\right\}\\
\nonumber&&\!\!+\frac{g}{16}\dgamma^0\!\left((\gamma^2\!-\!1)D_0^\mathrm{ad}
+(\gamma^2\!+\!1)\bs{v}\!\cdot\!\bs{D}^\mathrm{ad}\right)(\bs{v}\!\cdot\!\bs{E})\\
\label{eq:V_2}
\end{eqnarray}
and
\begin{align}
\nonumber \left\{U,\:O_1\right\}=&\left\{\frac14(\gamma^2-1)O_0+\frac i2\dgamma^0
\:\bs{v}\!\cdot\!\bs{D},\:O_1\right\}\\
\nonumber =& \left\{\frac i2\dgamma^0\:\bs{v}\!\cdot\!\bs{D},\:O_1\right\}
-\left\{-\frac14(\gamma^2-1)O_1, \:O_0 \right\}\\
\nonumber =& \left\{\frac i2\dgamma^0\:\bs{v}\!\cdot\!\bs{D},\:O_{(1)1}\right\}
-\left\{\frac i2\dgamma^0\:\bs{v}\!\cdot\!\bs{D},\:\left\{U,\:O_0\right\}\right\}\\
\nonumber&-\left\{-\frac14(\gamma^2-1)O_1, \:O_0 \right\}\\
\nonumber =& \left\{\frac i2\dgamma^0\:\bs{v}\!\cdot\!\bs{D},\:O_{(1)1}\right\}
+\left[U,\:\left[\frac i2\dgamma^0\:\bs{v}\!\cdot\!\bs{D},\:O_0\right]\right]\\
\nonumber&-\left\{\left\{\frac i2\dgamma^0\:\bs{v}\!\cdot\!\bs{D},\:U\right\}
-\frac14(\gamma^2-1)O_1, \:O_0 \right\}.\\
\label{eq:V_3}
\end{align}
Let us now consider the nested commutator in (\ref{eq:V_3}):
\begin{align*}
&\left[U,\:\left[\frac i2\dgamma^0\:\bs{v}\!\cdot\!\bs{D},\:O_0\right]\right]
=\left[U,\:\left[\frac i2\dgamma^0\:\bs{v}\!\cdot\!\bs{D},\:i\dgamma^0 D_0\right]\right]\\
&\hspace{4ex}=\left[\frac i4\dgamma^0\left((\gamma^2-1) D_0
+(\gamma^2+1)\bs{v}\!\cdot\!\bs{D}\right), \frac{ig}{2}\bs{v}\cdot\bs{E}\right]\\
&\hspace{4ex}=-\frac{g}{8}\dgamma^0\left((\gamma^2-1)D_0^\mathrm{ad}
+(\gamma^2+1)\bs{v}\!\cdot\!\bs{D}^\mathrm{ad}\right)(\bs{v}\cdot\bs{E})\,.
\end{align*}
We conclude from (\ref{eq:O12}), (\ref{eq:V_1}), (\ref{eq:V_2}) and (\ref{eq:V_3}) that
\begin{eqnarray}
\nonumber V&=&-\frac{g}{8}\gamma\:\epsilon_{jkl}\Sigma^j\Lambda^0_{\:\:\:k} E'_l
+\left\{\frac i2\dgamma^0\:\bs{v}\!\cdot\!\bs{D},\:U\right\}\\
&&-\frac14(\gamma^2-1)O_1-U^2 \label{eq:V}
\end{eqnarray}
and
\begin{eqnarray*}
O_{(2)2}&=&\frac{g}{8}\gamma\:\dgamma^0\bigg(D^\mathrm{ad}_\mu u_\nu F^{\mu\nu}
+i\epsilon_{jkl}\Sigma^j\Lambda^m_{\:\:\:k}\left\{D_m,\: E'_l\right\}\\
&&\hspace{10ex}-i\epsilon_{jkl}\Sigma^j\Lambda^0_{\:\:\:k}
\left\{\bs{v}\!\cdot\!\bs{D},\: E'_l\right\}\bigg)\\
&&-\frac{g}{16}\dgamma^0\!\left((\gamma^2-1)D_0^\mathrm{ad}
+(\gamma^2+1)\bs{v}\!\cdot\!\bs{D}^\mathrm{ad}\right)(\bs{v}\cdot\bs{E})\\
&&+\frac i4\dgamma^0\left\{\bs{v}\!\cdot\!\bs{D},\:\bs{D}^2-(\bs{v}
\!\cdot\!\bs{D})^2+g\bs{\Sigma}\!\cdot\!\bs{B'}\right\}\\
&&\\
&=&\frac{g}{8}\gamma^2\dgamma^0 \left(\bs{D}^\mathrm{ad}\cdot\bs{E}
-\bs{v}\cdot(\bs{D}^\mathrm{ad}\times\bs{B})\right) \\
&&+\frac{ig}{8}\gamma\dgamma^0\bs{\Sigma}\cdot
\left(\bs{D}\times{\bs{E'}}-{\bs{E'}}\times\bs{D} \right) \\
&&-\frac{ig\gamma^2}{8(1+\gamma)}\dgamma^0\left\{\bs{v}\!\cdot\!\bs{D},
\:\:\bs{\Sigma}\cdot(\bs{v}\times{\bs{E'}}) \right\}\\
&&+\frac{i}{4}\dgamma^0\left(\left\{\bs{v}\!\cdot\!{\bs{D}},\:\bs{D}^2\right\}
-2(\bs{v}\!\cdot\!\bs{D})^3\right)\\
&&+\frac{ig}{4}\dgamma^0\left\{\bs{v}\!\cdot\!\bs{D},
\:\:\bs{\Sigma}\cdot{\bs{B'}}\right\}\\
&&+\frac{(2-\bs{v}^2)g\gamma^2}{16}\dgamma^0 \left(D^{\mathrm{ad}}_0-\bs{v}
\!\cdot\!\bs{D}^{\mathrm{ad}}\right)\left(\bs{v}\cdot\bs{E}\right).
\end{eqnarray*}
%
\section{Lattice derivatives and field strength}
\label{app:latt_deriv}
In this section we give explicit expressions for the discretized
derivatives we use in our lattice action,
Eqs.~(\ref{eq:kinetic_energy_operator}), (\ref{eq:dH_full}). All
expressions are constructed from the elementary forward, backward and
symmetric derivatives
\begin{eqnarray*}
\Delta^{+}_\mu{\psi}(x)&=&U_\mu(x){\psi}(x+\hat{\mu})-{\psi}(x), \\
\Delta^{-}_\mu{\psi}(x)&=&{\psi}(x)-U_{-\mu}(x){\psi}(x-\hat{\mu}), \\
\Delta^{\pm}_\mu{\psi}(x)&=&\frac12\left[U_\mu(x){\psi}(x+\hat{\mu})
-U_{-\mu}(x){\psi}(x-\hat{\mu})\right].
\end{eqnarray*}
For performance reasons, we construct higher-order operators to be maximally
local by balancing the occurrence of these three types. We also symmetrize the expressions.

\noindent \emph{Unimproved derivatives:}
%
\begin{eqnarray*}
\Delta^{(2)}&=&\sum_{j=1}^3\Delta^{+}_j\Delta^{-}_j\\
\Delta_v^{(2)}&=&\frac12\sum_{j,k=1}^3v^jv^k\left(\Delta^{+}_j\Delta^{-}_k
+\Delta^{-}_j\Delta^{+}_k \right)\\
\Delta_v^{(3)}&=&\frac12\sum_{j,k,l=1}^3v^jv^kv^l
\left(\Delta^{+}_j\Delta^{\pm}_k\Delta^{-}_l
+\Delta^{-}_j\Delta^{\pm}_k\Delta^{+}_l \right)\\
\Delta_v^{(4)}&=&\frac12\sum_{j,k,l,m=1}^3\!\!\!v^jv^kv^lv^m
\big(\Delta^{+}_j\Delta^{-}_k\Delta^{+}_l\Delta^{-}_m\\
&&\hspace{20ex}+\Delta^{-}_j\Delta^{+}_k\Delta^{-}_l\Delta^{+}_m \big)
\end{eqnarray*}
%

\noindent \emph{Improved derivatives:}
%
\begin{eqnarray*}
\tilde{\Delta}^\pm_j&=&\Delta^{\pm}_j
-\frac{1}{6}\Delta^{+}_j\Delta^{\pm}_j\Delta^{-}_j\\
\tilde{\Delta}^{(2)}&=&\Delta^{(2)}
-\frac{1}{12}\sum_{j=1}^3\Delta^{+}_j\Delta^{-}_j\Delta^{+}_j\Delta^{-}_j\\
\tilde{\Delta}_v^{(2)}&=&\Delta_v^{(2)}+\frac14\sum_{j,k=1}^3v^jv^k
\Delta^{+}_j\Delta^{-}_j\Delta^{+}_k\Delta^{-}_k\\
&&-\frac{1}{12}\sum_{j,k=1}^3v^jv^k
\left(\Delta^{+}_j\Delta^{-}_j\Delta^{+}_j\Delta^{-}_k
+\Delta^{-}_j\Delta^{+}_j\Delta^{-}_j\Delta^{+}_k\right.\\
&&\hspace{14ex}\left.+\Delta^{+}_j\Delta^{-}_k\Delta^{+}_k\Delta^{-}_k
+\Delta^{-}_j\Delta^{+}_k\Delta^{-}_k\Delta^{+}_k\right)
\end{eqnarray*}
%

\noindent \emph{Unimproved adjoint derivative:}
%
\begin{eqnarray*}
\Delta^{\mathrm{ad}}_\mu \tilde{F}_{\rho\sigma}(x)&=&\frac{1}{2}
\bigg[U_\mu(x) \tilde{F}_{\rho\sigma}(x+\hat{\mu})U^\dag_\mu(x)\\
&&\hspace{4ex}-\:\:U_{-\mu}(x) \tilde{F}_{\rho\sigma}
(x-\hat{\mu})U^\dag_{-\mu}(x)\bigg]
\end{eqnarray*}
%

\noindent \emph{Improved field strength tensor:}
%
\begin{eqnarray*}
\tilde{F}_{\mu\nu}(x) &=&\frac53F_{\mu\nu}(x)\\
&&-\frac{1}{6}\bigg(\hspace{2ex}U_\mu(x)F_{\mu\nu}
(x\!+\!\hat{\mu})U^\dagger_\mu(x)\\
&&\hspace{4.7ex}+\:\:U_{-\mu}(x)F_{\mu\nu}(x\!
-\!\hat{\mu})U^\dagger_{-\mu}(x)\\
&&\hspace{4.7ex}-\:\:(\mu \leftrightarrow\nu)\hspace{2ex}\bigg),
\end{eqnarray*}
where
\begin{eqnarray}
 \nonumber F_{\mu\nu}(x) &\!=\!& \frac{-i}{2g}
\left( \Omega_{\mu\nu}(x)-\Omega^\dagger_{\mu\nu}(x)\right),
\label{curly_F}\\
\nonumber \Omega_{\mu\nu}(x) &\!=\!&  \sfrac{1}{4}\!\!\!\!
\sum_{\{(\alpha,\beta)\}_{\mu\nu}}\!\!\!\!\!\!U_\alpha(x)
U_\beta(x\!+\!\hat{\alpha})U_{\text{-}\alpha}(x\!+\!\hat{\alpha}\!
+\!\hat{\beta})U_{\text{-}\beta}(x\!+\!\hat{\beta})
\end{eqnarray}
with
\begin{equation*}
\{(\alpha,\beta)\}_{\mu\nu}=\{(\mu,\nu),(\nu,\text{-}\mu),
(\text{-}\mu,\text{-}\nu),(\text{-}\nu,\mu)\}\:\:\:{\rm for}
\:\:\:\mu\neq\nu
\end{equation*}

%
%
%
\section{Tadpole improvement}
\label{app:tadpoles}
In the perturbative calculation it is possible to explicitly work out
every path appearing in the evolution and cancelling the tadpole factors
which appear in every instance of $U_\mu(x) U_\mu^\dagger(x)$.
Here we give analytical expressions of the tadpole improvement corrections 
for this case for the full $\mathcal{O}(1/m^2,\vnr^4)$ action.

Numerical results for $m=2.8$ and $n=2$ can be found in Table
\ref{tab:OmegaMF_full_action_complete} and should be compared to
Table \ref{tab:OmegaMF_full_action}.
\begin{eqnarray*}
\hat{\Omega}^{(\mathrm{tadpole})}_0 &=& - \hat{\Omega}^{(\mathrm{tadpole})}_1\\
 &=& u_0^{(2)}\bigg(1-\frac{v^{2}}{3}-\frac{19v^{4}}{768}-\frac{v^{6}}{1024}\\
&&\!+\;\;\frac{2688-852v^{2}+11v^{4}-13v^{6}}{768\gamma m}\\
&&\!-\;\;\frac{3456-4920v^{2}+2497v^{4}-264v^{6}+15v^{8}}{3072\gamma^{2}m^{2}}\\
&&\!-\;\;\frac{516-1264v^{2}+1058v^{4}+275v^{6}-15v^{8}}{768\gamma^{3}m^{3}}\\
&&\!-\;\;\frac{-591+1460v^{2}-1358v^{4}+448v^{6}+5v^{8}}{256\gamma^{4}m^{4}}\\
&&\!-\;\;\frac{81-216v^{2}+246v^{4}-128v^{6}+25v^{8}}{64\gamma^{5}m^{5}}
\bigg)
\end{eqnarray*}
\begin{eqnarray*}
\hat{\Omega}^{(\mathrm{tadpole})}_2 &=& u_0^{(2)}\Big(-\frac{5}{3}
+\frac{7v^{2}}{32}+\frac{13v^{4}}{512}+\frac{v^{6}}{2048}\\
&&\!+\;\;\frac{-10880+4480v^{2}-215v^{4}+35v^{6}}{3072\gamma m}\\
&&\!+\;\;\frac{-12480\!+\!10288v^{2}\!+\!4321v^{4}\!-\!360v^{6}\!+\!15v^{8}}
{6144\gamma^{2}m^{2}}\\
&&\!+\;\;\frac{2412-4864v^{2}+3974v^{4}+311v^{6}-15v^{8}}{1536\gamma^{3}m^{3}}\\
&&\!+\;\;\frac{-879+2100v^{2}-1982v^{4}+640v^{6}+5v^{8}}{512\gamma^{4}m^{4}}\\
&&\!+\;\;\frac{81-216v^{2}+246v^{4}-128v^{6}+25v^{8}}{128\gamma^{5}m^{5}}
\Big)
\end{eqnarray*}
\begin{eqnarray*}
\hat{\Omega}^{(\mathrm{tadpole})}_v &=& u_0^{(2)}\Big(-\frac{5}{3}+\frac{11v^{2}}{48}
+\frac{29v^{4}}{1536}+\frac{v^{6}}{2048}\\
&&\!+\;\;\frac{-5440+1860v^{2}-51v^{4}+16v^{6}}{1536\gamma m}\\
&&\!-\;\;\frac{12480+712v^{2}-3521v^{4}+320v^{6}-15v^{8}}{6144\gamma^{2}m^{2}}\\
&&\!+\;\;\frac{2412-3016v^{2}+2306v^{4}+299v^{6}-15v^{8}}{1536\gamma^{3}m^{3}}\\
&&\!+\;\;\frac{-879+1812v^{2}-1614v^{4}+544v^{6}+5v^{8}}{512\gamma^{4}m^{4}}\\
&&\!+\;\;\frac{81-216v^{2}+246v^{4}-128v^{6}+25v^{8}}{128\gamma^{5}m^{5}}
\Big)
\end{eqnarray*}

\begin{table}
\begin{tabular}{ccccccc}
\hline\hline
 \\[-2ex]
$v$ && $\hat{\Omega}^{(\mathrm{tadpole})}_0/u_0^{(2)}$
&& $\hat{\Omega}^{(\mathrm{tadpole})}_2/u_0^{(2)}$
&& $\hat{\Omega}^{(\mathrm{tadpole})}_v/u_0^{(2)}$
\\ \\[-2ex]\hline
$0.00$ && $2.10610$ && $-3.14336$ &&  --- \\
$0.01$ && $2.10600$ && $-3.14319$ && $-3.14321$ \\
$0.10$ && $2.09592$ && $-3.12639$ && $-3.12898$ \\
$0.20$ && $2.06507$ && $-3.07557$ && $-3.08573$ \\
$0.30$ && $2.01267$ && $-2.99112$ && $-3.01321$ \\
$0.40$ && $1.93722$ && $-2.87358$ && $-2.91083$ \\
$0.50$ && $1.83666$ && $-2.72404$ && $-2.77787$ \\
$0.60$ && $1.70836$ && $-2.54438$ && $-2.61351$ \\
$0.70$ && $1.54900$ && $-2.33743$ && $-2.41672$ \\
$0.75$ && $1.45625$ && $-2.22476$ && $-2.30557$ \\
$0.80$ && $1.35355$ && $-2.10633$ && $-2.18513$ \\
$0.85$ && $1.23910$ && $-1.98182$ && $-2.05402$ \\
$0.90$ && $1.10911$ && $-1.84923$ && $-1.90880$ \\
$0.95$ && $0.95262$ && $-1.70037$ && $-1.73901$ \\
\hline\hline
\end{tabular}
\caption[Tadpole improvement corrections for the full
$\mathcal{O}(1/m^2,\vnr^4)$ action.]  {Tadpole improvement corrections
$\hat{\Omega}^{(\mathrm{tadpole})}_j$ for the full $\mathcal{O}(1/m^2,\vnr^4)$
action and cancellation of $U_\mu U_\mu^\dagger$ as described in the main
text. The heavy quark mass is $m=2.8$ and the stability parameter $n=2$.  Note
that $\hat{\Omega}^{(\mathrm{tadpole})}_1 = -
\hat{\Omega}^{(\mathrm{tadpole})}_0$.}
\label{tab:OmegaMF_full_action_complete}
\end{table}
It should be noted that the expressions for partial cancellation are
significantly simpler. Numerically we find that the difference is of the order
of $10\%$ in the one-loop coefficient, see Tables \ref{tab:OmegaMF_full_action}
and~\ref{tab:OmegaMF_full_action_complete}.  We conclude that it is sufficient
to avoid multiplying $U_\mu U_\mu^\dagger$ by $1/u_0^2$ within $H_0$ and
$\delta H$ separately.
%
%
\section{Further Perturbative results}
\label{app:results-simple-actions}

In this appendix, we present one-loop perturbative results for the
renormalization of the mNRQCD propagator for various simpler forms of
the mNRQCD action.
%
%
\subsection{Simplest case}
\label{sec:mnrqcd_simple}

We considered the simplest, unimproved mNRQCD action, \textit{i.e.}\
\begin{eqnarray*}
H_0&=& - i \bs{v}\cdot\bs{\Delta}^\pm-\frac{\Delta^{(2)}
-\Delta_v^{(2)}}{2\gamma m},\\
\delta H &=& 0
\end{eqnarray*}
coupled to the Wilson gluon action. The gluon propagator in Feynman gauge is
\begin{align*}
D^{-1}(k) & = 4\sum_\mu\sin^2\frac{k_\mu}{2}+\lambda^2 
\\
& = 
2-w-w^{-1}+4\sum_j\sin^2\frac{k_j}{2}+\lambda^2
\label{eq:gluon_prop}
\end{align*}
with $w=e^{ik_4}$. The gluon mass was set to $\lambda^2=10^{-6}$.

The case $\delta H=0$ is very simple, as all propagators and
vertices are diagonal in spinor and color space, and the calculations
can be performed in reasonable time on a workstation. We used
a heavy quark mass of~$m=2.8$ and the stability parameter is~$n=2$.

In Table \ref{tab:Omega_onlyH0_action} we list $\Omega_j$ for this
action before including mean-field corrections.
We only give the finite parts of the $\Omega_j$, the infrared
divergence $-2/(3\pi) \log \lambda^2$ is not included in the
results for $\Omega_1$, $\Omega_2$ and $\Omega_v$.

\begin{figure}
\centering

\ifpdf
\includegraphics[width=\linewidth]{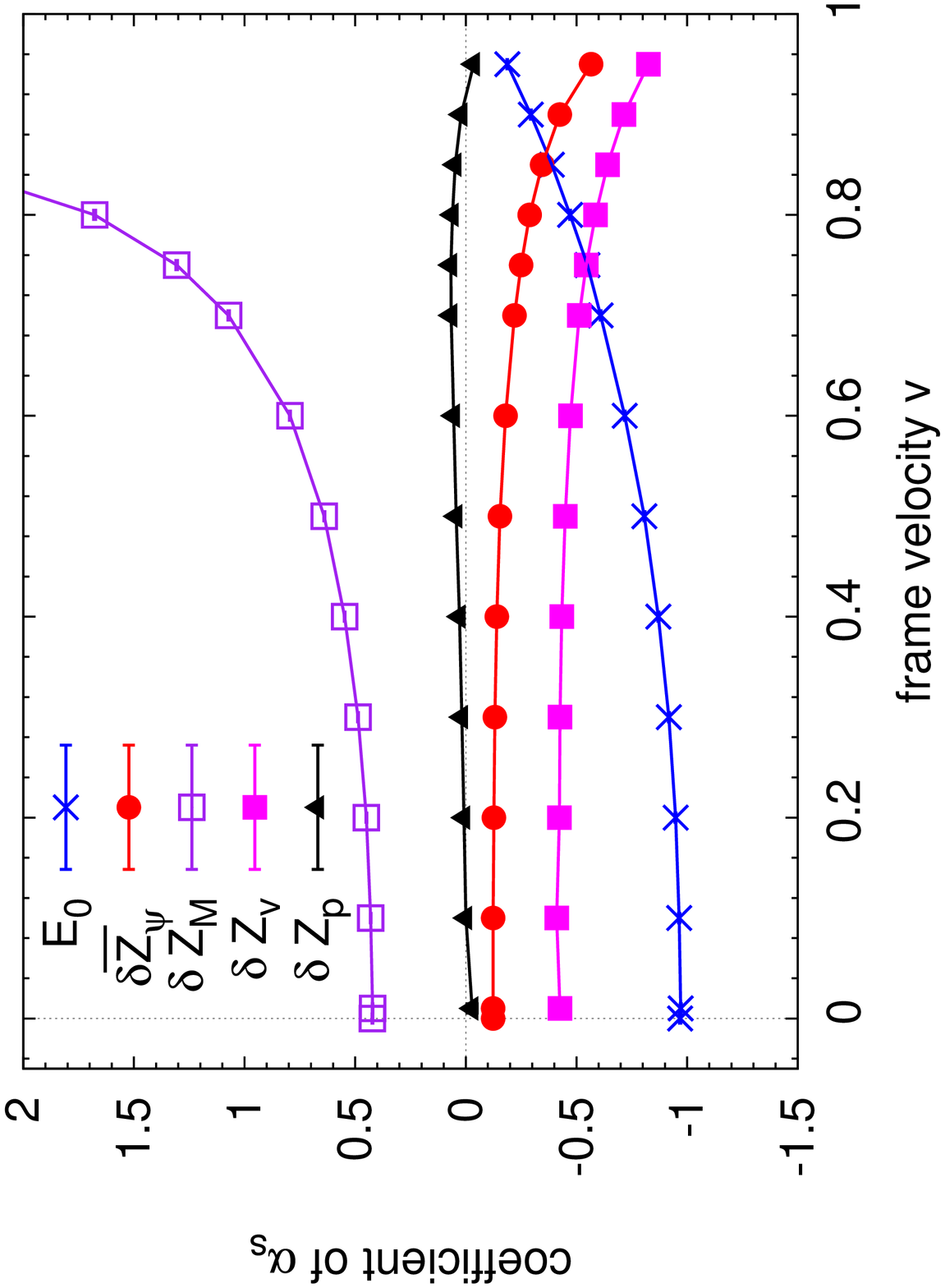}
\else
\includegraphics[height=\linewidth,angle=270]{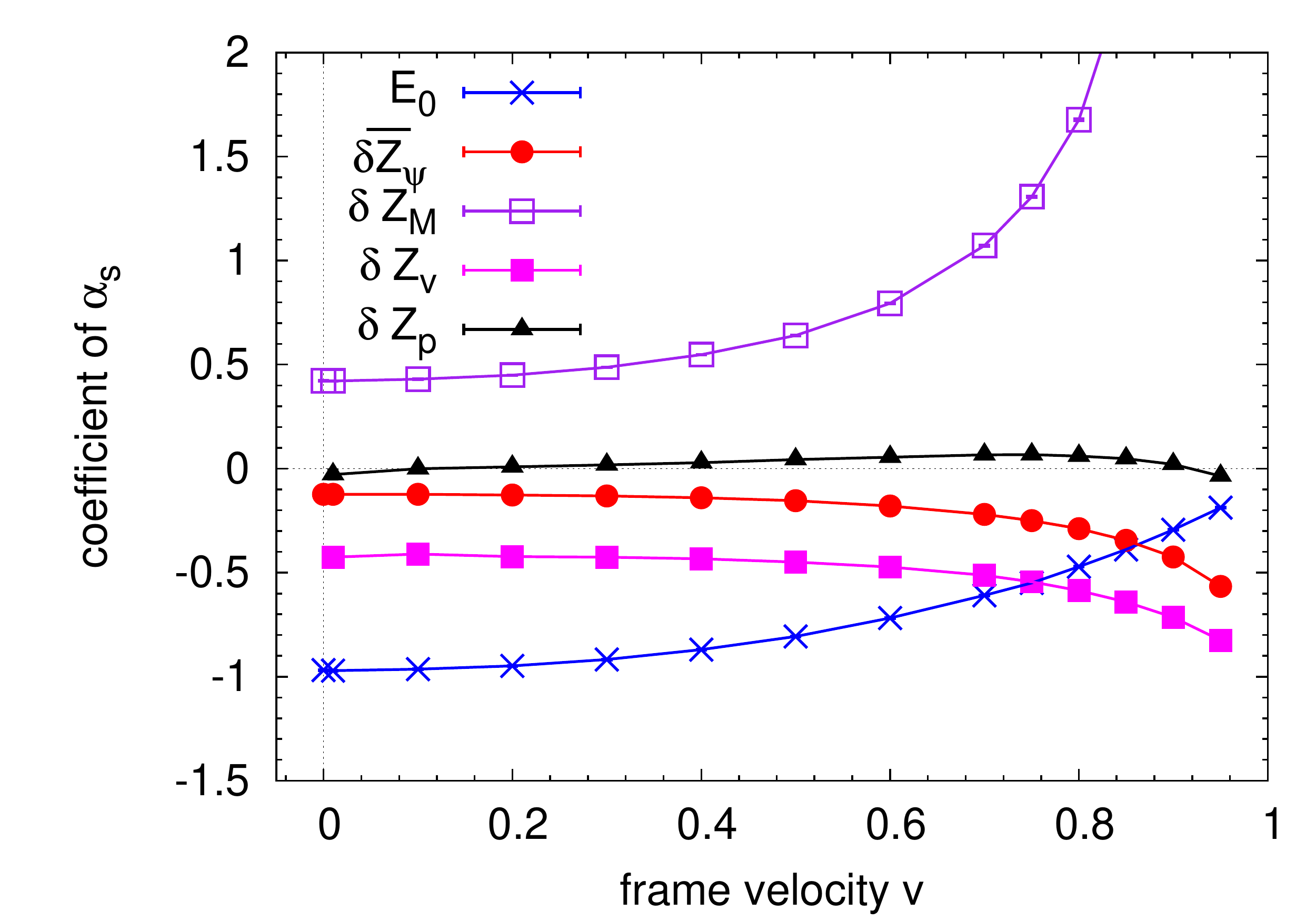}
\fi

\caption[Renormalization parameters for the simple action with
kinetic term $H_0$ only]{Heavy quark renormalization parameters
for the simple, unimproved action with kinetic term $H_0$ only.
The gluon action is the Wilson action with $\lambda^2 = 10^{-6}$
and we use $m=2.8$, $n=2$. All mean-field corrections are included
and the results are infrared-finite. Note $\delta Z_p$ is small due
to small violation of reparametrization invariance. Also note 
$\delta Z_m$ is large for this unimproved action.}
\label{fig:ren_parm_onlyH0_action}
\end{figure}

The mean-field corrections, cancelling $U_\mu U_\mu^\dagger$ factors as
described in the main text are
\begin{eqnarray}
\Omega^{(\mathrm{tadpole})}_0 &=& - \Omega^{(\mathrm{tadpole})}_1
= u_0^{(2)}\Big(1
+\frac{3-v^{2}}{\gamma m}\Big)\notag
\\[2ex] 
\Omega^{(\mathrm{tadpole})}_2 &=& \Omega^{(\mathrm{tadpole})}_v
= -u_0^{(2)}\Big(2\notag\\
&&\qquad+\;\;\frac{2n-1}{2n}\frac{3-v^{2}}{\gamma m}
\Big) \label{eq:MFcorrections_onlyH0_action}
\end{eqnarray}
whereas the corresponding expressions for tadpole cancellation described
as in Appendix~\ref{app:tadpoles} are ($n=2$)
\begin{eqnarray}
\hat{\Omega}^{(\mathrm{tadpole})}_0 &=& -\hat{\Omega}^{(\mathrm{tadpole})}_1
= u_0^{(2)}\Big(1-\frac{v^{2}}{8}
+\frac{3-v^{2}}{\gamma m}\notag\\&&\qquad\quad
-\;\;\frac{3-2v^{2}+v^{4}}{8\gamma^{2}m^{2}}
\Big)\notag
\\[2ex] 
\hat{\Omega}^{(\mathrm{tadpole})}_2 &=& \hat{\Omega}^{(\mathrm{tadpole})}_v
= u_0^{(2)}\Big(-2+\frac{v^{2}}{16}\notag\\&&\quad
+\;\;\frac{-9+3v^{2}}{4\gamma m}
+\frac{3-2v^{2}+v^{4}}{16\gamma^{2}m^{2}}
\Big).
\end{eqnarray}
The renormalization parameters of the heavy quark action (including mean-field
corrections) are plotted in Fig.~\ref{fig:ren_parm_onlyH0_action}.
For the one-loop coefficient of $u_0$ we use $u_0^{(2)} = 0.9735$
\cite{Hart:2004jn} and, as for the full action, we use 
cancellation of $U_\mu U_\mu^\dagger$ described in the main text.

\begin{table}

\begin{tabular}{ccccccccc}
\hline\hline
   \\[-2ex]
  $v$ && $\Omega_0$ && $\Omega_1$ && $\Omega_2$ && $\Omega_v$ \\
     \\[-2ex]\hline
  $0.00$ && $-2.9851(24)$ && $2.8619(24)$ && $3.9967(29)$ &&  --- \\
  $0.01$ && $-2.9879(24)$ && $2.8645(24)$ && $3.9987(29)$ && $4.003(23)$ \\
  $0.10$ && $-2.9721(24)$ && $2.8483(25)$ && $3.9889(29)$ && $3.9741(39)$ \\
  $0.20$ && $-2.9299(23)$ && $2.8033(24)$ && $3.9567(29)$ && $3.9474(31)$ \\
  $0.30$ && $-2.8564(23)$ && $2.7252(24)$ && $3.9022(29)$ && $3.8826(29)$ \\
  $0.40$ && $-2.7490(22)$ && $2.6092(23)$ && $3.8218(29)$ && $3.7898(28)$ \\
  $0.50$ && $-2.6085(22)$ && $2.4540(22)$ && $3.7104(30)$ && $3.6702(27)$ \\
  $0.60$ && $-2.4260(20)$ && $2.2462(21)$ && $3.5651(33)$ && $3.5087(27)$ \\
  $0.70$ && $-2.2057(18)$ && $1.9859(20)$ && $3.3833(39)$ && $3.3157(25)$ \\
  $0.75$ && $-2.0832(18)$ && $1.8335(20)$ && $3.2742(45)$ && $3.2110(26)$ \\
  $0.80$ && $-1.9371(17)$ && $1.6482(19)$ && $3.1333(57)$ && $3.0851(26)$ \\
  $0.85$ && $-1.7790(16)$ && $1.4343(20)$ && $3.0029(80)$ && $2.9447(26)$ \\
  $0.90$ && $-1.5992(15)$ && $1.1742(22)$ && $2.820(13)$ && $2.7790(29)$ \\
  $0.95$ && $-1.3887(13)$ && $0.8223(29)$ && $2.480(29)$ && $2.5639(36)$ \\\hline\hline
\end{tabular}

\caption[$\Omega_j$ for the simple action with $H_0$ only]{Infrared-finite
part of $\Omega_j$ for the unimproved action with kinetic term $H_0$ only, as
described in Appendix~\ref{sec:mnrqcd_simple}. The gluon action is the Wilson
action with $\lambda^2 = 10^{-6}$ and we use $m=2.8$, $n=2$. mean-field
corrections are not included, the errors shown are statistical due to the
\vegas\ integration.}\label{tab:Omega_onlyH0_action}
\end{table}

We have also computed the renormalization parameters for the action discussed
in \cite{Dougall:2004hw,Foley:2004rp},
\begin{eqnarray*}
H_0&=& - i \bs{v}\cdot\bs{\Delta}^\pm-\frac{\Delta^{(2)}
-(\bs{v}\cdot\bs{\Delta}^\pm)^2}{2\gamma m},\\
\delta H &=& 0,
\end{eqnarray*}
where a different (less local) discretization of the operator
$(\bs{v}\cdot\bs{D})^2$ is used. We used exactly the same simulation
parameters as given there, $m=2.0$, $n=2$ and found agreement with their
results for $\Omega_j$ within statistical errors.

\subsection{More improved case}
\label{sec:mnrqcd_sigmaB}

We now consider a more improved action, including the
spin-dependent $\bs{\sigma}\cdot\bs{\tilde{B}'}$-term and the spatial
and temporal lattice spacing improvement:
\begin{eqnarray*}
H_0&=& - i \bs{v}\cdot\bs{\Delta}^\pm-\frac{\Delta^{(2)}-\Delta_v^{(2)}}{2\gamma m},\\
\delta H&=&-\frac{g}{2\gamma m}\bs{\sigma}\cdot\bs{\tilde{B}'}+\delta H_\mathrm{corr},
\end{eqnarray*}
where $\delta H_\mathrm{corr}$ is the same as in (\ref{eq:final_corr}).
We use the Symanzik-improved gluon action so that the Landau gauge mean
link one-loop coefficient is $u_0^{(2)} = 0.750$~\cite{Nobes:2001tf}.

\begin{figure}
\centering

\ifpdf
\includegraphics[width=\linewidth]{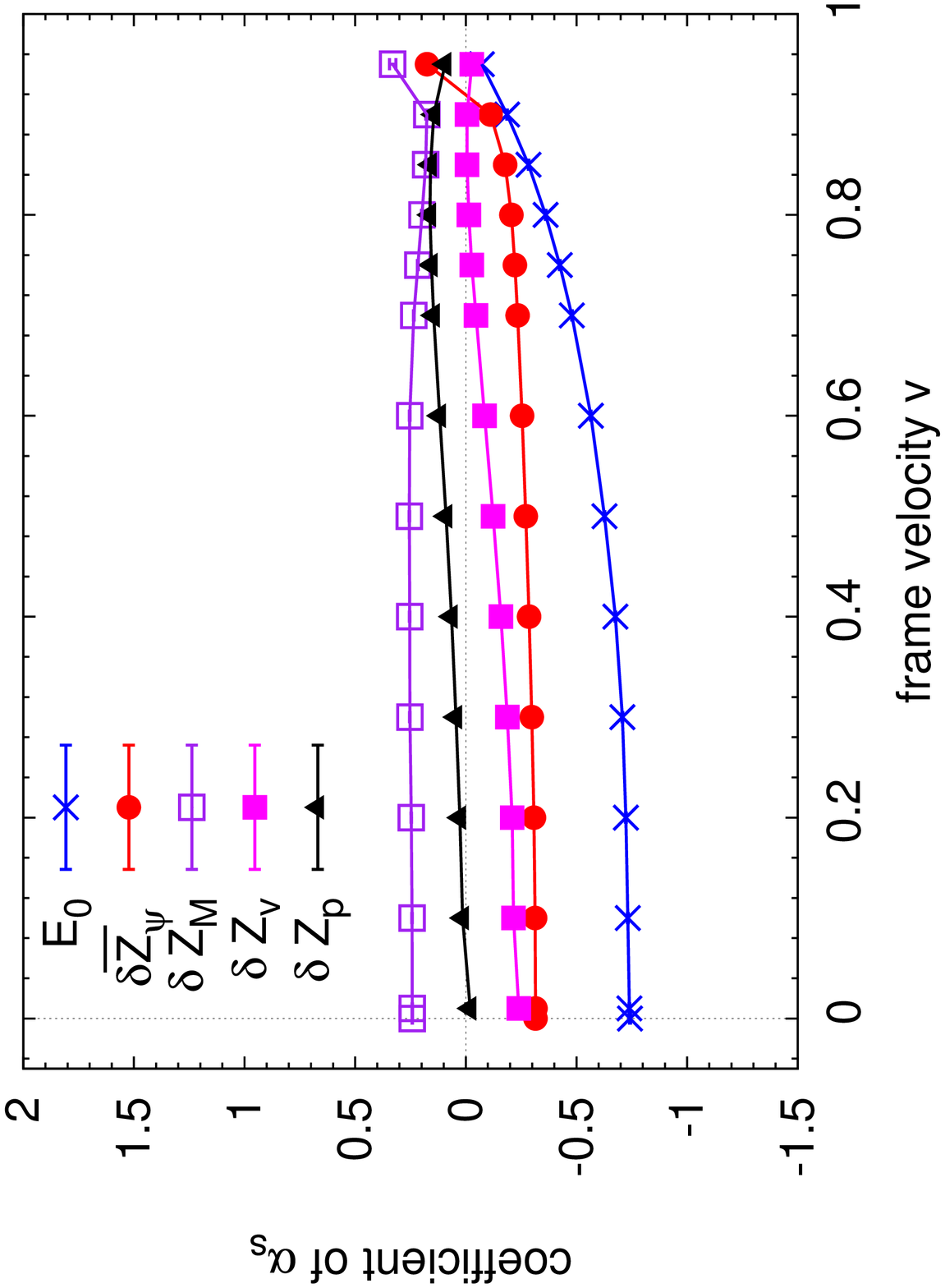}
\else
\includegraphics[height=\linewidth,angle=270]{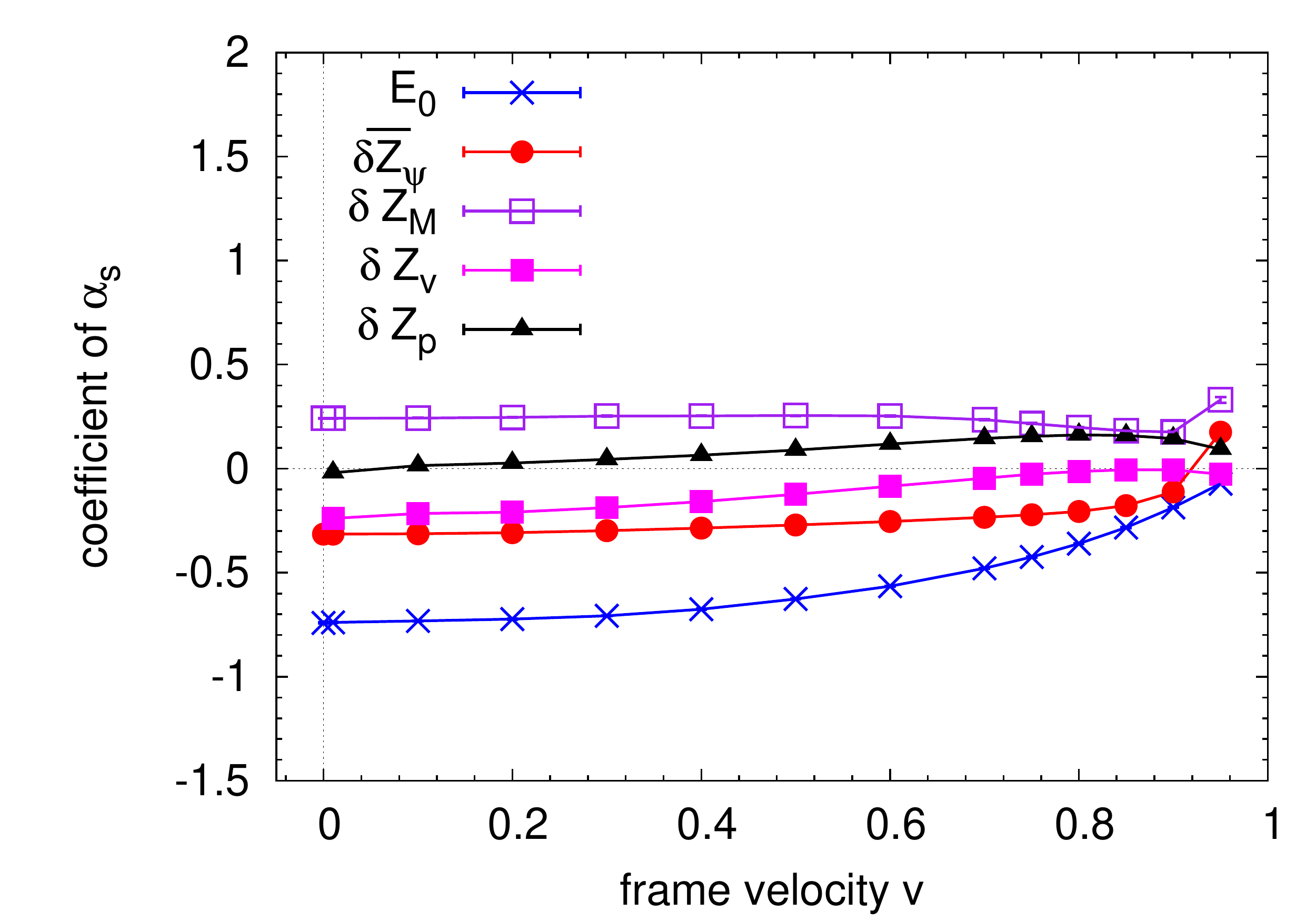}
\fi

\caption[Renormalization parameters for the $\mathcal{O}(1/m)$ action
with chromomagnetic interaction term $\delta H$]{Heavy quark
renormalization parameters for the $\mathcal{O}(1/m)$ action
with chromomagnetic interaction term $\delta H$.
The gluon action is the Symanzik-improved action with
$\lambda^2 = 10^{-6}$ and we use $m=2.8$, $n=2$. All mean-field
corrections are included.  Note improvement drastically reduces
$\delta Z_m$.}
\label{fig:ren_parm_sigmaB_action}
\end{figure}

Again, we considered $m=2.8$ and $n=2$, and as for the full
$\mathcal{O}(1/m^2,\vnr^4)$ action discussed in
the main text the gluon mass was taken to be $\lambda^2=10^{-6}$.
The results for the $\Omega_j$ obtained from the \vegas\ integration are given
in Table \ref{tab:Omega_sigmaB_action} and we show the renormalization
parameters (including mean-field corrections) as a function of the frame
velocity in Fig.~\ref{fig:ren_parm_sigmaB_action}.

\begin{table}
\begin{tabular}{ccccccccc}
\hline\hline
    \\[-2ex]
   $v$ && $\Omega_0$ && $\Omega_1$ && $\Omega_2$ && $\Omega_v$ \\
     \\[-2ex]\hline
$0.00$ && $-2.3938(19)$ && $2.0790(20)$ && $2.8211(23)$ &&  --- \\
$0.01$ && $-2.3910(19)$ && $2.0761(20)$ && $2.8180(23)$ && $2.816(20)$ \\
$0.10$ && $-2.3751(19)$ && $2.0621(20)$ && $2.8039(23)$ && $2.7780(32)$ \\
$0.20$ && $-2.3403(19)$ && $2.0327(20)$ && $2.7728(23)$ && $2.7437(25)$ \\
$0.30$ && $-2.2813(19)$ && $1.9830(19)$ && $2.7230(23)$ && $2.6729(23)$ \\
$0.40$ && $-2.1895(18)$ && $1.9033(18)$ && $2.6367(22)$ && $2.5670(22)$ \\
$0.50$ && $-2.0624(17)$ && $1.7912(17)$ && $2.5208(22)$ && $2.4222(20)$ \\
$0.60$ && $-1.9070(16)$ && $1.6525(17)$ && $2.3810(23)$ && $2.2484(19)$ \\
$0.70$ && $-1.7105(14)$ && $1.4765(16)$ && $2.1985(26)$ && $2.0375(18)$ \\
$0.75$ && $-1.5932(14)$ && $1.3716(15)$ && $2.0890(29)$ && $1.9160(18)$ \\
$0.80$ && $-1.4613(13)$ && $1.2555(15)$ && $1.9695(35)$ && $1.7893(18)$ \\
$0.85$ && $-1.3094(12)$ && $1.1314(15)$ && $1.8435(46)$ && $1.6597(18)$ \\
$0.90$ && $-1.1285(11)$ && $1.0166(16)$ && $1.7173(76)$ && $1.5480(20)$ \\
$0.95$ && $-0.9135(11)$ && $1.0892(26)$ && $1.731(20)$ && $1.6439(30)$ \\
\hline\hline
\end{tabular}
\caption[$\Omega_j$ for the $\mathcal{O}(1/m)$ action with
chromomagnetic term]{Infrared-finite part of $\Omega_j$ for
the $\mathcal{O}(1/m)$ action with chromomagnetic term, as described in
Appendix~\ref{sec:mnrqcd_sigmaB}. The
gluon action is the Symanzik-improved action with $\lambda^2 = 10^{-6}$
and we use $m=2.8$, $n=2$. Mean-field corrections are not included,
the errors shown are statistical from the \vegas\ integration.}
\label{tab:Omega_sigmaB_action}
\end{table}

%
%
\section{Poles of the improved gluon propagator}
\label{app:improved_gluon_poles}
As the heavy quark action contains only first order time derivatives finding
the poles in the propagator is trivial. It is also straightforward to find the
poles of the simple, unimproved Wilson gluon propagator.  However this is not
the case for the Symanzik-improved gluon action. In this section we analyze
the position of poles in the Symanzik-improved gluon propagator described in
Ref.~\cite{Groote:2000jd}.

We restrict our discussion to Feynman gauge where the gluon two-point function
is given by
\begin{eqnarray}
  M_{\mu\nu} &=& \left(\sum_\rho q_{\mu\rho} \hat{k}_\rho^2 + \lambda^2
    \right) \delta_{\mu\nu}\\&&\qquad +\;\;
  (1-q_{\mu\nu})\hat{k}_\mu \hat{k}_\nu\notag
\end{eqnarray}
with $q_{\mu\nu} = 1+\frac{1}{12}(\hat{k}_\mu^2+\hat{k}_\nu^2)$ and
$\hat{k}_\mu = 2\;\sin(k_\mu/2)$.

To find the poles of the propagator, first we compute the determinant of
this matrix which is a polynomial in $\hat{k}_j^2$ and
$\omega = \hat{k}_0^2$.
For a given three-momentum $k_j\in[-\pi,\pi]$ the
zeros of this expression in the $z=e^{ik_0}$ plane can be obtained by
solving $\det M(\omega) = 0$ and then using
$\omega = 2-z-1/z$. It turns out that the determinant can be
factored as $\det M(\omega) = (\omega + \bs{\hat k}^2+ \lambda^2) 
\det \tilde{M}(\omega)$, with $\bs{\hat{k}}^2 = 4 \sum_{j=1}^3 \sin^2(k_j/2)$,
so that one solution coincides with the root of the na\"{\i}ve propagator.
Numerically, for small $a^2\lambda^2$ also one of the solutions of
$\det \tilde{M}(\omega) = 0$ is very close to the na\"{\i}ve solution.
Note that the solutions come in pairs, $(z_+,z_-)$ with $z_+z_- = 1$, so
one of them lies inside the unit circle and the other outside.

\begin{figure}
\centering
\includegraphics[width=\linewidth]{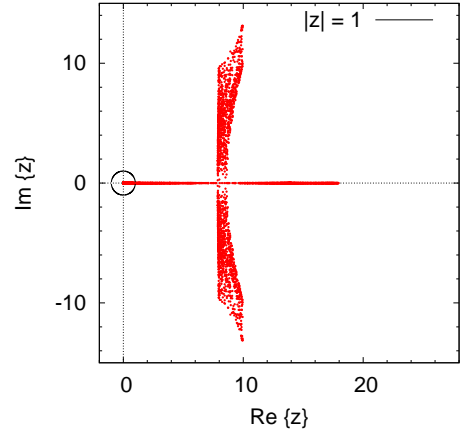}
\caption{Poles of the Symanzik-improved propagator in the complex $z$
plane.}\label{fig:gluon_improved_poles}
\end{figure}

For a given spatial momentum there are 14 solutions.
In Fig.~\ref{fig:gluon_improved_poles} these are plotted in the complex
$z$-plane for 1000 randomly chosen $k_j$. For the gluon mass a
value of $\lambda^2 = 10^{-6}$ was chosen.

\begin{figure}
\centering
\includegraphics[width=\linewidth]{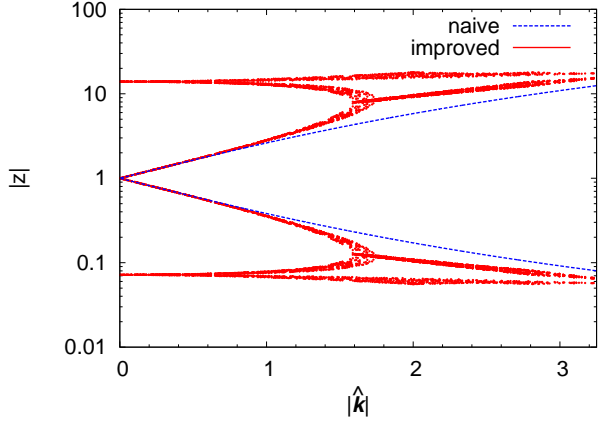}
\caption{Absolute value of poles in the na\"{\i}ve and Symanzik-improved gluon
  propagator as a function of $|\bs{\hat k}| = 2 \sqrt{\sum_{j=1}^3
    \sin^2(k_j/2)}$.}
\label{fig:gluon_abs_poles}
\end{figure}

To compare the poles in the improved propagator to the na\"{\i}ve poles,
their absolute value
is computed and it is compared to that of the na\"{\i}ve poles given by
\begin{eqnarray}
  z_{\pm}^{(\text{na\"{\i}ve})} &=& \frac{1}{2}\Big(
    2 + \bs{\hat k}^2 + \lambda^2 \\
    && \pm \sqrt{
      (\bs{\hat k}^2+\lambda^2)(\bs{\hat k}^2+\lambda^2+4) 
    } 
  \Big)\notag \,.
\end{eqnarray}
In Fig.~\ref{fig:gluon_abs_poles} these absolute values are plotted for
the same random three momenta. As can be seen from this plot
the absolute value of an improved pole is either larger than
$z_+^{(\text{na\"{\i}ve})}$ or smaller than $z_-^{(\text{na\"{\i}ve})}$
but it never lies between these values.

We performed a similar analysis for the propagator in Coulomb gauge and find
that also in this case the poles of the Symanzik-improved propagator always
lie outside the band defined by
$z_-^{(\text{na\"{\i}ve})} < |z| < z_+^{(\text{na\"{\i}ve})}$.

Hence, it is legitimate to use the position of the na\"{\i}ve
poles when deforming the integration contour in the determination of the
heavy quark renormalization parameters.

\bibliographystyle{h-physrev4}
\bibliography{hpqcd_mnrqcd}

\end{document}